\documentclass[aps,prb,twocolumn,groupedaddress,draft,showpacs,intlimits,amsmath,amssymb,floatfix]{revtex4}
\newcommand{\bk}{\textbf{k}}
\newcommand{\bq}{\textbf{q}}
\newcommand{\bp}{\textbf{p}}
\newcommand{\bn}{\textbf{n}}
\newcommand{\br}{\textbf{r}}
\usepackage{bm}
\usepackage{color}
\usepackage[final]{graphicx}
\usepackage{epsfig}
\begin{document}

\title{A Systematic Study of Electron-Phonon Coupling to Oxygen Modes Across the Cuprates}

\author{S. Johnston$^{1}$}
\author{F. Vernay$^2$}
\author{B. Moritz$^{3,4}$}
\author{Z.-X. Shen$^{3,5,6}$}
\author{N. Nagaosa$^{7,8}$}
\author{J. Zaanen$^9$}
\author{T. P. Devereaux$^{3,5}$}
%\affiliation{$^1$Department of Physics and Astronomy, University of Waterloo, Waterloo, Ontario, N2L 3G1, Canada.}
\affiliation{$^1$IFW Dresden, P.O. Box 27 01 16, D-01171 Dresden, Germany}
\affiliation{$^2$LAMPS, Universite de Perpignan Via Domitia, 66860 Perpignan Cedex, France}
\affiliation{$^3$Stanford Institute for Materials and Energy Science, SLAC National Accelerator Laboratory and 
Stanford University, Stanford, CA 94305, USA}
\affiliation{$^4$Department of Physics and Astrophysics, University of North Dakota, Grand Forks, ND 58202, USA}
\affiliation{$^5$Geballe Laboratory for Advanced Materials, Stanford University, Stanford, CA 94305, USA}
\affiliation{$^6$Department of Physics and Applied Physics, Stanford University, CA 94305, USA}
\affiliation{$^7$Department of Applied Physics, University of Tokyo, Bunkyo-ku, Tokyo 113-8656, Japan}
\affiliation{$^8$Cross-Correlated Materials Research Group (CMRG) and Correlated Electron 
Research Group (CERG), RIKEN-ASI, Wako 351-0198, Japan}
\affiliation{$^9$Leiden Institute of Physics, Leiden University, 2333CA Leiden, The Netherlands}

\date{\today}

\begin{abstract}
The large variations of T$_c$ across the
cuprate families is one of the major unsolved puzzles in condensed matter
physics, and is poorly understood. Although there appears to be a great deal of universality in the 
cuprates, several orders of magnitude changes in T$_c$ can be achieved through changes in the chemical composition and 
structure of the unit cell.  In this paper we formulate 
a systematic examination of the variations in electron-phonon coupling to oxygen phonons in the
cuprates, incorporating a number of effects arising from several aspects of
chemical composition and doping across cuprate families. 
It is argued that the electron-phonon coupling is a very sensitive probe of 
the material-dependent variations of chemical structure, 
affecting the orbital character of the band crossing the Fermi level, 
the strength of local electric fields arising from structural-induced symmetry breaking, 
doping dependent changes in the underlying band structure, and ionicity of the 
crystal governing the ability of the material to screen $c$-axis
perturbations. Using electrostatic Ewald calculations and known experimental structural data, we  
establish a connection between the material's maximal T$_c$ at optimal doping  
and the strength of coupling to $c$-axis modes.  We demonstrate that 
materials with the largest coupling to the out-of-phase bond-buckling (``$B_{1g}$") oxygen phonon 
branch also have the largest T$_c$'s.  In light of this observation we present model 
T$_c$ calculations using a two-well 
model where phonons work in conjunction with a dominant pairing interaction, presumably due to spin fluctuations, indicating how 
phonons can generate sizeable enhancements to T$_c$ 
despite the relatively small coupling strengths. Combined, these results can provide a 
natural framework for understanding the doping and material dependence of T$_c$ across the 
cuprates.  
\end{abstract}

%possible packs 
%1 - cuprate superconductivity 74.72.-h
%2 - electron-phonon interactions, electronic structure of solids, 71.38.-k
%3 - 71.38.-k - polarons & el-ph interactions.
%4 - 74.72.Gh - hole doped cuprates
\pacs{74.72.Gh, 71.38.-k, 74.78.-w} \maketitle
\section{Introduction}
Due to the extensive studies on the physical properties of the cuprates, many constraints 
on the pairing mechanism of their high-temperature superconductivity (HTSC) have been 
accumulated.  There is no doubt that the strong Coulomb interaction and the resultant strong 
electron correlations play crucial roles.  This effect is believed to be described by single-band 
Hubbard or $t$-$J$ models in 2D, and the magnetic mechanism 
for superconductivity has been proposed with the focus 
on the short range antiferromagnetism or spin singlet formation.\cite{ScalapinoBook}  These models 
have achieved great success in explaining many of the physical properties, such as the pseudogap, 
generalized magnetic susceptibility observed by neutron scattering, and the single particle Green's 
function found in angle-resolved photoemission spectra (ARPES).  However, these models are not successful 
in explaining the variation of in superconducting transition temperature T$_c$ from material to material 
and other material dependent properties.   
For example, the famous T-linear resistivity within the plane is universally observed among various 
cuprates, while T$_c$'s differ by two orders of magnitude.\cite{Resistivity,KomiyaPRB2002}  
It is surprising that the resistivity, 
which is one of the most representative physical observables of the electronic states in solids, is 
irrelevant to T$_c$.  The 2D Hubbard and $t$-$J$ models contain only a few parameters, such as hopping 
parameters $t$, $t^\prime$, $t^{\prime\prime}$ and interactions $U$ and $J$.  One possibility is 
that the range of the hopping and magnitude of $t^\prime$ and $t^{\prime\prime}$ are key factors determining 
T$_c$, which is determined by the structure and chemical composition perpendicular to the CuO$_2$ 
plane.\cite{Pavarini}  However, studies on the $t$-$J$ model have found 
that finite $t^\prime$
suppresses superconducting correlations.\cite{Spanu}  Recent investigations of 
the single-band Hubbard model using cluster 
dynamical mean field theory calculations\cite{KhatamiPRB2008}
do not show increased 
tendencies towards pairing for larger $t^\prime$ but variational studies do,\cite{ShihPRL2004} although 
the latter may be less controlled.  

From a structural point of view, the only known empirical rule of T$_c$ is that it increases at optimal doping 
as the number of CuO$_2$ layers $n$ is increased for $n < 3$.  Anderson noticed this $n$-dependence at a 
early stage, and proposed the interlayer mechanism of superconductivity.\cite{AndersonScience1995}  The idea 
is that the single-particle interlayer hopping is suppressed by strong correlations within the layer, 
while the two-particle hopping is not.  The onset of the latter below T$_c$ leads to the condensation energy 
of superconductivity. Experimentally it is found that the $c$-axis electron hopping is actually 
suppressed and there is no coherent band formation perpendicular to the plane.\cite{KomiyaPRB2002}
There is no plasmon 
observed in the normal state, and below T$_c$ the Josephson plasmon appears in the low energy 
region ($\sim 10$ meV).\cite{vanderMarel}   
The idea of an interlayer mechanism has also been criticized in 
light of the $c$-axis oscillator strength, and has subsequently been abandoned.\cite{Tsvetkov}

Increasing evidence for the importance of out-of-plane effects continues to accrue, which points 
to limitations of intrinsic planar models for the cuprates.  For example, the correlation of 
out-of-phase oxygen dopant ions in Bi$_2$Sr$_2$CaCu$_2$O$_{8+\delta}$ (Bi-2212) with features in the 
tunneling density of states \cite{McElroyScience2005} has been interpreted in terms of a local 
increase of the superconducting pair potential.\cite{Nunner}  In addition, the rapid suppression of T$_c$ with 
out-of-plane cation dopants compared to in-plane dopants is surprising given that the former 
do not appreciably affect in-plane resistivities.\cite{Fujita, GaoPRB2009}  
T$_c$ has also been empirically 
correlated with the Madelung energy difference between apical and planar oxygen atoms.\cite{OhtaPRB1991} 
This has been recently supported by studies on Ba$_2$Ca$_3$Cu$_4$O$_8$F$_2$, a compound which has 
vastly different transition temperatures by exchanging F with O at the apical site.\cite{IyoPhysicaC2003}
ARPES studies have inferred a pairing gap which is a factor of 2 larger on the bonding band in comparison 
to the anti-bonding band,\cite{ChenPRL2006} a trend consistent with other cuprates.\cite{DamascelliRMP}
Since the Fermi surface of the bonding band lies far away from either the antiferromagnetic 
reciprocal lattice zone boundary, or the van Hove points, linking the pairing mechanism with a 
purely electronic mechanism is not straightforward.  

\begin{figure}[t]
 \includegraphics[width=0.8\columnwidth]{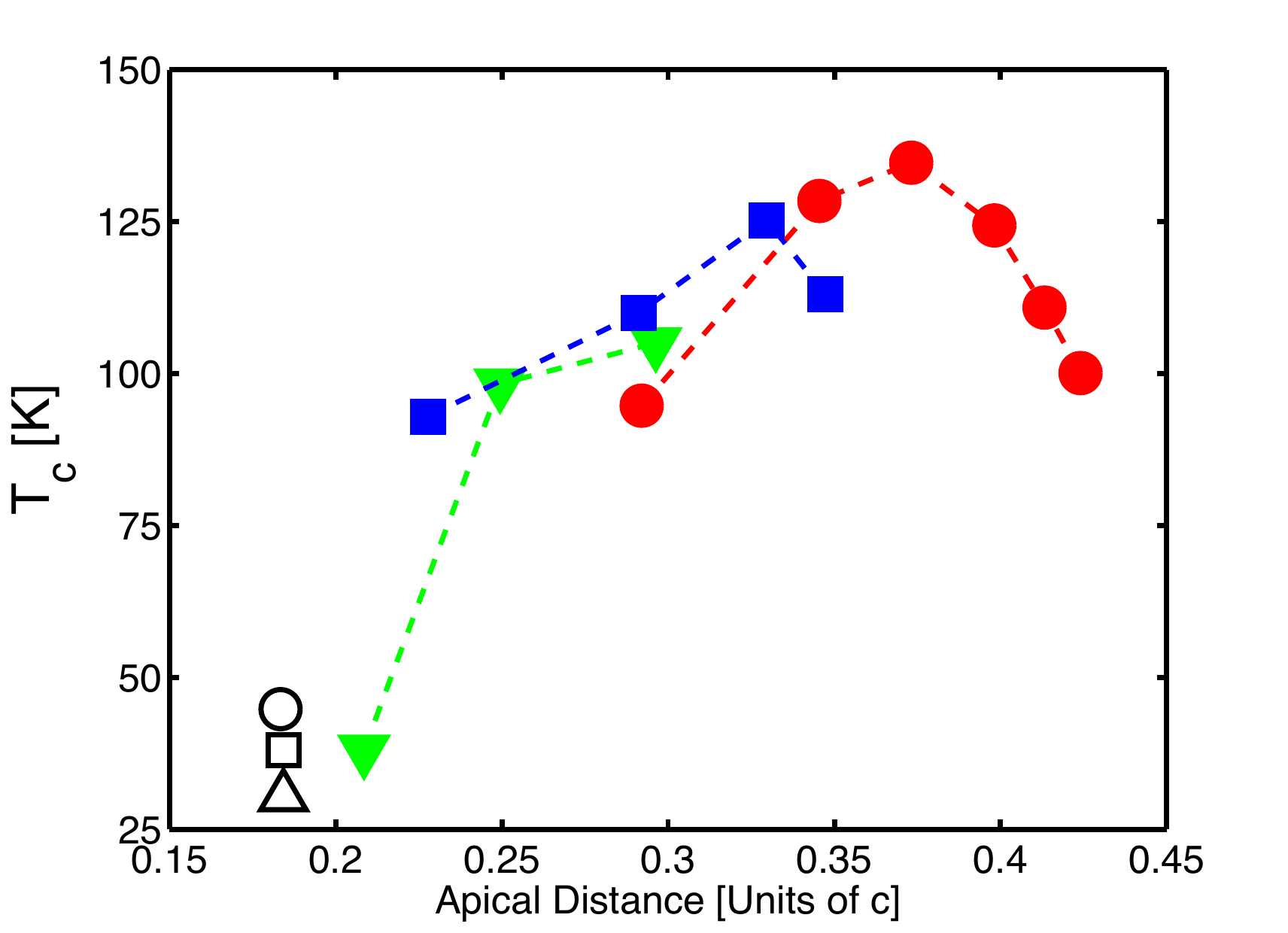}
 \caption{\label{Fig:Tc_vs_d}(Color online) The T$_c$ at optimal doping for many cuprate superconductors plotted 
 as a function of the distance between the apical oxygen site and the mirror plane located 
 at the center of the unit cell.  The Hg- (red) circles, Tl- (blue) squares and Bi-families 
 (green) triangles are shown as well as some LSCO and LBCO systems open (black).} 
\end{figure}

As discussed in Refs. \onlinecite{Pavarini} and \onlinecite{OhtaPRB1991}, the Madelung energy 
difference and $t^\prime$ are directly linked.  Pavarini {\it et al.} pointed out that the maximal 
T$_c$ in each family of cuprate materials scales with the next nearest hopping $t^\prime$.\cite{Pavarini} 
The energy of the Cu 4$s$ orbital relative to the $pd$-$\sigma^*$ band largely determines $t^\prime$ as the 
hybridization with the planar oxygen orbitals allows electrons to more effectively hop between 
2$p_{x,y}$ orbitals.  Strong apical 2$p_z$-4$s$ hybridization, determined by the Madelung energy 
difference, raises the energy of the 4$s$ orbital relative to the $pd$ band, reducing the 
effective hopping $t^\prime$.  Thus, the further away the apical oxygen is located from the CuO$_2$ plane 
the larger $t^\prime$.\cite{Pavarini}  
Similar empirical relations between T$_c$ and structural details can also be found.  
For example, 
there is an optimal distance between the apical oxygen site and the mirror plane of the unit cell 
for which T$_c$ takes on its maximum value, as shown in Fig. \ref{Fig:Tc_vs_d}. 
However, despite these observations, a direct connection between changes in the bandstructure and 
the pairing mechanism is lacking and to date, the connection between T$_c$ and 
these $c$-axis effects remains an empirical intrigue lacking a firm microscopic 
understanding.   

A phonon mechanism has also been pursued since the discovery of HTSC.\cite{Alexandrov}  The original 
intuition by Bednorz and M{\"u}ller was that the Jahn-Teller (JT) effect with high-frequency 
oxygen phonon modes leads to HTSC.\cite{BednorzMuller}  However, the degeneracy of the two $e_g$ orbitals is lifted 
considerably ($\sim 1$ eV) and the JT effect is now believed to be ineffective.  Furthermore, there are 
several experiments suggesting a minor role played by phonons.  These include the small isotope 
effect on T$_c$ at optimal doping, \cite{isotope}  
the absence of the phonon effect on the temperature dependence of the 
resistivity,\cite{Resistivity} and the absence of the phonon bottle-neck 
effect.\cite{Perfetti} This conclusion was also supported by 
early density functional calculations which did not give appreciably high values of
couplings and did not support a high T$_c$ in a BCS picture.\cite{SavrasovPRL1996,AndersenJLTP1996}

The high-T$_c$ cuprates, as strongly correlated systems, have attracted a great deal of
theoretical interest.  
\cite{SakaiPRB1997, CaponePRL2004, 
AlderPRL1997,SangiovanniPRL2005,SangiovanniPRL2006,WernerPRL2007,
RoschPRL2004,RoschPRB2004,BoncaPRB2008,HankePRB2003,ZeyherPRB1996,
IshiharaPRB2004,KochPRB2004,CitroPRB2005,MacridinPRL2006,AnisimovPRL1992,ZaanenPRB1994} 
Part of this interest has been focused on examining the renormalization of electron-phonon (el-ph)
interactions by strong electron-electron (el-el) correlations.
\cite{SakaiPRB1997, CaponePRL2004, 
AlderPRL1997,SangiovanniPRL2005,SangiovanniPRL2006,WernerPRL2007,
RoschPRL2004,RoschPRB2004,BoncaPRB2008,HankePRB2003,ZeyherPRB1996,
IshiharaPRB2004,KochPRB2004,CitroPRB2005,MacridinPRL2006} 
 For example, studies of the $t-J$ model incorporating an el-ph
interaction indicate that polaron crossover may occur at a weaker el-ph coupling strength than in the case
of a pure el-ph coupling model.~\cite{RoschPRL2004,RoschPRB2004,BoncaPRB2008}  Quantum Monte Carlo treatments
of the single-band Hubbard-Holstein model have shown that the renormalized el-ph vertex develops a strong forward
scattering peak with no substantial suppression of the el-ph vertex when the Hubbard U is large~\cite{HankePRB2003}
with similar results obtained for the $t$-$J$ model.~\cite{ZeyherPRB1996,IshiharaPRB2004} However, using
slave-boson approaches, the Hubbard model plus el-ph interaction was found to possess no significant forward
scattering peak together with an overall suppression of the el-ph vertex at low temperature.~\cite{KochPRB2004}
This same study observed an enhancement in the el-ph vertex for small $\bq$ scattering at high temperature,
which was linked to phase separation.  Cumulant expansion techniques~\cite{CitroPRB2005} have 
also found an enhancement of the el-ph vertex for small $\bq$, again interpreted in terms of 
incipient phase separation when approaching a critical value of $U$.  

The single-band Hubbard-Holstein model has also been studied within dynamical mean field theory (DMFT). 
In one study\cite{CaponePRL2004} the effect of the el-ph interaction was to stabilize the 
insulating state in the vicinity of the density-driven Mott transition.   
In a paramagnetic DMFT study \cite{SangiovanniPRL2005} the el-ph interaction was found to have 
little effect on the low energy physics produced by the Hubbard interaction while modifying the spectral 
weight associated with the upper and lower Hubbard bands at higher energy.  
From these results it was concluded that the primary effect of the el-ph interaction was to reduce 
the effective value of $U$.
Later work\cite{SangiovanniPRL2006} considered antiferromagnetic solutions in the presence of el-ph coupling 
and observed different behavior with strong polaronic effects.  In this case, the critical coupling for 
polaron formation was observed to shift to larger values as the system was doped away from half-filling.
The dichotomy between the paramagnetic and antiferromagnetic treatments\cite{SangiovanniPRL2005, SangiovanniPRL2006} 
indicates the possible importance of the magnetic order in considering the el-ph interaction in correlated systems.  
Finally, a DMFT study invoking a Lang-Firsov transformation for the lattice degrees of 
freedom has found evidence for a competition between the Mott insulating (metallic) and bipolaronic 
insulating phases near (away from) half-filling.\cite{WernerPRL2007}  

Using the dynamical cluster approximation, an extension of DMFT, examinations of the el-ph interaction within small Hubbard
clusters~\cite{MacridinPRL2006} find an overall suppression of $d$-wave superconducting T$_c$ with increasing
el-ph coupling.  This occurs despite an increase in the apparent pairing correlations within the $d_{x^2-y^2}$
channel.  The reduction in T$_c$ was attributed to polaron formation, which reduces quasi-particle weight at the
Fermi level and suppresses T$_c$ through the loss of carrier mobility.  The enhancement of pairing correlations
reported in Ref.~\onlinecite{MacridinPRL2006} indicates that the bare el-ph vertex has been renormalized in
favour of $d$-wave pairing, consistent with the observations of Ref.~\onlinecite{HankePRB2003}.  
Furthermore, exact diagonalization (ED) studies on the $t$-$J$ model, 
which include el-ph coupling to buckling and breathing vibrations,   
also show that the former enhance $d$-wave pairing while the latter suppress it.\cite{SakaiPRB1997}  
While these results provide no definitive interpretation of the effect of el-ph coupling in strongly correlated systems, there
is strong evidence that whatever impact strong correlations may have, the el-ph interaction may still play a
significant role that should not be overlooked in these systems.

From an experimental front, the role of phonons in HTSC has become more 
prominent in recent years.  While long-studied from Raman, infrared, 
and neutron measurements, recent ARPES and scanning tunneling microscopy (STM) 
measurements on several cuprates has reinvigorated the exploration of the role of phonons on HTSC.\cite{CukPSS2004}   
Experiments on Bi-2212,\cite{LanzaraNature2001,CukPRL2004,LeeNature2006,LeePRB2008} 
La$_{2-x}$Sr$_x$CuO$_4$ (LSCO),\cite{ZhouPRL2005} Bi$_2$Sr$_2$CuO$_{6+\delta}$ (Bi-2201),
\cite{MeevasanaPRL2006} Ba$_2$Ca$_3$Cu$_4$O$_8$F$_2$,\cite{YChenPRL2009} and Tl families
Tl$_2$Ba$_2$CaCu$_2$O$_8$ (Tl-2212), TlBa$_2$Ca$_2$Cu$_3$O$_9$ (Tl-1223),
and Tl$_2$Ba$_2$CuO$_6$ (Tl-2201)\cite{WSLeePRL2009} 
have revealed kinks in the energy dispersion of these materials.  These 
kinks have been interpreted as Hubbard renormalizations,\cite{Byczuk} coupling to the neutron 
resonance and/or spin continuum,\cite{Eschrig} and Engelsberg-Schrieffer renormalizations\cite{Engelsberg} 
due to coupling of electrons to a collection of optical phonons.  These phonons include the out-of-phase 
$c$-axis oxygen buckling modes and the in-plane Cu-O bond-stretching modes.  
The dispersion kink observed in the nodal region, ($0,0$) - ($\pi/a,\pi/a$), and the 
peak-dip-hump structure observed in the anti-nodal region, $(0,\pi/a)$-($\pi/a,\pi/a$),  
clearly shows 
that the electrons are interacting with bosons of a well defined energy $\sim 70$ and 
$\sim 36$ meV, respectively. For the nodal region, 
it has been convincingly argued that this structure is due to the oxygen bond-stretching phonon 
as the kink is observed independent of superconductivity and of the presence/absence of 
the spin resonance peak.  As for the anti-nodal region, it has been claimed that the kink 
appears only below T$_c$, and hence it is attributed to the spin resonance mode at 41 meV.\cite{Spin}  
However, an extensive study conducted more recently has found the evidence for the kink structure in the 
normal state and over a wide range of momentum space.\cite{CukPRL2004}  Further, 
contrasting single and multilayer cuprate ``kinks'', and materials known to have 
a neutron resonance, also indicates that the observed renormalizations are most
likely due to optical phonons, although this is still controversial.\cite{LeePRB2007,WSLeePRL2009,JohnstonCondmat}
  
It is well known that the $c$-axis phonons show some of the most dramatic lineshape changes 
with doping and temperature compared to any phonons observed via neutron\cite{Pintschovius} and 
Raman\cite{Krantz} scattering.  For example, the apical phonon frequency shifts by as much as 
20 cm$^{-1}$ with doping and temperature in a number of compounds: 
La$_{2-x}$Sr$_x$CuO$_{4}$, HgBa$_2$Ca$_{n-1}$Cu$_n$O$_{4n+\delta}$ ($n = 1-4$) and 
Bi$_2$Sr$_2$Ca$_2$Cu$_3$O$_{10+\delta}$.\cite{Pintschovius, Krantz} Moreover, recent ARPES data on Bi-2201 have shown 
kinks in the energy range of the $c$-axis phonons which are weaker in overdoped compounds in  
comparison to optimal doped compounds.\cite{MeevasanaPRL2006}  This has also been interpreted 
in terms of increased screening of the el-ph interaction with increasing hole concentrations.  
In addition, the anomaly of the Raman $A_{1g}$-polarized mode due to the onset of 
superconductivity is observed in three- and four-layer compounds.\cite{MunzarPRL2003}  
This has been successfully analyzed in terms of the internal electric field produced by the 
interlayer Josephson plasmon and its coupling to the phonon.
This means that the system behaves 
as an ionic crystal along the $c$-axis in the normal state, and suddenly turns into a superconductor.   

While the role of the neutron resonance and phonons remains controversial, it is of relevant 
interest whether these signatures in ARPES may be used as an angle-resolved 
analogy to the tunnelling ripples in conventional superconductors,\cite{McMillan} thus providing   
information on the pairing mechanism in the cuprates.   
In order to connect el-ph coupling to a possible pairing mechanism a systematic study of coupling 
across families of cuprate materials is desirable.  In Ref. \onlinecite{tpdPRL2004} ARPES 
observed renormalizations of the band were interpreted as due to the $B_{1g}$ branch for anti-nodal 
electrons and the bond-stretching branch for nodal electrons.  While the latter coupling is of a deformation 
type, the coupling constructed for the $B_{1g}$ branch involves a charge-transfer between planar oxygens 
due to a modulation of the electrostatic or Madelung energies of the planar oxygen sites.  A local 
crystal field, generated by a mirror plane symmetry breaking, allows for a coupling at first order 
in atomic displacements.\cite{tpdPRB1995}  (We note here that the $A_{1g}$/$B_{1g}$ nomenclature 
only holds for Raman $\bq = 0$ momentum transfers.  However, throughout this work we denote 
the entire out-of-phase branch as ``$B_{1g}$" and the entire in-phase branch as ``$A_{1g}"$.)  
Since the cuprates are poor conductors along the $c$-axis, 
the electrostatic interaction can be thought to be largely unscreened.  Calculations based on Ewald's 
method have been performed on YBa$_2$Cu$_3$O$_7$ (YBCO)\cite{LiSSC1995} and large crystal fields have 
been obtained and the resulting coupling matches well with the coupling determined from Fano 
lineshape analysis of Raman data.\cite{tpdPRB1995}

Recently, the issue of whether the el-ph coupling in the cuprates is strong enough to explain 
the observed band renormalizations has been revisited via density functional (LDA) calculations,
\cite{BohnenEPL2003,HeidPRL2008,GiustinoNature2008} updating previous estimates.\cite{SavrasovPRL1996,
AndersenJLTP1996}  While many efforts have been made to extract bosonic
coupling from ARPES renormalizations in the cuprates, there is no widely
accepted way to uniquely determine the strength of the coupling at kink
energies, and thus reliable comparisons of calculations with experiment must
be viewed with some caution.\cite{Damascelli2010} While LDA calculations have
provided remarkably good agreement with phonon dispersions, there are a number of
facets of LDA calculations which may only provide part of the story of
el-ph coupling. As LDA calculations overestimate 
the itinerancy of the electrons, they describe the cuprates as good metals
even at half-filling. Moreover, the obtained interlayer transfer 
integral and hence the $c$-axis plasmon frequency is 
both coherent and much larger than the experimental observation.  
Both of these factors serve to overestimate the 
screening ability of the cuprates, especially in the underdoped region, and
thus underestimates the strength of the el-ph interaction.  
This may be one of the reasons why 
LDA predicts smaller linewidths for the half-breathing oxygen bond-stretching modes and the 
apical oxygen modes, sometimes by more than one order of magnitude.\cite{Reznik}  
It is therefore not clear if these findings indicate that the el-ph 
coupling is small or that DFT-based approaches alone 
are inadequate for describing the physics of the cuprates.  
As has been found in STM experiments,\cite{McElroyScience2005} a nano-scale 
inhomogeneous structure on a length scale of 
15 $\AA$ exists universally in Bi-2212 and YBCO.  This length scale cannot be larger than the 
screening length and we can conclude that the screening length within the CuO$_2$ plane is not shorter  
than 15 $\AA$, much longer than the Thomas-Fermi screening length of the typical metal ($\sim 0.5$ \AA).  
Furthermore, charge transfer between the layers is almost prohibited. As a
result of the 
transfer integral between the layer is proportional to $[\cos(k_xa)-\cos(k_ya)]^2$,  
the opening of the pseudogap in the $(\pi,0)$ and $(0,\pi)$ regions strongly suppresses the interlayer hopping.  
Considering these discrepancies between LDA and experiments, one can imagine that the el-ph coupling 
is in reality much stronger than LDA predicts. Focusing on the buckling modes, theoretical 
consideration have been limited to the two-dimensional plane and the inter-layer Coulomb 
interaction has been neglected.  This is usually justified in the metal since the screening length 
is much shorter than the interlayer distance.  This is not the case in the cuprates.  

In this 
paper we provide a comprehensive and self-contained story on el-ph coupling to oxygen phonons 
as a function of doping across the cuprate families.
We formulate a theory for el-ph coupling in the cuprates taking into account 
the local environment around the CuO$_2$ planes, the poor screening of charge fluctuations out of the 
plane, doping-dependent band character variations, and structural differences across the cuprate
superconductors.  

The organization of this paper is as follows.  In section II we present a general 
discussion of el-ph coupling to $c$-axis oxygen phonons.  After summarizing prior 
work on the out-of-plane planar oxygen modes, we then provide a derivation of the 
coupling to modes involving $c$-axis apical oxygen motion.  
In section III we then discuss the anisotropy of the bare couplings 
and examine the total strength of the bare el-ph coupling and its  
contribution to the single-particle self-energy 
and $d$-wave anomalous self-energy.  

In section IV we develop the formalism for poor screening and 
examine its implications for the anisotropy and overall magnitude of 
the renormalized el-ph vertices.  Due to the poor $c$-axis conductivity  
we find that the $c$-axis phonons cannot be effectively screened for small 
in-plane momentum transfer $\bq_{2D}$. This effect becomes more pronounced as the 
crystal becomes more ionic and screening becomes increasingly inoperable 
in the underdoped side of the phase diagram.   
However, in the case of the $B_{1g}$ modes 
the coupling is anomalously anti-screened producing an enhancement of the 
coupling in the anti-nodal region.  In terms the projected $d$-wave couplings, the 
small $\bq_{2D}$ behavior of the screened vertices produces an enhancement in the 
total phonon contribution to pairing.  Therefore, the combined effects of poor 
screening reduces the total el-ph coupling and enhances the $d$-wave projected 
coupling with doping.  This has important implications for the doping 
dependence of the el-ph self-energies probed by ARPES as well as any contribution 
to pairing mediated by phonons.  

In section V we turn to materials trends and a systematic examination of the Madelung 
potential and crystal field strengths across the Bi-, Tl- and Hg families of cuprates 
is presented.   
Here, using an ionic point charge model and the Ewald summation technique, we 
identify systematic trends in the strength of the crystal fields which mirror 
trends in the material's T$_c$ at optimal doping.  Through this 
observation we link the structure and chemical composition to the 
strength of the coupling to the $c$-axis modes and discuss 
how this can be used to understand the large variations in T$_c$ 
observed across the cuprates.  We also present considerations for doping-induced 
changes to the value of the crystal field in Bi-2212.  

In light of these findings, section VI presents a 
simple two-channel model for pairing in the cuprates, which includes  
a dominant, $d$-wave pairing, high-energy bosonic mode and a weaker phonon mode.  
Using this model, we demonstrate that phonons can provide a sizeable 
enhancement to T$_c$ which is in excess of the T$_c$ that would be obtained from phonons alone.  
Furthermore, due to the dominant bosonic mode, the resulting value of the isotope exponent $\alpha$ 
is small ($\alpha < 0.15$) despite the large enhancement of T$_c$ ($\sim 40$ K).  This 
calculation, in combination with the materials and doping dependent trends 
identified in the previous sections, shows that a phonon assisted pairing model 
provides a natural framework for understanding trends observed for T$_c$ across 
the cuprates. Finally, in section VII we conclude by summarizing our findings and 
discuss open questions concerning el-ph coupling in strongly correlated systems.  

In addition to the treatment outline above, in the appendix we explore how el-ph 
coupling to $c$-axis modes is modified by strong correlations in the half-filled 
parent insulators using exact diagonalization of small multi-band Hubbard clusters.  
Specifically we address how el-ph coupling modifies the properties of the Zhang-Rice 
singlet (ZRS).  Here we find that static lattice displacements have a strong 
influence on the ZRS hoppings, energy and antiferromagnetic exchange energy J.  
These results have a direct impact on the use of down-folded models such as the 
$t$-$J$ model as they indicate that the effects of the el-ph coupling cannot be 
simply cast as modulations of a single parameter such as $t$, $J$, or the energy of the ZRS.  

\section{General Electron-Phonon Considerations in the Cuprates}\label{SectionTwo}
In this section we present a review of some generic considerations for
electrons coupling to oxygen motions in- and out- of the CuO$_2$ plane. Since
many of the derivations have appeared before, we can be brief, with the main
aim to generalize previous results to a five band model in order to include off-axis 
orbitals and apical oxygen phonon modes.  

We begin by considering an ideal CuO$_2$ plane isolated from its environment.  Since hopping 
integrals are modulated to second order in atomic displacements along the $c$-axis, 
the el-ph coupling due to this mechanism is weak.  However, if the same plane is placed 
in an asymmetric electrostatic environment a local crystal field, which breaks 
mirror plane symmetry, provides a coupling linear in displacement.  The plane must then 
spontaneously buckle in a pattern where the oxygen (copper) atoms are displaced away from 
(towards) the mirror plane.

In a three band model,\cite{tpdPRB1995} the local field coupling was used to construct a charge-transfer 
el-ph vertex for coupling to the Raman active out-of-phase and in-phase $c$-axis 
oxygen vibrations.  The in-phase phonon modulates charge transfer between planar oxygen and 
copper orbitals while the out-of-phase phonon modulates charge transfer between only the 
planar oxygen orbitals.  A single-band el-ph coupling was obtained:
\begin{equation}\label{Eq:interaction}
H_{el-ph} = \frac{1}{\sqrt{N}}\sum_{\bk,\bq,\sigma}|g(\bk,\bq)|^2 
c^\dagger_{\bk-\bq,\sigma}c_{\bk,\sigma}\left[b^\dagger_{\bq} + b_{-\bq} \right], 
\end{equation} 
where $c^\dagger_{\bk,\sigma}$ ($c_{\bk,\sigma}$) creates (annihilates) an electron in the 
partially filled antibonding band with momentum $\bk$, energy $\epsilon_\bk$, and 
spin $\sigma$, and $b^\dagger_\bq$ ($b_\bq$) creates (annihilates) a phonon of 
energy $\Omega_\bq$ and wavevector $\bq$.  

Considering a modulation of the electrostatic coupling of the charge density at the oxygen 
sites coupled to the local on-site potential $\Phi_{ext}$, the Hamiltonian is of the form 
\cite{tpdPRB1995}
\begin{equation}
H^\prime_{site} = -e\sum_{{\bf n},\sigma} p^\dagger_{{\bf n},\sigma} p_{{\bf n}, \sigma}
\Phi_{ext}[{\bf u}(a{\bf n})], 
\end{equation}
where ${\bf u}(a{\bf n})$ is the oxygen displacement vector in the unit cell at lattice site 
{\bf n}, $e$ is the electron charge and $p^\dagger_{\bf{n},\sigma}$ ($p_{\bf{n},\sigma}$) 
is the creates (annihilates) an electron at site ${\bf n}$, which can include both planar and 
apical oxygen orbitals.  This coupling mechanism differs from the deformation coupling considered in Ref. 
\onlinecite{AndersenJLTP1996}.  Expanding for small displacements $H^\prime_{site}  = 
H_{site} + H_{el-ph} + O(u^2)$ where $H_{site}$ includes the Madelung contribution 
to the site energies and the term linear in $\bf{u}$ generates the el-ph interaction
\begin{equation}\label{Eq:el-ph}
H_{el-ph} = -e\sum_{\bf{n},\sigma} p^\dagger_{\bf{n},\sigma}p_{\bf{n},\sigma}
{\bf E}_{\bf{n}}\cdot {\bf u}(a{\bf n}).
\end{equation}
$\bf{E}_{\bf{n}}$ is the local crystal field at the oxygen site provided this field is finite, 
which occurs at locations of broken mirror symmetry in the unit cell.  

In order to derive the form of the coupling $g(\bk,\bq)$ Eq. (\ref{Eq:el-ph}) must be 
rewritten in the form of Eq. (\ref{Eq:interaction}). To do so, the el-ph Hamiltonian is 
Fourier transformed to momentum space and the oxygen operators are replaced by 
band representation operators $p_{\bk,\delta,\sigma} = \phi_\delta(\bk)c_{\bk,\sigma}$.  Here, 
$\phi_\delta(\bk)$ is the oxygen ($\delta = x,y$ for planar oxygen and $\delta = a$ for apical oxygen)  
eigenfunction for the pd-$\sigma^*$ band,  
which is obtained from a tight-binding model for the CuO$_2$ plane. 

\begin{figure}
 \includegraphics[width=0.7\columnwidth]{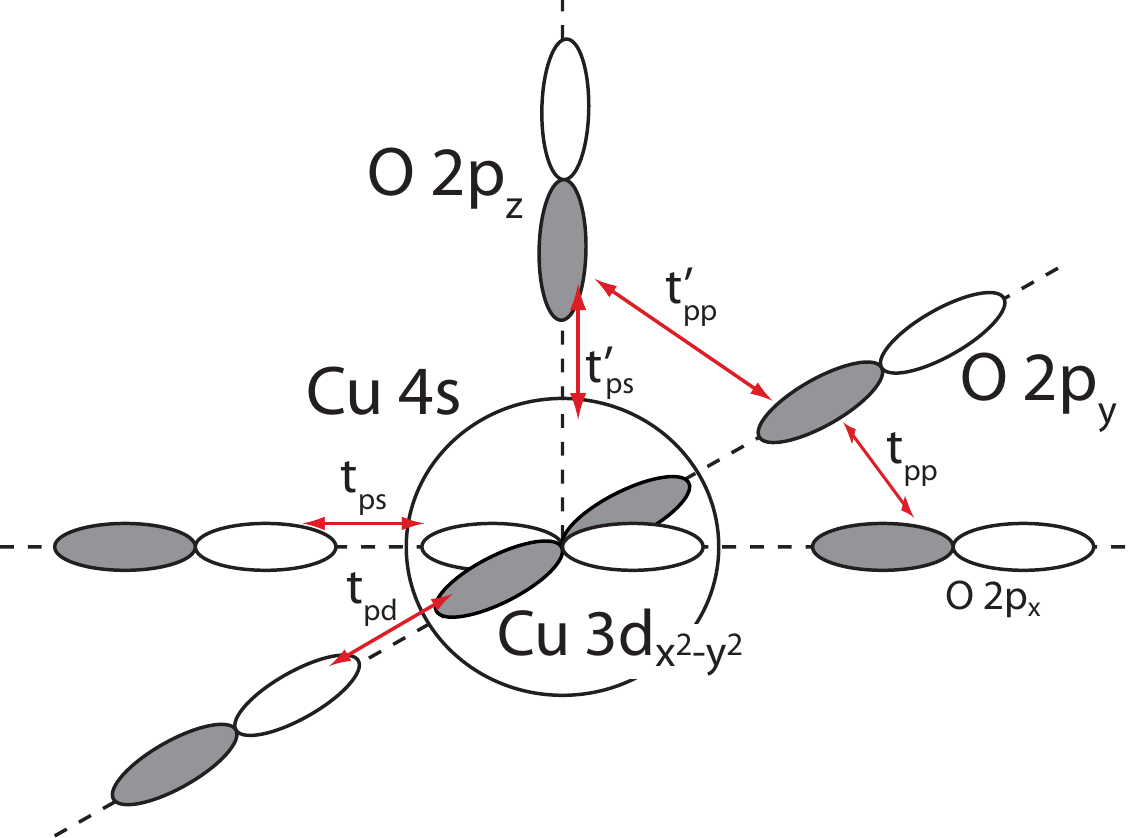}
 \caption{\label{Fig:5band} (Color online) The five-band model used to derive the form of the el-ph 
 couplings $g(\bk,\bq)$.}
\end{figure}

\subsection{Multi-band Models}
Prior work focusing on the $A_{1g}$, $B_{1g}$ and breathing branches 
made use of a three-band model.\cite{OhtaPRB1991, tpdPRB1995, tpdPRL2004}  
In order to extend these works to include the apical oxygen modes this model must be extended to 
a five-band model as shown in Fig. \ref{Fig:5band}.   
The basis set of this model contains a 4$s$ ($s_{\bn,\sigma}, s^\dagger_{\bn,\sigma}$) 
and 3$d_{x^2-y^2}$ ($d_{\bn,\sigma}$,$d^\dagger_{\bn,\sigma}$) orbital on each copper site 
$\bn$, two planar oxygen 2$p_{x,y}$ orbitals ($p_{\bn,\delta,\sigma}$,$p^\dagger_{\bn,\delta,\sigma}$) 
with $\delta = x,y$, and one apical oxygen 2$p_z$ orbital ($a_{\bn,\sigma}$,$a^\dagger_{\bn,\sigma}$).  
Here, we neglect the in-plane O 2$p$ orbitals oriented perpendicular to the Cu-O bonds.
These orbitals form weaker $pd$-$\pi$ bonds with the lower energy Cu $t_{2g}$ orbitals and do not 
contribute heavily to the character of the band crossing the fermi level.\cite{AndersenJLTP1996}   
Site energies are denoted by $\epsilon_{s,d,p,z}$, respectively.  Defining canonical fermions\cite{Shastry}
$\alpha$, $\beta$ from combinations of the planar oxygen orbitals via a Wannier transformation

\begin{equation}
\alpha_{\bk,\sigma}, \beta_{\bk,\sigma}  = \pm i \frac{s_{x,y}(\bk)p_{\bk,x,\sigma} \mp 
s_{y,x}(\bk)p_{\bk,y,\sigma}}{\mu_\bk},
\end{equation} 
where $s_{x,y} = \sin(k_{x,y}a/2)$ and $\mu^2_\bk = s_x^2(\bk) + s^2_y(\bk)$, the Hamiltonian can be 
written as $H = \sum_{\bk,\sigma} H_{\bk,\sigma}$:

\begin{eqnarray}\label{Eq:5band}\nonumber
H_{\bk,\sigma}&=&H_{\mathrm site} - 2t_{pp}\nu_\bk [\beta^\dagger_{\bk,\sigma}\beta_{\bk,\sigma} - 
\alpha^\dagger_{\bk,\sigma}\alpha_{\bk,\sigma} ] \\ \nonumber
&+&2\bigg[ t_{pd}\mu_{\bk,\sigma}d^\dagger_{bk,\sigma}\alpha_{\bk,\sigma}
         + t_{pp}\chi_\bk\alpha^\dagger_{\bk,\sigma}\beta_{\bk,\sigma} \\ \nonumber
&+&t_{ps}\kappa_{\bk}s^\dagger_{\bk,\sigma}\alpha_{\bk,\sigma} - t_{ps}\lambda_\bk
   s^\dagger_{\bk,\sigma}\beta_{\bk,\sigma} + \frac{t^\prime_{ps}}{2}s^\dagger_{\bk,\sigma}a_{\bk,\sigma} \\
&-&t^\prime_{pp}\kappa_\bk a^\dagger_{\bk,\sigma}\alpha_{\bk,\sigma} + t^\prime_{pp}\lambda_\bk a^\dagger_{\bk,\sigma}
  \beta_{\bk,\sigma} + h.c.
\bigg]
\end{eqnarray}

\noindent with $H_{\mathrm site}$ containing the site energies and $h.c.$ denoting  
the hermitian conjugate.  Finally, the basis functions are defined as: 
\begin{eqnarray}\label{Eq:basis}\nonumber
\nu_\bk&=&4\frac{s_x^2(\bk)s^2_y(\bk)}{\mu^2_\bk}\quad\quad
\kappa_\bk = \frac{s_x^2(\bk) - s_y^2(\bk)}{\mu_\bk} \\ 
\lambda_\bk&=&2\frac{s_x(\bk)s_y(\bk)}{\mu_\bk}   \quad\quad
\chi_\bk = \lambda_\bk\kappa_\bk.
\end{eqnarray}

In the limit where the apical and copper 4$s$ orbitals are removed ($t_{ps} = t^\prime_{ps} = 0$, etc.) 
there expressions recover prior work carried out using a three-band 
model.\cite{OhtaPRB1991, tpdPRB1995, tpdPRL2004}  

\subsection{Planar Oxygen $c$-Axis Modes}
For planar oxygen vibrations the el-ph coupling is given by ($\bp=\bk-\bq$)
\begin{eqnarray}\label{Eq:B1g} \nonumber
g_{B1g, A1g}(\bk,\bq) = eE_z\sqrt{\frac{\hbar}{2M_ON(\bq)\Omega_{B1g,A1g}}} \\ \nonumber
 \times\bigg[\phi^\dagger_x(\bk)\phi_x(\bp)e^{-iq_xa/2}(1+e^{-iq_yq})\mp \\
             \phi^\dagger_y(\bk)\phi_y(\bp)e^{-iq_ya/2}(1+e^{-iq_xa})\bigg]
\end{eqnarray}
where the minus (plus) sign is for coupling to the $B_{1g}$ ($A_{1g}$) branch.  Here 
$M_O$ denotes the oxygen mass, $E_z$ is the $c$-axis component of the local field at the planar 
oxygen site, $N^2(\bq) = 4[\cos^2(q_xa/2) + \cos^2(q_ya/2)]$ is the phonon eigenvector 
normalization, and $\Omega_{A1g, B1g}$ denote the assumed dispersionless frequencies of the 
$B_{1g}$ and $A_{1g}$ branches, respectively.  The motion of the heavier Cu atoms has been 
neglected in this treatment.  Eq. (\ref{Eq:B1g}) is a generic expression 
for the $A_{1g}$ and $B_{1g}$ vertices, independent of the underlying tightbinding model 
used to determine the band eigenfunctions $\phi_{x,y,Cu}(\bk)$.  In the three-band model, the 
band eigenfunctions are defined as\cite{tpdPRL2004}
\begin{eqnarray} \nonumber
\phi_{x,y}(\bk)&=&\mp \frac{i}{A(\bk)}[\epsilon(\bk)t_{x,y}(\bk) - t^\prime(\bk)t_{y,x}(\bk)] \\
\phi_{Cu}(\bk)&=& \frac{1}{A(\bk)}[\epsilon^2(\bk) - t^{\prime2}(\bk)], 
\end{eqnarray}
where $t_{x,y}(\bk) = t_{pd}s_{x,y}(\bk)$, $t^\prime(\bk) = -4t_{pp}\sin_x(\bk)\sin_y(\bk)$, 
the normalization is 
\begin{eqnarray} \nonumber
A^2(\bk)&=& [\epsilon^2(\bk) - t^{\prime2}(\bk)]^2 + [\epsilon(\bk)t_x(\bk) - t^\prime(\bk)t_y(\bk)]^2 \\
     &+& [\epsilon(\bk)t_y(\bk) - t^\prime(\bk)t_x(\bk)]^2,
\end{eqnarray}
and $\epsilon(\bk)$ is the bare dispersion, given in Ref. \onlinecite{tpdPRB1995}.  From Eq. 
(\ref{Eq:B1g}) it can be seen that the symmetry of the phonon is implanted into the el-ph coupling 
to provide substantial momentum anisotropy.  For the $B_{1g}$ phonon branch, $g(\bk,\bq)$ changes sign for 
$k_x,q_x \rightarrow k_y,q_y$, while for the $A_{1g}$ phonon branch it does not.  As a result 
the coupling involves all fermionic states for the $A_{1g}$ branch, while for the $B_{1g}$ 
branch the antinodal states along the Brillouin zone (BZ) axes are weighted heavily and 
nodal states along the zone diagonal are projected away.   

\subsection{Apical Oxygen $c$-Axis Modes}

Since apical phonons show some of the strongest renormalizations in La- and Hg-cuprates,\cite{Pintschovius, Krantz} 
and since the apical oxygen atoms do not lie in a mirror plane symmetry even in single layer cuprates, 
they are included as an extension of our previous work.  
Early on the apical phonon was thought to be quite anharmonic and related 
to the Jahn-Teller mechanism in YBa$_2$Cu$_3$O$_7$,\cite{Mustre} although 
many confusing results were found.\cite{Sasha}
More recently the coupling is thought to be electrostatic\cite{Zhong,RoschPRL2005} in nature and 
in some treatments weakly momentum dependent.\cite{Zhong} 
It should be emphasized that formally a Holstein (momentum independent) 
coupling in any model with long-range 
Coulomb interactions will be screened out by backflow due to charge conservation and therefore 
the coupling is expected to be very small.  This will be discussed in Section IV.  Here instead we place focus 
on a strongly momentum dependent coupling arising from 
charge transfer mechanisms between apical and planar oxygen orbitals, similar to the 
electronic pathways involved in $c$-axis tunnelling.\cite{AndersenJCP}

An apical orbital 
displacement modulates the Madelung energy as in Eq. (\ref{Eq:el-ph}) and the resulting el-ph 
coupling is of the form of Eq. (\ref{Eq:interaction}), with 
\begin{equation}\label{Eq:gapex0}
g_{apex}(\bk,\bq) = g^{apex}_0\phi^\dagger_{a}(\bk)\phi_{a}(\bk-\bq)\epsilon^z_\mathrm{a}(\bq). 
\end{equation}
Here, $\epsilon^z_a(\bq)$ is the $c$-axis component of the eigenvector for the apical branch coming 
from the atomic displacement in Eq. (\ref{Eq:el-ph}),  
%\begin{equation}
$g^{apex}_0 = eE^a_z\sqrt{\hbar/2M_O\Omega_{a}}$, and 
%\end{equation}
$\phi_a(\bk)$ is the apical eigenfunction, obtained from diagonalizing  
Eq. (\ref{Eq:5band}).  For our purpose however, we are primarily interested in the 
leading order momentum dependence of the coupling $g(\bk,\bq)$.  Starting from Eq. 
(\ref{Eq:5band}), the L{\"o}wdin down-folding procedure\cite{Lowdin} is applied to determine the 
apical character of the resulting partially filled band crossing the Fermi level.  
The resulting form for the apical eigenfunction $\phi_a(\bk)$ is then
\begin{equation*}
\phi_a(\bk) = 2t_{pz}\frac{\kappa_\bk}{\epsilon_\bk - \epsilon_a} 
\end{equation*}
with $\kappa_\bk$ defined in Eq. (\ref{Eq:basis}). 
Finally, for simplicity we neglect the momentum dependence of the apical phonon's eigenvector 
and set $\epsilon^z_\mathrm{a}(\bq) = 1$.   
More complicated models in which the apical phonon involves the motion of the in-plane 
oxygens can be treated accordingly.  

The local crystal field $E_z^a$ at the apical oxygen 
site modulates a charge transfer between the apical oxygen and the planar orbitals.  This 
mechanism of charge transfer is analogous to the charge transfer mechanism yielding bi-layer 
splitting.\cite{AndersenJCP}  From Eqs. (\ref{Eq:5band}) and (\ref{Eq:gapex0}), the resulting 
momentum dependence of the coupling via this transfer, $g(\bk,\bq) \sim [\cos(k_xa) - \cos(k_ya)]
[\cos(p_xa) - \cos(p_ya)]$ with $\bp = \bk - \bq$, is strongest for anti-nodal electrons, 
and has a form factor similar to $c$-axis hopping $t_\perp(\bk)$.\cite{AndersenJCP}  
Although this coupling has an anisotropy similar to that of the coupling to the $B_{1g}$ branch, it 
does not contribute to $d$-wave pairing due to its phase and momentum dependence at large $q$.

\begin{widetext}
\subsection{In-plane Bond-Stretching Modes}

For completeness, we also consider the coupling to the planar Cu-O bond stretching modes, the 
so-called breathing modes, within the framework of the three-band model, as  
derived in Refs. \onlinecite{RoschPRB2004}, \onlinecite{IshiharaPRB2004} and \onlinecite{tpdPRL2004}.   
The bond-stretching modes couple to electrons via both a direct modulation of the hopping integral $t_{pd}$ as 
well as electrostatic changes in the Madelung energies as the orbitals are displaced. 
\cite{IshiharaPRB2004}  As done in Ref. \onlinecite{RoschPRB2004} we consider only the 
overlap modulation.  The derivation is briefly sketched here.  

To obtain the form of the el-ph coupling the overlap integral $t_{pd}$ is taken to be 
site dependent $t^{\bf{n}}_{pd}$.  It is then assumed that the Cu and O atomic displacements, 
$u_{\bf n}^{Cu}$ and $u_{\bf{n},\alpha}^O$, about their equilibrium positions, $R_{\bf n}$ and 
$R_{\bf n} + \hat{\delta}a/2$, where $\hat{\delta}$ are basis vectors for the CuO$_2$ plane, 
are small. The overlap integral is expanded and only the first order term is retained
\begin{equation}
t^{\bf n}_{pd} = t_{pd}^0 + \sum_{\hat{\delta}=\pm \hat{x},\pm \hat{y}} 
%\frac{\partial t^{\bf{n}}_{pd}}{\partial r}\bigg|_{r=R} 
\vec{\nabla} t_{pd}^{\bf{n}}\big|_{r=R_{\bf n}}
\cdot \left({\bf u}_{\bf n}^{Cu} - {\bf u}^O_{{\bf n}+a\hat{\delta}/2} \right) + O(u^2)
\end{equation}

The modulation of the hopping integrals provides the el-ph coupling Hamiltonian 
\begin{equation}
H_{el-ph}^{br} = \sum_{n,\sigma,\delta}P_\delta 
%\frac{\partial t_{n,\delta}}{\partial r} 
\vec{\nabla} t_{pd}^{\bf{n}}\big|_{r=R_{\bf n}} \cdot 
\left[{\bf u}_{\bf n}^{Cu} - {\bf u}^O_{{\bf n}+a\hat{\delta}/2} \right] 
\left[d^\dagger_{n,\sigma}p_{n,\sigma,\delta} +h.c. \right], 
\end{equation}
with $P_{x,y} = \pm 1 = -P_{-x,-y}$ denoting the phase of the Cu-O overlap.  
Following Ref. \onlinecite{tpdPRL2004}, we neglect the Cu vibration and set 
$\partial t^{\bf n}_{pd}/\partial x|_{R_{\bf n}} = -Q_\delta g_{dp}$, where $Q_{\pm x} = 
Q_{\pm y} = \pm 1$ and $g_{dp}$ is a scalar function that depends on the equilibrium 
Cu-O distance.  The el-ph coupling Hamiltonian can then be simplified to
\begin{equation}
H_{el-ph}^{br} = g_{dp}\sum_{n,\delta,\sigma} P_\delta Q_\delta u^O_{\delta}(n)
\left[d^\dagger_{n,\sigma}p_{n,\sigma,\delta} + h.c. \right], 
\end{equation}  
as obtained in Ref. \onlinecite{tpdPRL2004}.  Here $u^0_\delta$ denotes the 
displacement of the oxygen atom $\delta$ along the Cu-O bond.  
By introducing the Fourier transform the el-ph coupling may be re-written as:

 \begin{equation}
  H_{el-ph}^{br} = \frac{g_{0}^{br}}{\sqrt{N}}\sum_{\bk,\bq}\sum_{\sigma,\delta=x,y} P_\delta
  \cos(k_\alpha a/2) \epsilon^O_\delta(\bq)[d^\dagger_{\bk,\sigma}p_{\bk-\bq,\sigma,\delta} 
  + p^\dagger_{\bk+\bq,\sigma,\delta}d_{\bk,\sigma}](b^\dagger_\bq + b_{-\bq}). 
 \end{equation}
Here, $\epsilon_\delta(\bq) = 
 \sin(q_\delta a/2)/\sqrt{\sin^2(q_xa/2) + \sin^2(q_ya/2)}$ is the component of 
 the phonon eigenvector for oxygen $\delta$ parallel to the Cu-O bond. 
 Finally, the electronic 
 eigenfunctions of the planar O $\phi_{x,y}(\bk)$ and Cu 3d$_x^2-y^2$ orbitals $\phi_{Cu}(\bk)$  
 for the pd-$\sigma^*$ band,  
 with operators $c$, $c^\dagger$,  are 
 introduced. The resulting Hamiltonian reduces to the form of Eq. (\ref{Eq:el-ph}) with 
 \begin{equation}
  g_{br}(\bk,\bq) = g^{br}_0 \sum_{\alpha=x,y}P_\alpha \epsilon_\alpha(\bq) \left[ 
  \cos(p_\alpha a/2)\phi_{Cu}(\bp)\phi_\alpha(\bk) - 
  \cos(k_\alpha a/2)\phi_{Cu}(\bk)\phi_\alpha(\bp) 
  \right]
 \end{equation}
\end{widetext}
Here $\bp = \bk - \bq$, $g_0^{br} = g_{dp}\sqrt{\hbar/2M_O\Omega_\bq}$.  We note that generally 
$t_{pd}(d_{Cu-O}) \sim d^\beta_{Cu-O}$ and thus $g_{pd} = \beta t_{pd}/d_{Cu-O}$.  Since 
typically $\beta = 3.5$,\cite{RoschPRB2004} an estimate for the strength of the coupling 
is obtained from $g_{pd} \sim 2$ eV/\AA\space for $t_{pd} = 1.1$ eV.  For $\Omega_\bq = 70$ meV, this gives 
$g_{br} = 86$ meV. 

\section{K,Q-Momentum Dependence of the Bare Vertices}

\subsection{Momentum dependence throughout the Brillouin Zone}
In the previous section it was shown how the explicit form for el-ph 
coupling to oxygen modes is determined by the
nature of the charge transfer modulated by the  
lattice displacement, the local environment surrounding the CuO$_2$ plane, as
well as the orbital content of a single downfolded band crossing the Fermi level. 
The relevant parameters - magnitude of orbital hybridization,
local crystal field, the charge-transfer energy, the shape of the Fermi surface, 
and the density of states at the Fermi level - all contribute in setting the overall magnitude of the coupling
as well as the full fermionic {\bf k} and bosonic {\bf q} momentum dependence of the 
coupling $g({\bf k,q})$. The band character enters through the band eigenvectors 
$\phi$ which further depend on the complexity of the unit cell. At the BZ center the
wavefunctions are atomic and the character of the band is unique. 
Large momentum variations 
of the band character then occur for increasing momentum, and a very strong momentum 
dependence of the overall el-ph coupling can occur.  This strong momentum dependence 
has indeed been observed in recent LDA treatments \cite{HeidPRL2008}. 
 
We remark that we are first interested in the magnitude and anisotropy of the 
bare couplings in the absence of charge screening in order to determine 
possible discrepancies with LDA treatments, which treat correlations on the 
mean field level and give 3D metallic screening. 
In order to estimate general tendencies, momentum dependencies as well as magnitudes, 
in this section we explore some simplifications. 

We begin by assuming that $2t_{pd}$ 
is much greater than any relevant energy scale in the system, keeping in mind that the charge transfer
energy $\Delta = \epsilon_p - \epsilon_d = \sim 0.8$ eV is much reduced from its bare value $\sim 3.5$ eV  
when treating correlations in mean field approaches like LDA.\cite{AndersenJLTP1996} In this 
limit the $\phi$ functions can be represented
as 
\begin{equation*}
\phi_{x,y}({\bf k})= \pm i \frac{\sin(k_{x,y}a/2)}{\sqrt{\sin^2(k_{x}a/2)+\sin^2(k_{y}a/2)}} + O(\Delta/2t_{pd})^2
\end{equation*}
and 
\begin{equation*}
\phi_{Cu} = \frac{\Delta/2t_{pd}}{\sqrt{\sin^2(k_{x}a/2)+\sin^2(k_{y}a/2)}}. 
\end{equation*}
To the same order, the denominators in these expressions are 
constant over constant energy contours. Therefore, since we will restrict ourselves largely to the 
Fermi surface, we represent 
the band functions $\phi_{x,y}(\bk) = A_O\sin(k_\alpha a/2)$, $\phi_b(\bk) = A_{Cu}$ and 
$\phi_a(\bk) = A_ad_k$, with $d_k = [\cos(k_xa/2)-\cos(k_ya/2)]/2$ and the coefficients determined by 
$A^2_O = \langle\phi^2_{x,y}(\bk)\rangle/\langle\sin^2(k_{x,y}a/2)\rangle$, 
$A_{Cu}^2 = \langle\phi^2_b(\bk)\rangle$ and 
$A_a^2 = \langle \phi_a^2(\bk)\rangle/\langle d^2_k\rangle$, respectively.  
Here $\langle\dots\rangle$ denotes a Fermi surface average: 
$\langle A\rangle = \sum_\bk A_\bk \delta(\epsilon_\bk)/\sum_\bk \delta(\epsilon_\bk)$. 
In this way, the overall coupling anisotropy can be simplified without loss of generality. 

As a consequence, the fermionic momentum dependence of the coupling to the breathing modes disappears:
\begin{equation*}
g_{br}(\bk,\bq) = g^{br}_0A_{Cu}A_{O}\sum_{\alpha=x,y}P_\alpha e_\alpha(\bq)\sin(q_\alpha a/2).
\end{equation*}
Substituting the phonon eigenvectors, the coupling to the breathing modes becomes
\begin{equation}\label{Eq:gbr}
g_{br}(\bq) = g^{br}_0A_{Cu}A_{O}\sqrt{\sin^2(q_xa/2) + \sin^2(q_ya/2)}. 
\end{equation}
This form has also been obtained in a $t$-$J$ approach \cite{RoschPRB2004} however, in this case, the oxygen 
and copper character have been explicitly retained through $A_O$ and $A_{Cu}$, respectively.\cite{tpdnote}  

\begin{widetext}
 The el-ph vertex for the $A_{1g}$ and $B_{1g}$ modes can be likewise simplified
 \begin{eqnarray}\label{Eq:gb1g} \nonumber
  g_{A1g, B1g}(\bk,\bq)&=&eE_z\sqrt{\frac{2\hbar}{M_ON(\bq)\Omega_{B1g,A1g}}}A^2_Oe^{-i(q_x+q_y)a/2} \\
 & &\times \left[\sin(k_xa/2)\sin(p_xa/2)\cos(q_ya/2) \pm \sin(k_ya/2)\sin(p_ya/2)\cos(q_xa/2)\right]
 \end{eqnarray}

 These expressions recover the Raman form factors in the limit $\bq \rightarrow 0$ for each mode
 and they obey the symmetry conditions for momentum reflections about $45^\circ$ as discussed 
 previously.  The fermionic momentum dependence cannot be neglected in either of these expressions, where 
 in particular, a strong fermionic momentum dependence of the 
coupling to the $B_{1g}$ modes occur, preferentially  weighting anti-nodal 
states with small momentum transfers.  

 Lastly, the momentum structure of the apical coupling simplifies considerably in the same manner: 
 \begin{equation}\label{Eq:gapex}
  g_{apex}(\bk,\bq) = g^{apex}_oA^2_{a}[\cos(k_xa) - \cos(k_ya)][\cos(p_xa) - \cos(p_ya)]/4.
 \end{equation}
 Once again, a substantial fermionic momentum dependence emerges from 
 the $c$-axis charge transfer pathways and the apical character of the band. 
\end{widetext}

Before proceeding further a few comments are in order.  
In using Eqs.(\ref{Eq:gbr})-(\ref{Eq:gapex}) we have simplified the 
el-ph couplings while explicitly retaining the 
role of the band character in determining the overall strength of the couplings.  
Since the eigenfunctions enter to the fourth power for $|g|^2$, the total 
coupling strengths determined in this approach may change considerably 
when adjusting multi-band parameters.  However, this method has the advantage that  
the materials dependence of the coupling, parameterized by $A_O$, $A_a$ and $A_{Cu}$, can be 
calculated using a variety of methods such as exact diagonalization, quantum Monte Carlo 
or LDA.  We also emphasize that the total coupling strengths (calculated in the next section) that 
we obtain using this formalism are similar to those obtained from LDA treatments, 
even though the latter includes the effects of screening.
\cite{HeidPRL2008, BohnenEPL2003, SavrasovPRL1996} 

\begin{figure}[t]
 \includegraphics[width=\columnwidth]{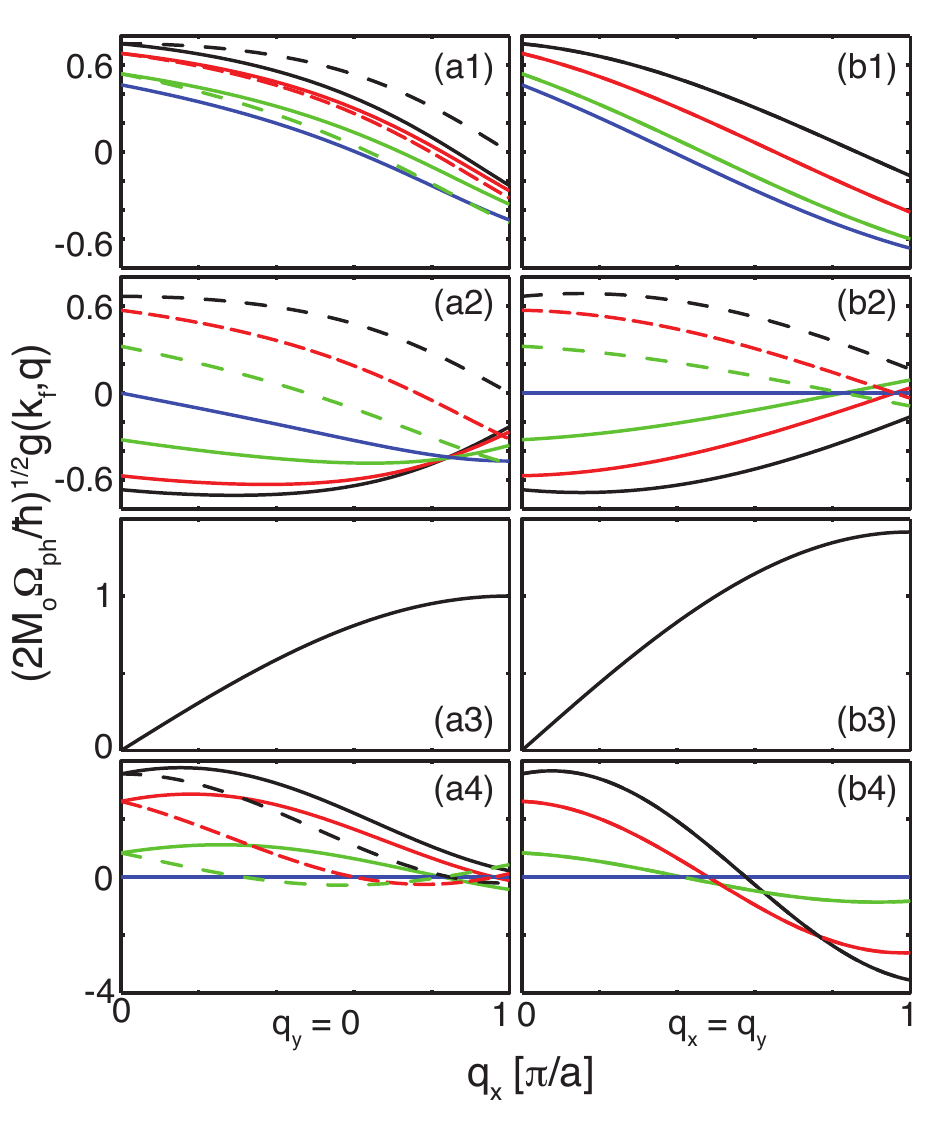}
 \caption{\label{Fig:gkq} (Color online) 
 Plots of the el-ph coupling constant $g(\bk_F,\bq)$ for 
 fermionic momentum on the Fermi surface as a function of transferred momentum 
 $(q_x,q_y=0)$ (a1-a4) and $q_x=q_y$ (b1-b4), respectively.  a1,b1 (a2,b2) plot 
 coupling to the $A_{1g}$ ($B_{1g}$) branches, respectively, a3, b3 plot 
 coupling to the breathing branch and a4,b4, plots coupling to the apical 
 branch.  The colors denote angles from the corner of the BZ (shown in the inset 
 of Fig. \ref{Fig:lambda_fs}) given by solid black - $0^\circ$, solid red - $15^\circ$, 
 solid green - $30^\circ$, solid blue - $45^\circ$ (nodal), dashed green - 
 $60^\circ$, dashed red - $75^\circ$ and dashed black - $90^\circ$ (anti-nodal).   
 }
\end{figure}

%Another advantage of this approach is that it allows 
This approach also allows 
for the use of a renormalized 
bandstructure, while retaining the explicit band character of the original five-band model. 
This is important since the overall strength of the el-ph couplings 
scales with the density of states at the Fermi level $N_F$ (see Eq. (\ref{Eq:LambdaFS})); 
narrow bandwidth systems will exhibit larger coupling in 
comparison to large bandwidth systems with the same vertex $g(\bk,\bq)$.  
With an appropriate choice in parameters, the five-band model given in section \ref{SectionTwo} 
reasonably reproduces the bandwidth (and $N_F$) determined by LDA  
calculations.\cite{MarkiewicsPRB2005}  However, as has been noted,\cite{Reznik}  
LDA over predicts the total bandwidth (and consequently $N_F$ is under predicted) 
in comparison with experiment.  
Therefore, we expect that the total couplings will be underestimated using the 
five-band model with parameters chosen to match LDA.   A simple rescaling of the 
five-band model bandstructure in conjunction with the full form of the $\phi$-functions 
is insufficient to correct this since this procedure would 
produce incorrect values for the $\phi$-functions and therefore generate errors in the 
character of the band as a function of $\bk$.   However, the use of the couplings defined by 
Eqs. (\ref{Eq:gbr})-(\ref{Eq:gapex}) allows us to resolve this issue.  Here, 
the correct band character is captured by calculating $A_O$, $A_{Cu}$ and $A_a$ using 
the five-band model but the Fermi surface and bandstructure are obtained from 
elsewhere in order to better match experiment.   
In this work, we adopt a five-parameter tightbinding model for Bi-2212 derived from fits to 
ARPES data.\cite{NormanPRB2003}   This approach allows us to capture the increased 
value of $N_F$ while simultaneously retaining estimates for the correct band character.  
We also note that the specific shape of 
the Fermi surface is not crucial to the overall anisotropy of the couplings.  

To visualize the momentum dependence of the coupling in more detail, we plot in Fig. \ref{Fig:gkq} 
$g(\bk_F,\bq)$, given by Eqs. (\ref{Eq:gbr}) - (\ref{Eq:gapex}), as a function of 
transferred momenta $\bq$ along two directions as indicated.  
The dependency on transferred momenta arises from the 
nature of the charge-transfer coupling of the different modes.  The $c$-axis modes 
(Figs \ref{Fig:gkq}a1-2,4 and \ref{Fig:gkq}b1-2,4), being electrostatic in nature, translate 
into stronger coupling for small momentum transfers, while the deformation-type coupling 
of the breathing branches gives stronger coupling at large $\bq$, and vanishes in the 
limit $\bq \rightarrow 0$ (Figs. \ref{Fig:gkq}a3 and \ref{Fig:gkq}b3).

Apart from the breathing modes, an appreciable fermionic wavevector dependence of the 
couplings to the $c$-axis modes is found as a consequence of the character of the 
underlying atomic vibrations.  For the case of the $A_{1g}$ and apical modes, for 
momentum transfers along the zone diagonal (Figs. \ref{Fig:gkq}b1 and \ref{Fig:gkq}b4, 
respectively), the fermionic dependence is symmetric with respect to reflections about 
$\pi/4$ while the $B_{1g}$ coupling (Fig. \ref{Fig:gkq}b2) changes sign.  Momentum 
transfers along the zone face 
(Figs. \ref{Fig:gkq}a1, \ref{Fig:gkq}a2, \ref{Fig:gkq}a4) do not obey any set selection 
rule, although the symmetric- or anti-symmetric-like character of the coupling is evident.  
Finally, the strong momentum dependence of the charge transfer along the $c$-axis 
dictates that the apical coupling vanishes for any fermion momentum along the 
zone diagonal (Fig. \ref{Fig:gkq}b4).  

\subsection{\label{Section:Lambda}Momentum dependence on the Fermi surface}

\begin{figure}[t]
 \includegraphics[width=\columnwidth]{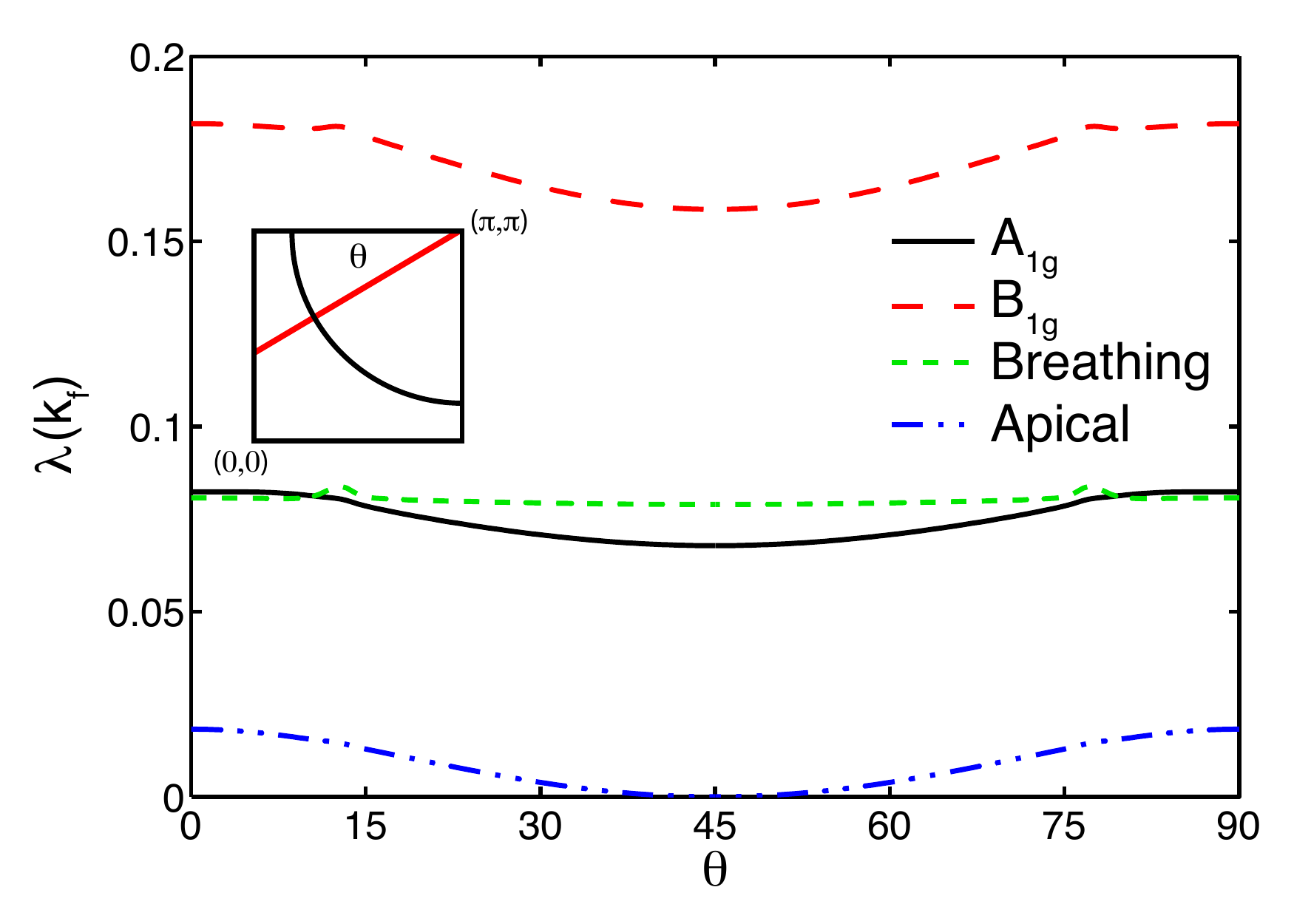}
 \caption{\label{Fig:lambda_fs} 
 (Color online) 
 Plots of $\lambda_\nu(\bk_F)$ for momentum points along the Fermi surface.  
 Inset: the definition of the angle $\theta$ used in Figs. \ref{Fig:gkq} and 
 \ref{Fig:lambda_fs}. The parameters used are defined in the text.}
\end{figure}

The strong dependence of $g(\bk,\bq)$ on both $\bk$ and $\bq$ leads to anisotropic coupling between 
electrons and phonons.  As the most relevant scattering processes involve those states 
near the Fermi level, the explicit momentum dependence as seen in ARPES is most clearly 
envisioned by calculating $k$-dependent self-energies, in terms of el-ph coupling $\lambda$.
We take the modes to be
dispersionless $\Omega_{\bq,\nu} = \Omega_\nu$, which couple to electrons via 
the Fock piece of the electron self-energy $\lambda_\nu(\bk)$ in lowest order\cite{mahan}:
\begin{eqnarray}\label{Eq:LambdaFS}\nonumber
\lambda_\nu(\bk)&=&\frac{2}{N\Omega_\nu}\sum_\bp |g(\bk,\bq)|^2 \delta(\xi_\bp) \\
&=&\frac{2N_F}{\Omega_\nu}\langle |g(\bk,\bq)|^2\rangle_{\bp_{FS}}
\end{eqnarray}
with $\xi_\bp = \epsilon(\bp) - \mu$, $\mu$ is the chemical potential, 
$N$ is the number of momentum points and $N_F = \frac{1}{N}\sum_\bk \delta(\xi_\bk)$ 
is the density of states at the Fermi level.   
The delta function restricts the sum to initial and final fermion states $\bk$, $\bp$, 
that lie on the Fermi surface, with scattering between them governed by the 
transferred phonon momentum $\bq$.  The resulting $\lambda_\nu(\bk_F)$ for the 
four modes are shown in Fig. \ref{Fig:lambda_fs}.  
The Fermi surface contour 
and bandstructure are again determined from a five-parameter tightbinding bandstructure.\cite{NormanPRB2003}  
The el-ph coupling vertices are evaluated using Eqs. (\ref{Eq:gbr}) - 
(\ref{Eq:gapex}), along with the 
conventional parameter set (in eV): $t_{pd} = 1.1$, $t_{pp} = 0.5$, 
$t_{pz} = 0.29$, $t_{sp} = 2$, $t_{sz} = 1.5$, $\epsilon_{d-p} = 0.8$, 
$\epsilon_{d-z} = 1.00$, $\epsilon_{d-s} = -7$. \cite{OhtaPRB1991, SiteEnergy}
For these parameters we obtain $A^2_O = 0.446$, $A^2_{Cu} = 0.592$ and $A^2_a = 4.52\times10^{-2}$. 
Additionally, estimates of the local field strength at the planar and apical oxygen 
sites for Bi-2212 are used: $E^{plane}_z = 3.56$, $E_z^{apex} = 16.33$ eV/\AA, 
respectively.  A systematic derivations of these E field values across the cuprate families is 
given in Section \ref{Madelung}.

A strongly varying $\lambda(\bk)$ is obtained for the $c$-axis modes, largely weighting 
anti-nodal fermion states, arising from both the fermion dependence of the bare couplings 
as well as the small momentum transfers connecting anti-nodal points on the Fermi 
surface, as noted in prior treatments \cite{tpdPRL2004} as well as more recent LDA-
projected studies.\cite{HeidPRL2008} The anisotropy is particularly strong for the 
apical coupling because of the strong fermionic momentum dependence of $g(\bk,\bq)$.  On the 
other hand, coupling to the breathing modes has a much weaker anisotropy along the 
Fermi surface, due to large momentum transfers ($\pi,\pi$) connecting anti-nodal 
portions of the Fermi surface, as well as ($\pi,0$) and ($0,\pi$) transfers connecting 
nodal points.   Inclusion of a variation of the copper character across the Fermi 
surface yields a more anisotropic coupling.\cite{tpdPRL2004}  However, the character 
of the anisotropy is not as strong as that observed in Ref. \onlinecite{HeidPRL2008}, 
where the breathing branches couple more strongly to anti-nodal states.  

The full coupling $\lambda_z$, renormalizing the single particle self-energy, is given by a 
sum over all modes averaged over the Fermi surface, $\lambda_z = \sum_\nu \langle 
\lambda_\nu(\bk) \rangle$, and can be visualized most easily as an average over the 
curves shown in Fig. \ref{Fig:lambda_fs}.  The renormalization $\lambda_\phi$ for a 
$d_{x^2-y^2}$-wave superconductor gives the el-ph contribution to the anomalous 
self-energy and is given by a $d$-wave projected average 
\begin{equation}\label{Eq:lambdad}
\lambda_\phi = 2\sum_\nu \frac{\sum_{\bk,\bp}d_\bk d_\bp|g_\nu(\bk,\bq)|^2
\delta(\xi_\bk)\delta(\xi_\bp) }
{N\omega_\nu\sum_\bk d^2_\bk\delta(\xi_\bk)}
\end{equation} 
where $\bq = \bp-\bk$ and $d$-wave basis function $d_\bk = [\cos(k_xa/2)-\cos(k_ya/2)]/2$, 
as before.  A positive (negative) $\lambda_\phi$ denotes an attractive 
(repulsive) pair interaction in the 
$d_{x^2-y^2}$ channel.  The $\lambda$ values obtained for the four phonon branches 
considered in this work are given in table \ref{Tbl:Lambda}. 
As noted before, 
the $A_{1g}$ and $B_{1g}$ modes enhance $d$-wave pairing, the breathing modes suppress 
it and the apex modes gives no contribution.  
We note that these values for the $\lambda_z$ are similar to those obtained from recent LDA results
for YBa$_2$Cu$_3$O$_7$\cite{HeidPRL2008,BohnenEPL2003} 
but are smaller than those calculated for doped 
CaCuO$_2$.\cite{SavrasovPRL1996,AndersenJLTP1996} We again remark 
that important consequences of screening are not considered here, 
which can affect both the magnitudes and anisotropies of the couplings as will be discussed in 
Sec. \ref{Sec:Screening}.

\begin{table}[tr]
 \begin{center}
  \begin{tabular}{|c|c|c|}
   \hline
    Branch & $\lambda_z$ & $\lambda_\phi$ \\
   \hline
    $A_{1g}$ & $7.74\times10^{-2}$ & $4.42\times10^{-2}$ \\
    $B_{1g}$ & $0.17$ & $0.12$ \\
    Breathing & $8.27\times10^{-2}$ & $-4.75\times10^{-2}$ \\
    Apical & $1.11\times10^{-2}$ & $0$ \\ \hline
    Total & $0.341$ & $0.1149$ \\ \hline
  \end{tabular}
 \end{center}
 \caption{\label{Tbl:Lambda} Tabulated values of $\lambda_{z,\phi}$ for 
  the four phonon branches considered in this work. }
\end{table}

\subsection{Kinematic constraints}

It is important to note that in ARPES there is a strong kinematic constraint that also 
governs which fermion states can couple to each mode.  Due to energy conservation of 
scattering from dispersionless phonons, the normal state ARPES spectra may show mixing of 
electron and phonon states if the band energy along a particular cut crosses the phonon 
mode energy.  This gives rise to a kink in the dispersion, with trailing intensity 
observed asymptotically to the phonon mode energy.  Otherwise, if the band along 
a particular cut lies above the mode energy, the spectra may show level repulsion in the 
form of a flattened band bottom along that cut.  

\begin{figure}[t]
 \includegraphics[width=0.8\columnwidth]{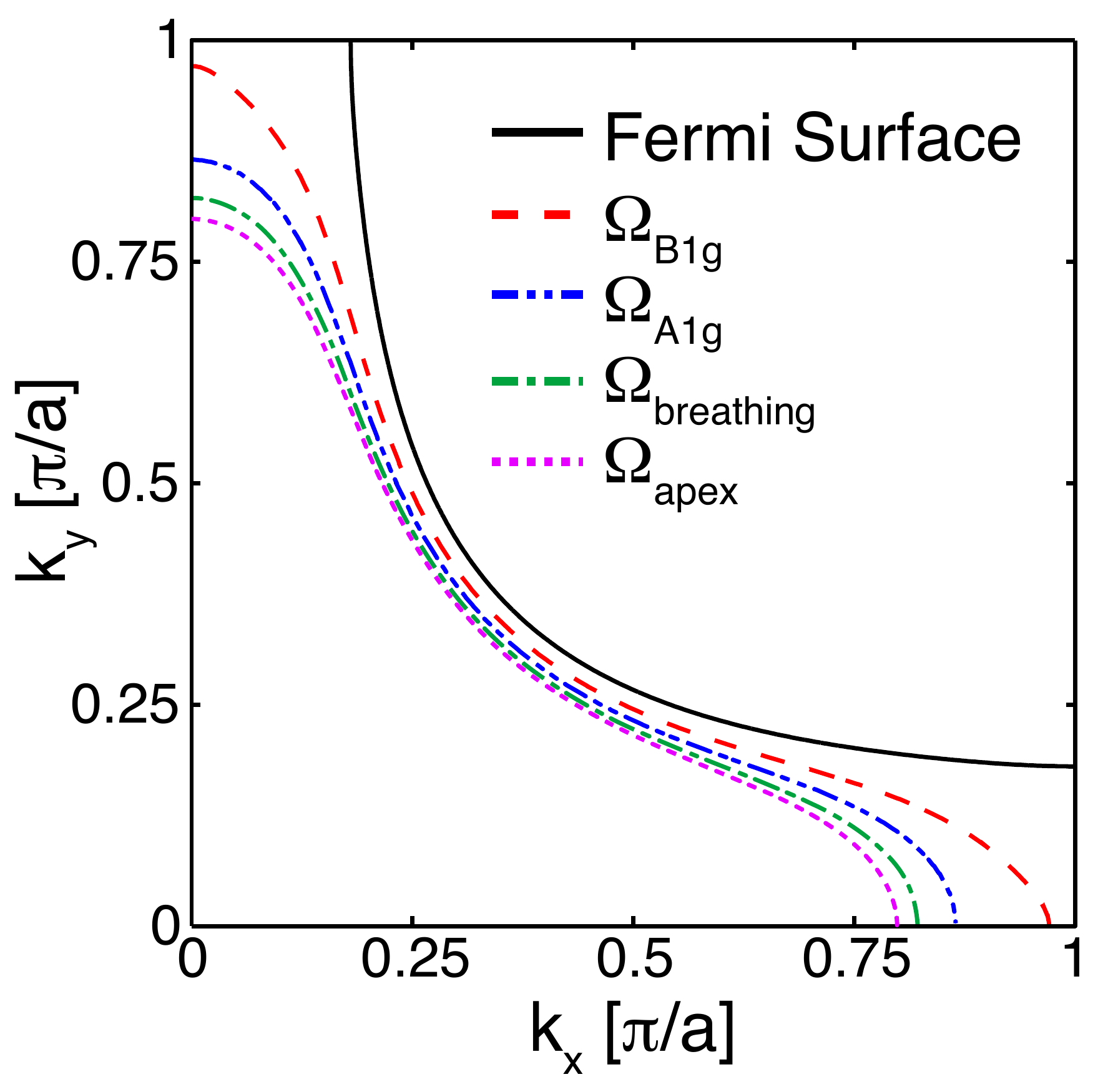}
 \caption{\label{Fig:BandContour} (Color online) A contour plot of the 
 band structure of optimal doped Bi-2212 obtained from the five-parameter tightbinding model 
 of Ref. \onlinecite{NormanPRB2003}. The outermost contour (solid black) corresponds to 
 the FS.  The remaining contours correspond to $\epsilon_\bk-\mu = \Omega_\nu$ 
 for the four phonon branches considered in this work. Reading from the FS 
 to the $\Gamma$-point, the contours correspond to the $B_{1g}$, $A_{1g}$, breathing 
 and apical branches, respectively.}
\end{figure}

In the cuprates, the shallow band 
near the anti-nodal parts of the BZ limits the couplings of those states to only modes with 
energies less than $\xi(0,\pi)$. Thus, quite generally, we would expect coupling to the 
modes to disappear from the spectra when $\Omega_\nu < \xi(\bk)$ along the entire 
cut.  The contours $\xi(\bk) = \Omega_\nu$, which define this kinematic constraint,  
are plotted in Fig. \ref{Fig:BandContour} at optimal doping.  
The Fermi level is also shown for reference and corresponds to the outermost contour (solid black).     
In the case of the apical mode, although the coupling may be largest for $\bk_{AN}$, 
a kink effect is prevented kinematically.  Moreover, the breathing modes will be observable 
only for near-nodal cuts due to the form of the coupling as well as the energies of the 
modes which are large 
when compared to the $A_{1g}$/$B_{1g}$ modes and the saddle point energy $\xi(0,\pi)$.  
In fact, for optimally doped 
Bi-2212, the $B_{1g}$ modes lie right at $\xi(0,\pi)$, and can be seen most 
clearly in the anti-nodal directions, although its coupling should be felt across the 
entire Fermi surface as shown in Fig. \ref{Fig:lambda_fs}. 

In the superconducting state, the energy contours shift by the maximum 
value of the energy gap and are set by $\Omega_\nu+\Delta_0$, and the kinematic 
range shrinks for coupling to the higher energy modes.\cite{tpdPRL2004,BickersPRB2004}

\section{Poor Screening Along the $c$-axis}\label{Sec:Screening}
\subsection{Overall Approach}
Screening - responsible for reducing the net el-ph coupling in conventional good metals - 
must be included for all $c$-axis phonons.  In the usual treatment of screening,\cite{Abrikosov} 
the screened el-ph coupling $\bar{g}$ is 
\begin{equation}
\bar{g}(\bk,\bq,\Omega) = g(\bk,\bq) + \frac{V(q)\Pi_{g,1}(q,\Omega)}{1-V(q)\Pi_{1,1}(q,\Omega)}
\end{equation}
where $V(q) = 4\pi e^2/q^2$ is the 3D Coulomb interaction and $\Pi_{a,b}(q,\Omega)$ is the 
frequency-dependent polarizability calculated with vertices $a$, $b$, respectively.  
Since the plasmon frequency $\Omega_{pl}$ is usually much larger than the phonon frequencies 
in 3D metals, in the limit $\bq \rightarrow 0$ the effective coupling is $\bar{g}(\bk,\bq\rightarrow0) 
-\delta g$ with $\delta g$ denoting the average value of $g(\bk,\bq)$ over the Fermi surface. 
Thus the bare el-ph coupling is normally well screened. 

Turning now to the cuprates, from the point of view of Coulomb interactions, the materials are 3D with 
$q^2 = q_{2D}^2 + q_z^2$ and $\bq_{2D} = (q_x, q_y)$ .  However, due to the largely incoherent 
$c$-axis transport observed across the phase diagram (apart from the overdoped side), the 
polarizability is largely determined by the planar conduction electrons and  
$\Pi_{a,b}(q,\Omega) = \Pi^{2D}_{a,b}(q_{2D}, \Omega)$.  As a result, important changes to 
the effects of screening occur for small $\bq_{2D} \ll \bq$.

A qualitative feel for the effects of poor screening can be obtained by considering 
a coupling which is independent of Fermion momentum $\bk$ and in the limit of small 
in-plane momentum transfers, where $v_Fq_{2D} \ll \Omega_{ph}$.  This limit is most 
relevant for Raman active phonons and $\Pi_{1,1}(q,\Omega_{ph}) = nq^2_{2D}/m\Omega_{ph}^2$, 
with $n/m = 2/V_{cell}\sum_\bk f(\epsilon_\bk)\frac{\partial^2\epsilon_\bk}{\partial k_x^2}$ 
and $\Pi_{g,1}(q,\Omega_{ph}) = \frac{nq^2_{2D}}{m_{g}(q)\Omega^2_{ph}}$, with 
$n/m_{g}(q) = 2/V_{cell}\sum_\bk f(\epsilon_\bk)g(\bq)\frac{\partial^2\epsilon_\bk}{\partial k_x^2}$ 
where $V_{cell}$ is the unit cell volume and $f$ is the Fermi distribution. 
The screened el-ph interaction is then
\begin{equation}\label{Eq:gbar}
\bar{g}(\bk,\bq,\Omega_{ph})=g(\bq) - \frac{m}{m_{g}(q)}\frac{\Omega_{pl}^2(q)}{\Omega_{pl}^2(q) - \Omega_{ph}^2}, 
\end{equation}
with $\Omega^2_{pl}(q) = \Omega^2_{pl}\cdot(q_{2D}/q)^2$.  Thus $c$-axis Raman-active phonons 
in the cuprates, with couplings strongest for small momentum transfers $\bq$ and $\Omega_{pl}(q) < \Omega_{ph}$, 
should survive the effects of screening.  

\subsection{Considerations for the polarizability and $\lambda_\nu$}
To evaluate the polarizabilities in the cuprates, we use the standard Lindhard expression
\begin{eqnarray}\nonumber
\Pi_{a,b}(q,\Omega)&=&\frac{2}{V_{cell}}\sum_\bk a(\bk,\bq)b(\bk,\bq) \\
& &\times \frac{f(\epsilon_\bk)-f(\epsilon_{\bk+\bq})}
{\epsilon_\bk - \epsilon_{\bk+\bq} + \Omega + i\delta}, 
\end{eqnarray}
with all wavevectors being two-dimensional.  This form of the polarizability has been 
considered in previous works focusing on the role of 
screening and reduced dimensionality in the doping-dependent 
softening of the in-plane bond-stretching modes.\cite{Bauer, 
MukhinPRB2007}   
For the Coulomb interaction $V(q)$, more appropriately for the cuprates, 
we consider a layered system with charge density 
essentially confined to 2D layers, Coulombically coupled across the planes:\cite{Fetter, AristovPRB2006}
\begin{equation}\label{Eq:Vq}
V(q_{ab},q_z) = \frac{V}{q_{ab}a\tanh(q_{ab}/q_0)}\frac{1}{1+F_z^2}
\end{equation}
where $V = 2\pi e^2/a\sqrt{\epsilon_{ab}\epsilon_c}$, $q_0 = (4/c)\sqrt{\epsilon_c/\epsilon_{ab}}$ with 
$\epsilon_{ab,c}$ the in-plane, out-of-plane dielectric constants, respectively.  The $c$-axis 
lattice constant sets the in-plane momentum scale beyond which the planes become effectively 
uncoupled, and leads to an interpolation between the 2D and 3D form for the Coulomb interaction 
where $F_z = \sin(q_zc/4)/\sinh(q_{ab}/q_0)$ provides the $q_z$ dispersion of the interaction. 

\begin{figure}[t]
 \includegraphics[width=\columnwidth]{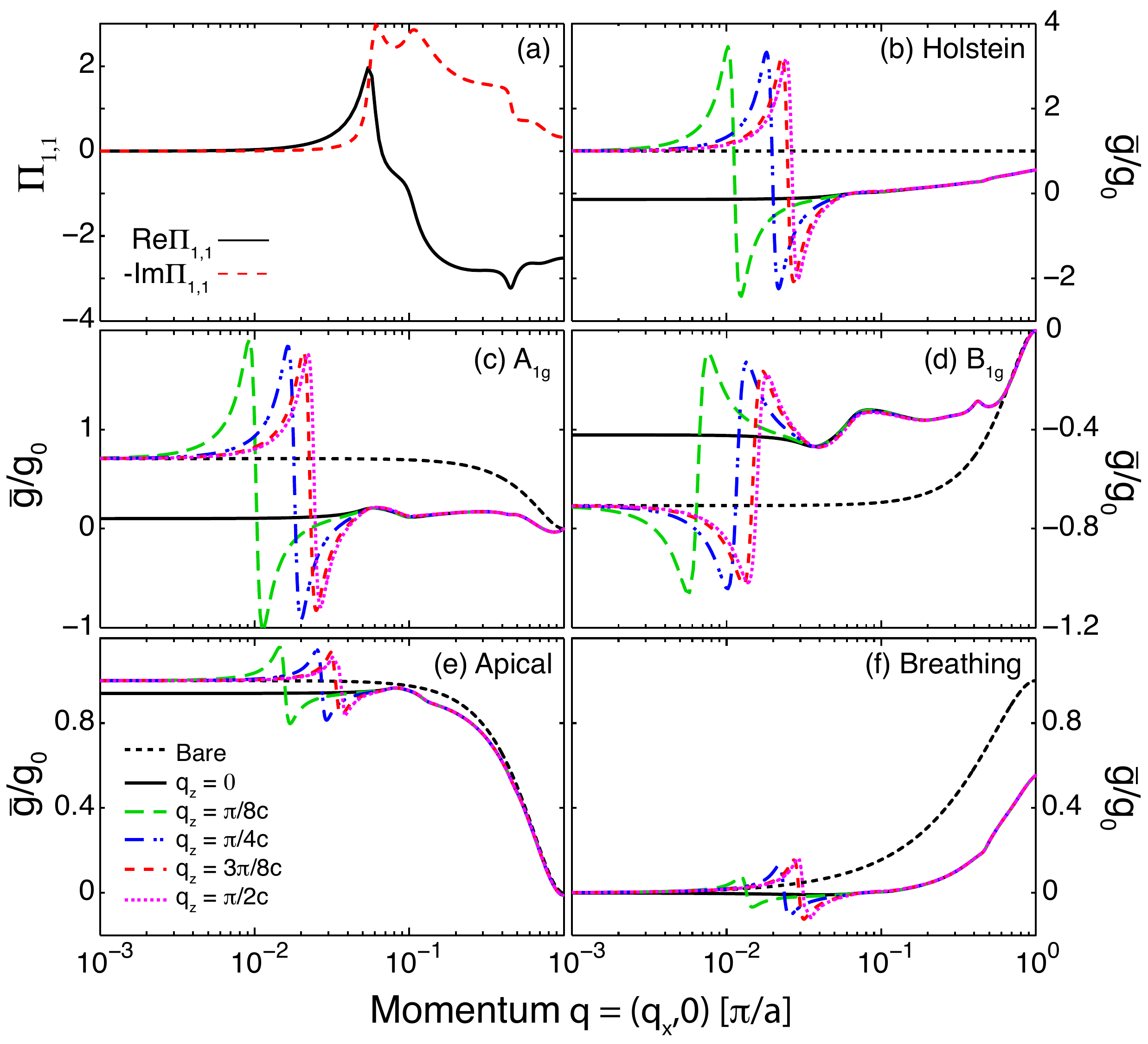}
 \caption{\label{Fig:pol}
 (Color online) 
 (a) Plots of the real (black) and imaginary ($-1\times$, red) parts of the charge susceptibility 
 $\Pi_{1,1}$ and (b)-(f) the renormalized el-ph coupling $\bar{g}(\bk,\bq)$ in arb. units for 
 $\bk = (0,\pi/a)$ and $\bq = (q_x,0)$ for different values of $q_z$ as indicated by the legend.  
 For (b)-(f) the black dashed line corresponds to the bare coupling $g(\bk,\bq)$.  A momentum-independent 
 bare coupling is considered in (b), followed by the $A_{1g}$, $B_{1g}$, apical and breathing modes 
 in (c)-(f), respectively.  In the case of the breathing modes a small $q_y$ component has been added 
 to visualize the effects of screening at small $\bq_{2D}$.  The $q_x$ axis 
 in all plots is on a logarithmic scale to highlight the small $\bq_{2D}$ region.   
 }
\end{figure}

The real and imaginary parts of the pure 
charge polarizability $\Pi_{1,1}$ are plotted in Fig. \ref{Fig:pol}a.  Here, $\delta = 5$ meV has 
been included to smooth singularities, $\epsilon(\bk)$ is taken for optimal doping in Bi-2212,
\cite{NormanPRB2003}  
$c = 30.52$ \AA, $a = 3.8$ \AA, and in-plane\cite{QuijadaPRB1999} and out-of-plane
\cite{KovalevaPRB2004} conductivities yield $\epsilon_{ab} \sim \epsilon_c = 4.8$ for the high-
frequency dielectric constants, which determine the plasma frequency.  For this choice in parameters 
we obtain $\Omega_{pl} = 0.914$ eV at optimal doping in agreement with Ref. \onlinecite{QuijadaPRB1999}.    
In order to calculate the screened el-ph vertex the static dielectric constants must be used in 
Eq. (\ref{Eq:Vq}).  Here, we take $\epsilon_{c}(0) = 10$, which is obtained from optical measurements, 
\cite{KovalevaPRB2004} and set $\epsilon_{ab}(0) = 4\epsilon_c(0)$.  This choice in ratio results in 
a factor two reduction in the breathing vertex at $\bq = (0,\pi)$ which is consistent with Ref. 
\onlinecite{RoschPRB2004}.  Additionally, 
we have set $\Omega = 60$ meV in Fig. \ref{Fig:pol}a.  The real part of $\Pi_{1,1}$ rises as 
$q^2$ before abruptly falling as the imaginary part rises when the condition $v_Fq = \Omega$ 
is satisfied.  These kinematic constraints give similar forms for the mixed polarizabilities 
$\Pi_{g,1}$, although the detailed $q$ dependence may be slightly different.   

In Fig. \ref{Fig:pol}b-f several renormalized (screened) el-ph vertices $\bar{g}(\bk_{AN},q_x,\Omega_{pl})$ 
are plotted for each of the modes considered in this work.  Here $\bk_{AN} = (\pi,0)$ and the momentum 
transfer is taken along the zone face $\bq = (q_x,0)$ and $\bar{g}$ is plotted for several values of 
$q_z$ as indicated.  To illustrate the effects of poor screening of the long-range Coulomb 
interaction, in Fig. \ref{Fig:pol}b, a momentum-independent bare coupling $g = 1$ (arb. units) 
is also shown. For a momentum-independent vertex $\Pi_{g,1} = g\Pi_{1,1}$.  In this case, for 
$\Omega \ll \Omega_{pl}(q)$ and $q_z = 0$ screening is perfect, leaving only a small 
el-ph interaction at large $\bq_{2D}$. 
This can be seen as the black line, which corresponds to $q_z = 0$ in Figs. \ref{Fig:pol}b-f.  For 
finite $q_z$ a small cone of wavevectors satisfies $\Omega \ge \Omega_{pl}(\bq)$ where screening 
is inoperable and the renormalized vertex recovers the bare value.   This 
enhancement of the small $q$ coupling occurs over a small range of $q$-vectors until $(\bq_{2D},q_z)$ 
satisfy $\Omega_{pl}(q) = \Omega_{ph}$, where a logarithmic divergence occurs which is cut-off by 
damping.  For increasing $q_z$ this condition moves to a progressively larger $\bq_{2D}$ and a 
window of momentum points opens where the vertex can be well-represented by its bare value.  

Likewise, Fig. \ref{Fig:pol}c-f plot the renormalized couplings for the $A_{1g}$, $B_{1g}$, apical 
and breathing modes, 
respectively.  For large in-plane momentum transfers $\bq_{2D}$ the $c$-axis couplings are 
largely unaffected by screening due to the fall-off of the polarizability at large $\bq_{2D}$.  However, 
for the breathing branch, where the bare coupling weights large momentum transfers, the growth of the 
mixed polarizability leads to an overall suppression of the screened vertex at large momentum transfers.  
For small in-plane momentum transfers - relevant for $c$-axis Raman active phonons considered here - 
screening is ineffective and the coupling is anomalously enhanced over large $q$ coupling.  Again, 
for the breathing branch, only small effects are noticed at small $\bq_{2D}$ due to the nature 
of the deformation-type coupling for these modes.  Since we have neglected any fermionic dependence on the bare 
breathing couplings, the mixed polarizability is equal to the pure charge polarizability in this case.

\begin{figure}[t]
 \includegraphics[width=\columnwidth]{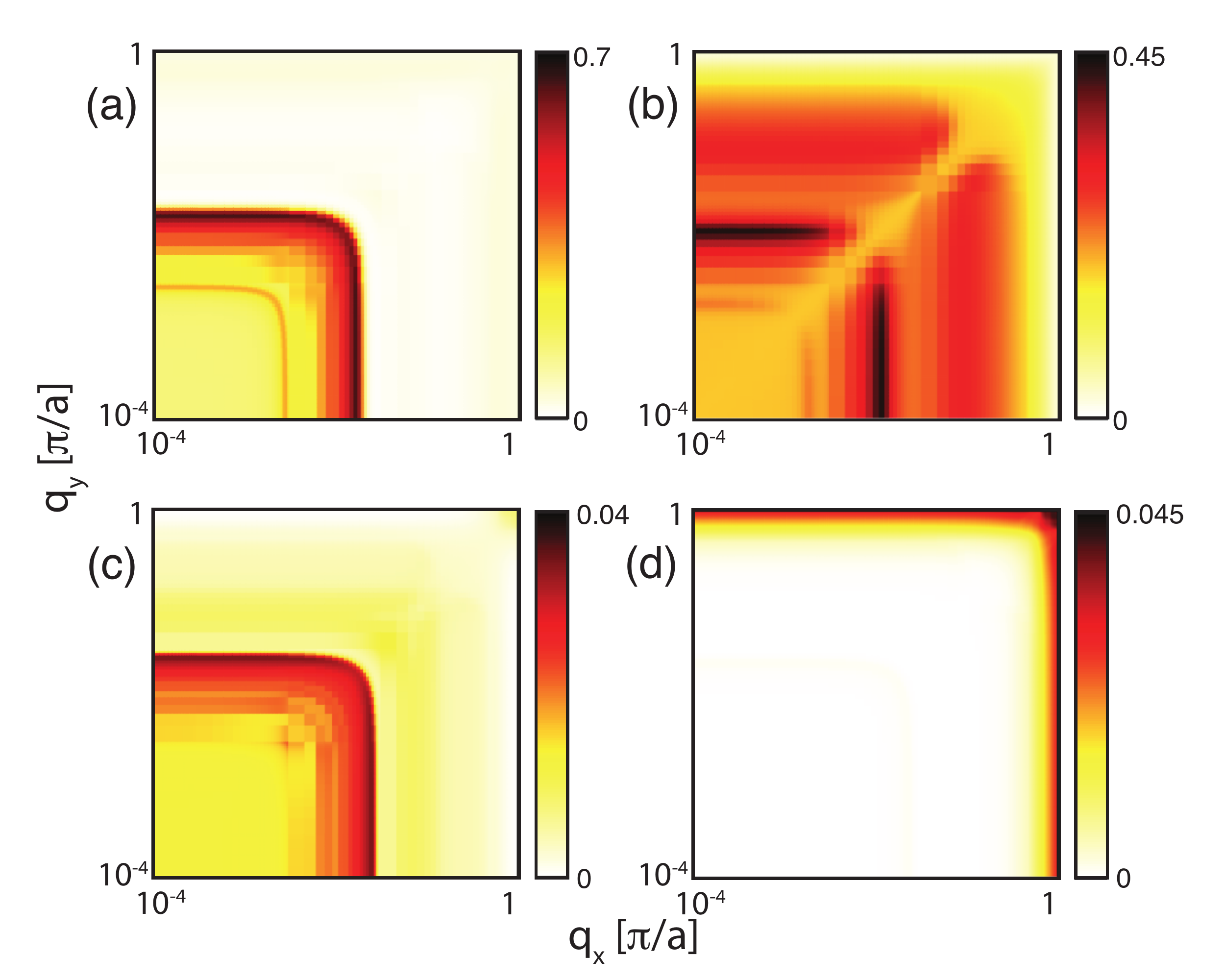}
 \caption{\label{Fig:Lambda_q2d} 
 (Color online) 
 Plots of the momentum dependent coupling $\lambda_{\nu}(\bq_{2D})$, Eq. (\ref{Eq:Lambda_q}), 
 for $\nu = $ $A_{1g}$ (a), $B_{1g}$ (b), apical (c) and bond-stretching (d) modes.  
 The coupling are plotted on a logarithmic scale for $\bq_{2D}$ to highlight the region 
 of poor screening.   } 
\end{figure}

Using Eq. (\ref{Eq:gbar}) the momentum-dependent el-ph coupling strength $\lambda_\nu(\bq_{2D})$ 
\begin{equation}\label{Eq:Lambda_q}
\lambda_\nu(\bq_{2D}) = \frac{2N_F}{N_c\Omega_\nu}\sum_{\bk,q_z} 
\langle|\bar{g}_\nu(\bk,\bq,\Omega_{pl})|^2 \rangle
\end{equation}
is plotted in Fig. \ref{Fig:Lambda_q2d} as a function 
of $\bq_{2D}$ (on a log scale to highlight the small $\bq$ behavior) 
for $\nu = $ $A_{1g}$, $B_{1g}$, apical and bond-stretching modes 
shown in panels Fig. \ref{Fig:Lambda_q2d}(a)-(d), respectively.  One can clearly see that when summed over all out-of-plane momentum 
transfers $q_z$, the net coupling is on the order of the bare coupling for small $\bq_{2D}$, while 
the coupling at large 
$\bq$ is suppressed.  The fermionic momentum dependence of the bare vertices for the $B_{1g}$ case 
noticeably alters $\lambda(\bq_{2D})$ for momentum transfers along the BZ diagonal where it is largely 
projected out.  The $A_{1g}$ and apical $\lambda(\bq_{2D})$ are quite similar even though the anisotropy of 
the bare interaction is substantially different and weights different regions of the BZ.  For all the 
$c$-axis modes, coupling is small for $\bq_{2D} = {\bf Q}_{AF} = (\pi/a,\pi/a)$.  Since we have 
neglected Hubbard short-range Coulomb repulsion, which further suppresses coupling at large momentum 
transfers,\cite{ZeyherPRB1996} Raman $c$-axis modes are thus not expected to appear in transport measurements.  
Likewise, the breathing modes are altered at small $\bq_{2D}$, but the overall coupling does not give much 
weight for these transfers.  Since the apical and $B_{1g}$ modes strongly favor coupling to the anti-nodal 
fermions, we may conclude that the underlying density of states, combined with the fermionic 
momentum dependence, strongly influences the magnitude of the couplings at small $\bq_{2D}$; the 
phonon and eigenvectors control the actual angular dependence while the nature of the coupling - 
electrostatic or deformation - controls the $|\bq|$-dependence.  

\begin{figure}[t]
 \includegraphics[width=\columnwidth]{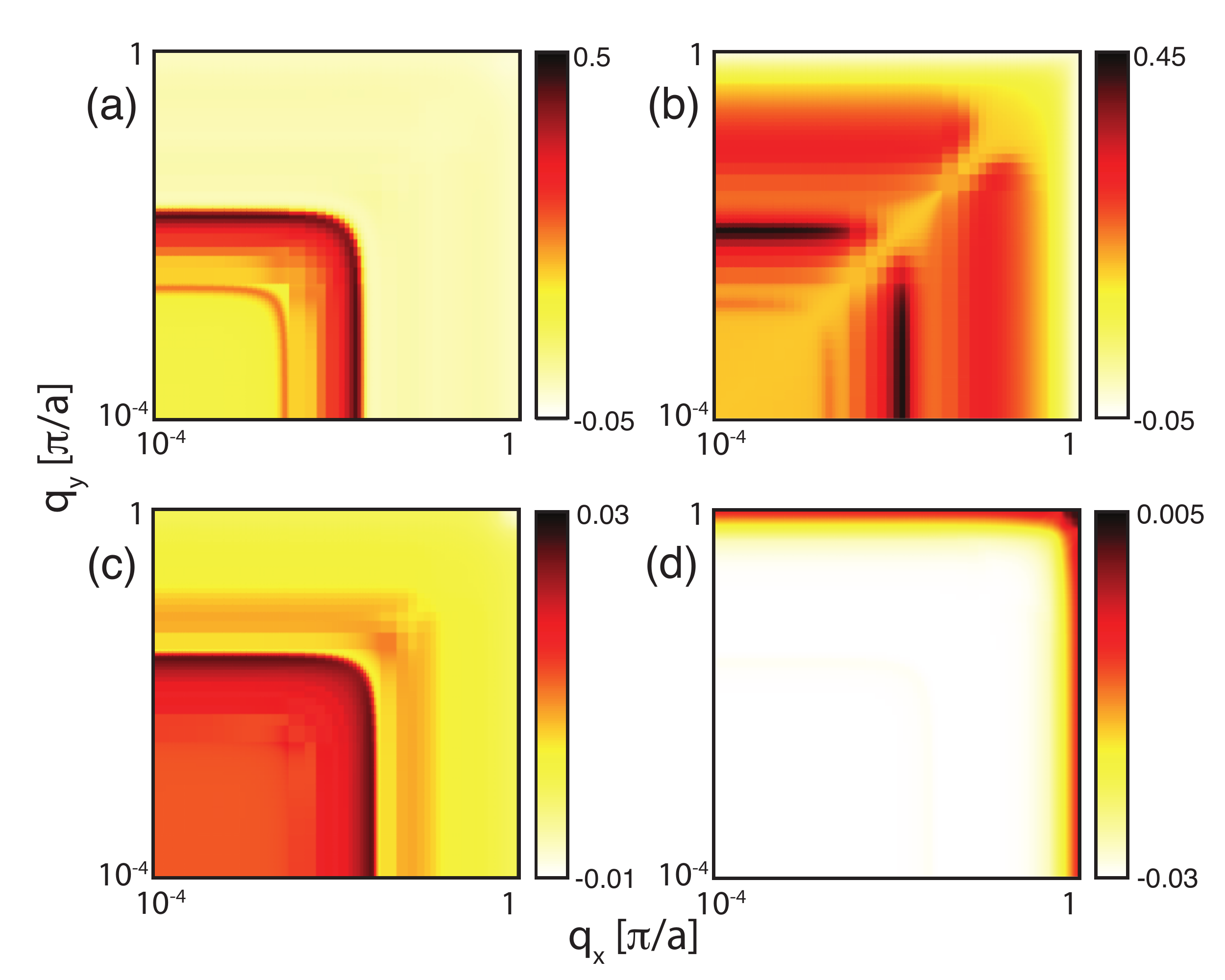}
 \caption{\label{Fig:Lambda_d_q2d} 
 (Color online) 
 Plots of the $d$-wave coupling $\lambda_{\nu,d}(\bq_{2D})$, Eq. (\ref{Eq:Lambda_q_d}), 
 for $\nu = $ $A_{1g}$ (a), $B_{1g}$ (b), apical (c) and bond-stretching (d) modes.  
 The coupling are plotted on a logarithmic scale for $\bq_{2D}$ to highlight the region 
 of poor screening.    
 }
\end{figure}

This $\bq$-dependence strongly affects the $d$-wave projected pair interaction as well.  In 
Fig. \ref{Fig:Lambda_d_q2d} we plot the $d$-wave projected coupling 
\begin{equation}\label{Eq:Lambda_q_d}
\lambda_{\nu,d}(\bq_{2D}) = \frac{2N^d}{N_c\Omega_\nu}\sum_{q_z} 
\langle|\bar{g}(\bk,\bq,\Omega_{pl})|^2d_\bk d_{\bk-\bq} \rangle^d_{\bk_{FS}}
\end{equation}
for the same modes considered in Fig. \ref{Fig:Lambda_q2d}.  As with Fig. \ref{Fig:Lambda_q2d}, 
$\lambda_{\nu,d}(\bq_{2D})$ has been plotted on a log scale in order to highlight behavior at 
small $\bq_{2D}$.  One can see straightforwardly 
that the phonons that strongly favor small $\bq$ scattering largely promote $d$-wave 
pairing (the $c$-axis modes) while phonons that favor large $\bq$-scattering, such as the 
bond-stretching modes, are detrimental.  Furthermore, for the bond-stretching modes, 
while a large region of $q$ space supports pairing, the large weight near $\bq = {\bf Q}$ 
dominates the coupling.  The consequence of poor screening for finite $q_z$ transfers, 
which accentuates small $q$ couplings, is to enhance the overall $d$-wave coupling 
compared to $q_z = 0$ as well as standard Debye screening ($\Omega_{pl} \gg \Omega_\nu$) 
for $c$-axis modes, and diminishes the repulsive part for the bond-stretching modes.  

\begin{figure}[t]
 \includegraphics[width=\columnwidth]{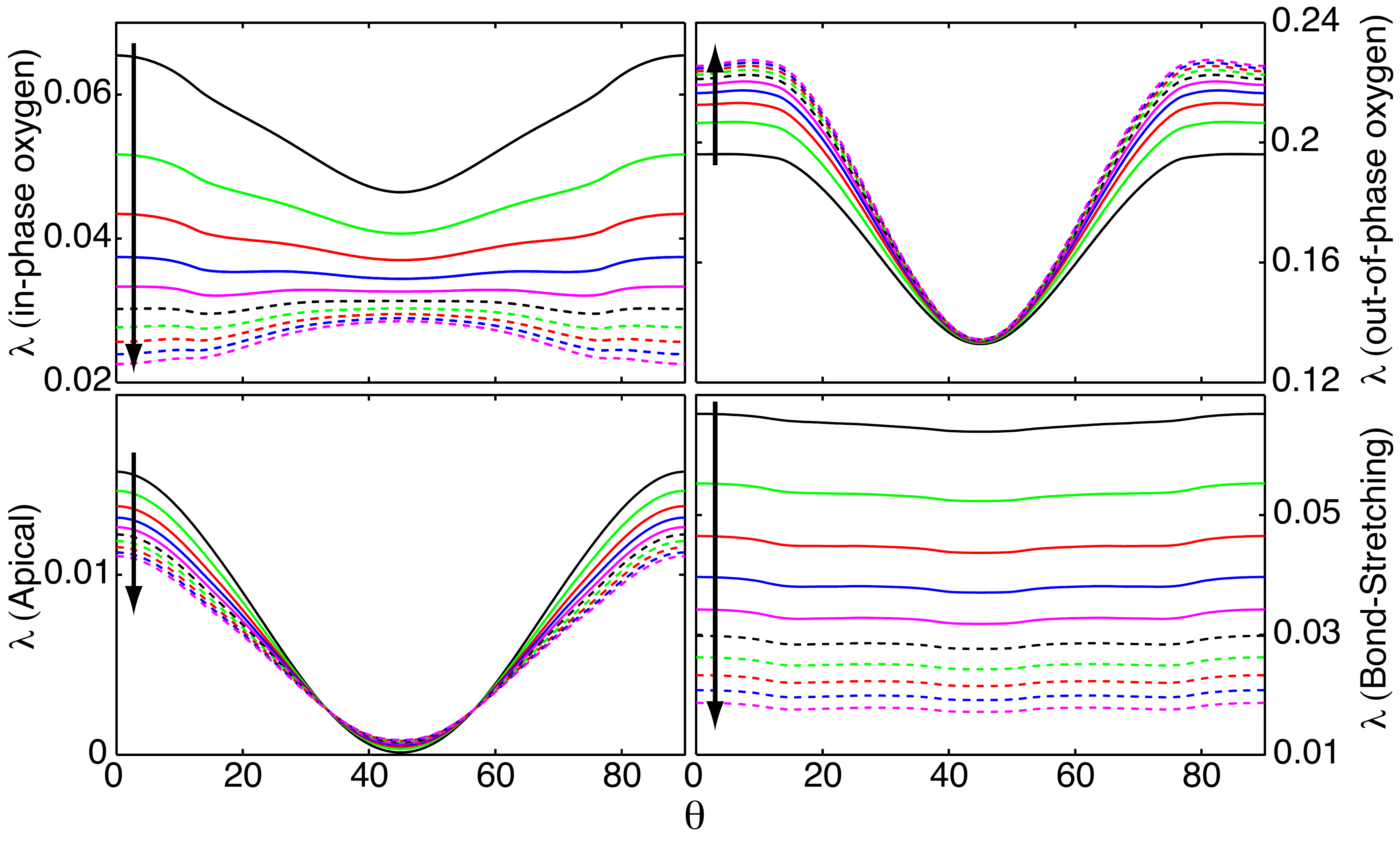}
 \caption{\label{Fig:lambda_fs_screened} 
 (Color online) Plots of the screened e-ph $\lambda(\bk_F)$ for fermions at the Fermi level as a function 
 of $\theta$ for several values of the plasma frequency $\Omega_{pl}$.  The value of 
 $\Omega_{pl}$ varies from 0.289 eV (black, solid) to 0.914 eV (purple, dashed).  
 The black arrows indicate the direction of increasing $\Omega_{pl}$.  
 The parameters used to generate this figure are defined in the text.   
 }
\end{figure}

\begin{figure}[t]
 \includegraphics[width=\columnwidth]{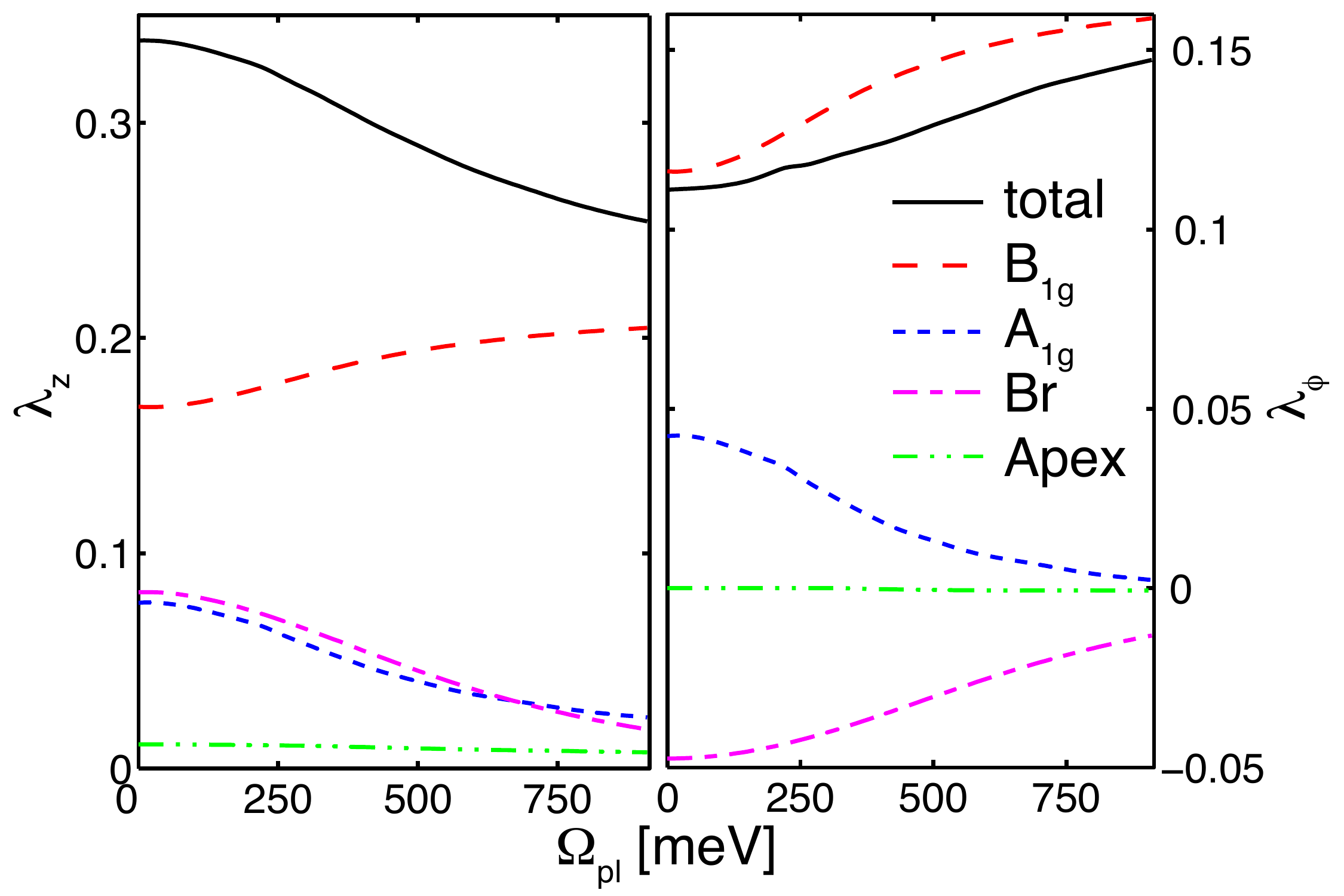}
 \caption{\label{Fig:lambda_total} 
 (Color online) 
 Plots of $\lambda_z$ (left) and $\lambda_\phi$ (right) as a function of the plasma frequency 
 $\Omega_{pl}$.  The parameters taken are defined in the text.}
\end{figure}

To visualize the effect of screening on the fermionic dependence of the coupling, we plot in 
Fig. \ref{Fig:lambda_fs_screened} the screened $\lambda_\nu(\bk_{FS})$, derived from 
Eq. (\ref{Eq:LambdaFS}) where $g$ has been replaced with $\bar{g}$ from Eq. (\ref{Eq:gbar}).  
Here, to mimic the effect of doping in the cuprates, we have varied the plasma frequency to 
smaller values in accordance with experiment, reflecting the increased insulating behavior 
and ionicity along the $c$-axis that occurs with underdoping in the phase diagram of hole-
doped cuprates.  The values for the total coupling $\lambda_z = \sum_\nu 
\langle\lambda_\nu(\bk)\rangle$ and $\lambda_\phi$ defined in Eq. (\ref{Eq:lambdad}) 
are shown in Fig. \ref{Fig:lambda_total}. 

Screening causes several noticeable effects compared to the unscreened case shown in Fig, 
\ref{Fig:lambda_fs}.  In order to mimic the effects of doping the screened el-ph vertex 
for the four branches is shown for $\Omega_{pl}$ ranging from 290 to 
914 meV.  (The black arrows indicate the direction of increasing $\Omega_{pl}$.) 
As the plasma frequency is increased the $A_{1g}$, apical and 
breathing branches can be screened more effectively and the overall vertex is lowered 
around the Fermi surface.  In the case of the $A_{1g}$ and apical $c$-axis branches, 
the largest effect occurs 
in the anti-nodal region where the bare vertex is largest. 
In the case of the $B_{1g}$ branch there is an anomalous anti-screening which 
occurs and the $B_{1g}$ vertex in the antinodal region is enhanced with increasing 
$\Omega_{pl}$.  This non-intuitive result is due to the out-of-phase oscillations of the 
oxygen modes since the only the difference between the bare vertices for 
the $A_{1g}$ and $B_{1g}$ branches is the phase of the phonon eigenvectors.  
From these results it is evident that the self-energy due to coupling to $c$-axis 
phonons should redistribute weight around the Fermi surface 
as the number of doped holes varies.  Thus, we infer that in the cuprates a window in 
$q$-space at small $\bq_{2D}$ occurs in which the el-ph interaction can still be quite large 
and avoid screening.  This would not occur if the material was fully conducting along the 
$c$-axis and this effect increases with underdoping as $\Omega_{pl}$ is reduced with decreasing hole 
concentration.  

In contrast, since poor screening does not affect large momentum transfers, the strength of the 
breathing coupling dramatically increases for smaller $\Omega_{pl}$.  This implies that the 
electron self-energy contribution from the breathing branches should grow with underdoping.  
Once again, however, we remark that this finding must be viewed 
with caution since large $\bq$-behavior is governed strongly by Coulomb interactions.  

Finally, in Fig. \ref{Fig:lambda_total} we plot the total coupling $\lambda_{z,\phi}$ resulting 
from the screened el-ph vertex as a function of $\Omega_{pl}$.  Following the trends in 
Fig. \ref{Fig:lambda_fs_screened}, the total coupling in both channels decreases with 
increasing $\Omega_{pl}$ for the in-plane 
breathing branch, as well as the apical and $A_{1g}$ branches, while the coupling to the 
$B_{1g}$ branch is enhanced.  In the case of the $d$-wave coupling, the attractive interaction 
of the $A_{1g}$ branch is largely cancelled by the repulsive interaction of the breathing 
branch.  Therefore, the total attractive interaction for $d$-wave pairing, which is primarily 
provided by the $B_{1g}$ branch, is expected to be enhanced with progressive overdoping. 

The transition from insulating to more metallic behavior, reflected in 
$\Omega_{pl}$, is one aspect of the doping 
dependence expected for el-ph coupling.  However, this is not the only change expected to 
occur with doping.  As a sample is doped away from half-filling the character of the 
$pd$-$\sigma^*$ band is expected to change as spectral weight redistributes itself 
within the quasiparticle and lower and upper Hubbard bands.\cite{MoritzNJP2009} 
This will be reflected in the 
parameters $A_O$, $A_a$ and $A_{Cu}$.  Since these parameters enter the el-ph vertices in the 
fourth power, doping-induced changes in the bandstructure 
can have a large impact on the strength of the el-ph coupling.  The value of the 
local crystal fields are also expected to vary with doping.  This was shown in 
Ref. \onlinecite{JohnstonEPL2009}, which examined the role of an interstitial oxygen dopant 
in the SrO/BiO layers of Bi-2212 and found that the dopant locally enhanced $E_z$ 
at the planar oxygen site by up to a factor of five.  
The structural details of the crystal can also induce changes to the character of the band at 
Fermi level ($A_O$, $A_{Cu}$ and $A_a$), through variations in the Madelung energies of the atoms 
and the degree of hybridization between the CuO$_2$ plane and off-plane atoms such as the apical 
oxygen.   Due to these considerations the overall strength of the el-ph coupling can have a complex 
dependence on both the carrier concentration of the CuO$_2$ plane as well as 
the composition and structure of the material.  In the next section we examine the materials 
dependence of the el-ph interaction by conducting a systematic examination of the role 
of structure in determining the strength of the interaction across various families of 
high-T$_c$ cuprates.  

\section{\label{Madelung}Madelung Energies and Local Fields}

The environment around the CuO$_2$ plane influences electron dynamics in the plane via 
the Madelung energies and local crystal fields.  As a result, local symmetry breaking plays an 
important role in lattice dynamics.  For example, X-ray measurements on YBa$_2$Cu$_3$O$_7$ \cite{xray}
have shown that the CuO$_2$ planes are statically buckled, and a linear el-ph coupling 
results from the breaking of local mirror plane symmetry.  LDA calculations for the infinite 
layer material CaCuO$_2$ also found evidence for static buckling of the plane due to the steric 
interactions among the oxygen 2$p_z$ orbitals which produces a substantial linear $B_{1g}$ el-ph 
coupling.\cite{AndersenJLTP1996} Among the different contributions to local symmetry breaking, 
this section is devoted to an investigation of the magnitude of the Madelung energies and 
local fields across families of the cuprates as a mechanism of local symmetry breaking.  

We remark that LDA investigations of el-ph coupling are limited mostly to stoichiometric YBCO, with virtual crystal
extensions to LSCO and CaCuO. LDA bandstructure, being metallic in nature, will tend to screen charge variations whereas
those charge variations, particularly off the CuO$_2$ plane, may be unscreened and result in strong local modifications of 
band parameters triggering nano-scale inhomogeneity as well as strong variations of el-ph coupling. 
This was demonstrated in Ref. \onlinecite{JohnstonEPL2009}, where it was shown that the presence of 
an interstitial oxygen dopant's unscreened charge in the SrO/BiO layers produced 
large local enhancements in the crystal field strength at the planar oxygen sites.  This resulted in a 
local increase in the strength of the coupling to the $A_{1g}$ and $B_{1g}$ branches,  which in turn reduced the energy gap and broadened 
the spectral features in the hole addition/removal spectrum as well as locally increased $J$ via the gain 
in lattice energy.  

Here, we focus our calculations in the ionic limit, where the ions are represented 
as point charges, and evaluate the electrostatic sums using Ewald's method.  We show that the Madelung potential 
landscape is sensitive to the details of the unit cell, producing material-dependent variations in the field 
strength at the planar and apical oxygen sites.  For the field at the oxygen site of the outermost CuO$_2$ plane, which determines 
the coupling to the $A_{1g}$ and $B_{1g}$ modes, this variation mirrors the observed variations in T$_c$ 
suggesting a link between these quantities. 
The use of formal valences in the electrostatic calculations reflects the stociometric compounds and therefore 
the estimates obtained from their use corresponds to the parent compounds.  However, in 
light of the findings of Ref. \onlinecite{JohnstonEPL2009}, we also address doping-induced changes in the crystal 
field strength in Bi-2212 in order to assess how these fields are expected to vary with doping.   Finally, 
the variations in $\lambda_{z,\phi}$ that arise as a function of material due to the  
electrostatic considerations presented here are also discussed.    
Here we focus on the $B_{1g}$ mode, which has the largest contribution to $d$-wave pairing and take in 
into account variations in the local field, relative shifts in site energies, and changes in the 
atomic character of the band, all of which are directly affected by the 
details of the crystal structure.    

\subsection{Formal Valances: The Parent Compounds}

\begin{figure}[t]
 \includegraphics[width=\columnwidth]{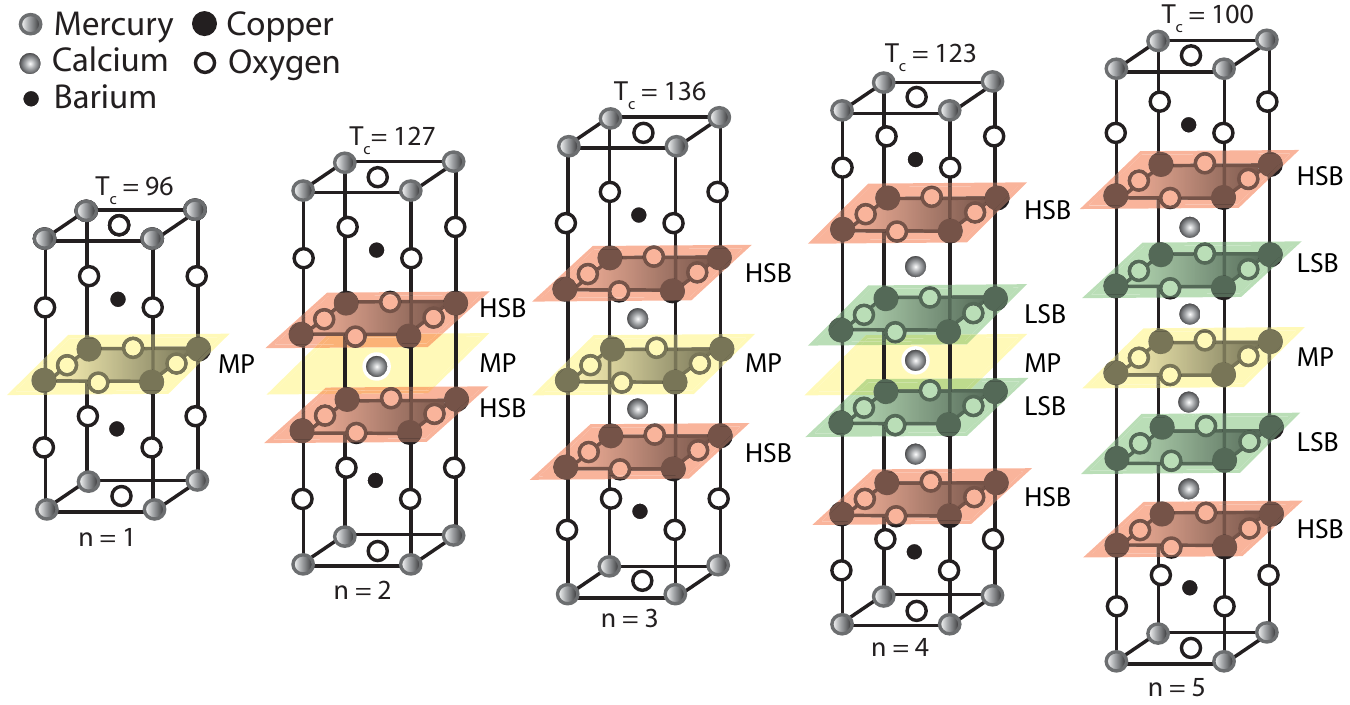}
 \caption{\label{Fig:Structure}
 (Color online)  
 The HgBa$_2$Ca$_{n-1}$Cu$_n$O$_{4n+\delta}$ ($n = 1-5$) unit cell showing the locations of 
 the mirror plane (yellow, MP), low symmetry breaking plane (green, LSB) and high symmetry breaking 
 planes (red, HSB).  The degree of symmetry breaking establishes the strength of the electric 
 field which couples to the $c$-axis phonons. 
 }
\end{figure}

A structural view of the lattice across the Hg family, as shown in Fig. \ref{Fig:Structure}. 
Starting with $n=1$ shows that the coupling to planar oxygen $c$-axis phonons must be weak, while 
couplings involving the apical oxygen phonon can be strong due to the asymmetry along the $c$-axis 
at the apical site (CuO$_2$ on one side and HgO on the other).  For $n > 1$ the outer layer 
does not lie in a mirror-plane and the coupling strength depends on the local asymmetry 
(in this case BaO on one side, Ca on the other).  However, the innermost plane may lie in a mirror plane 
when $n$ is odd, while the intermediate planes will generally have a weaker degree of symmetry 
breaking in comparison to the outermost plane.  Thus, in the context of el-ph coupling, 
the most strongly coupled modes will be those involving displacements of the apical and planar 
oxygen atoms in the outermost layers.  In Hg compounds the local asymmetry around the outermost 
plane is due to Ca and BaO structures on opposite sides of the CuO$_2$ plane. In YBCO it is 
due to Ca and SrO structures.  Such changes in composition will affect the magnitude 
of the crystal field and these differences can be quantified by examining the Madelung 
potential for the apical and planar oxygen sites for the Hg, Tl and Bi cuprate families.       

The electric field strength can be regarded as the first derivative of the Madelung potential, which 
consists of a contribution of all electrostatic interactions between ions in the solid.  For a 
test charge $q_t$, located at position ${\bf r}$, the Madelung potential $\Phi$ is defined as: 
\begin{equation}\label{Eq:Madelung}
\Phi = q_t\int d{\bf r}^\prime \frac{\rho({\bf r}^\prime)}{|\br - \br^\prime|}
\end{equation}
where $\rho$ is the charge density.  In the ionic limit the charge density is comprised of 
point charges $\rho(\br) = \sum_i\delta(\br-{\bf R}_i)$, where the sum $i$ is taken over the 
atoms of the crystal located at ${\bf R}_i$.  

\begin{figure}[t]
 \includegraphics[width=\columnwidth]{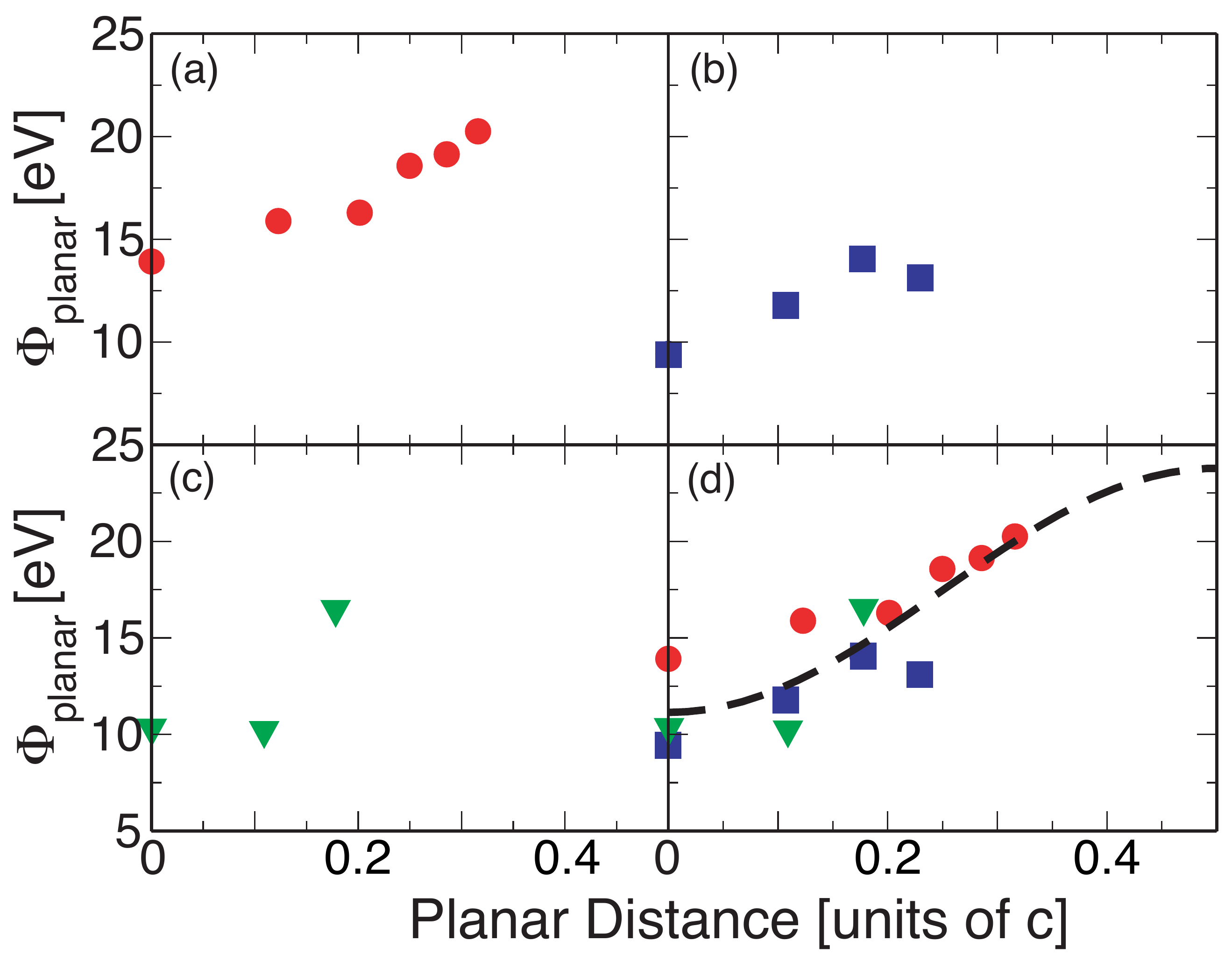}
 \caption{\label{Fig:MadelungB1g} (Color online) The Madelung
  potential at the CuO$_2$ planar oxygen site as function of
  distance from the plane of mirror symmetry: (a)
  HgBa$_{2}$Ca$_{n-1}$Cu$_{n}$O$_{2n+2}$ ($n$ = 1-6, red circles).
  (b) Tl$_{2}$Ba$_{2}$Ca$_{n-1}$Cu$_{n}$O$_{2n+2}$, ($n$ = 1-4, blue
  squares). (c) Bi$_{2}$Sr$_{2}$Ca$_{n-1}$Cu$_{n}$O$_{2n+2}$ ($n$ =
  1-3, green triangles). (d) An overlay of the Hg, Tl and Bi plots
  rescaled by the lattice spacing, $a$.  The black line is a fitted
  curve of the form $a\Phi = A + B\cos(2\pi z/c)$.}
\end{figure}

There are several common methods used to calculate the Madelung
potential. A general and powerful method was given by
Ewald,\cite{Ewald, LiLidik} which is used here to determine the Madelung
energies for the planar and apical oxygen sites in the outermost
CuO$_{2}$ planes in the unit cell for Bi (n=1-3), Tl (n=1-4) and
Hg (n=1-6) cuprates.  The atomic positions are obtained from the known 
structural data given in Refs. 
\onlinecite{LSCO1,LSCO2,LBCO,Bi2201,Bi2212,Bi2223,Hg1201,Hg1212,Hg1223,Hg1234,
Hg1245_Hg1256,Tl2201,Tl2212_Tl2223,Tl2234}.
In all cases formal valence charges for the ions have been
assumed. It should be noted that for the materials considered
herein, the surfaces are non-polar, as opposed to
YBa$_2$Cu$_3$O$_7$ (YBCO).\cite{LiLidik} Therefore the surface termination layer is
not critical for our calculations. 

In the case of YBCO, assuming that the surface is terminated between the 
BaO and CuO sublayers, a local field between 0.8 - 2.1 eV/\AA\space was obtained, arising 
from the asymmetry condition of Y$^{3+}$ and Ba$^{2+}$ on either side of the CuO$_2$ 
plane.\cite{LiLidik} This polar surface contribution overwhelms the periodic 
contribution from Ewald's method. Without the polar contribution, the coupling for 
example to the $B_{1g}$ mode is underestimated when compared to that obtained from periodic 
LDA methods. In Ref. \onlinecite{LiLidik} it was shown how deviations away from 
formal valences can give much different values for the local crystal fields. 
We have also found that the local electric fields
are more sensitive than the Madelung energy to small structural distortions and 
deviations away from formal valences. 
Therefore we conclude that while the overall field values should be viewed 
as being approximate, we assume that the variations across the different 
materials can be qualitatively compared.

\begin{figure}[t]
 \includegraphics[width=\columnwidth]{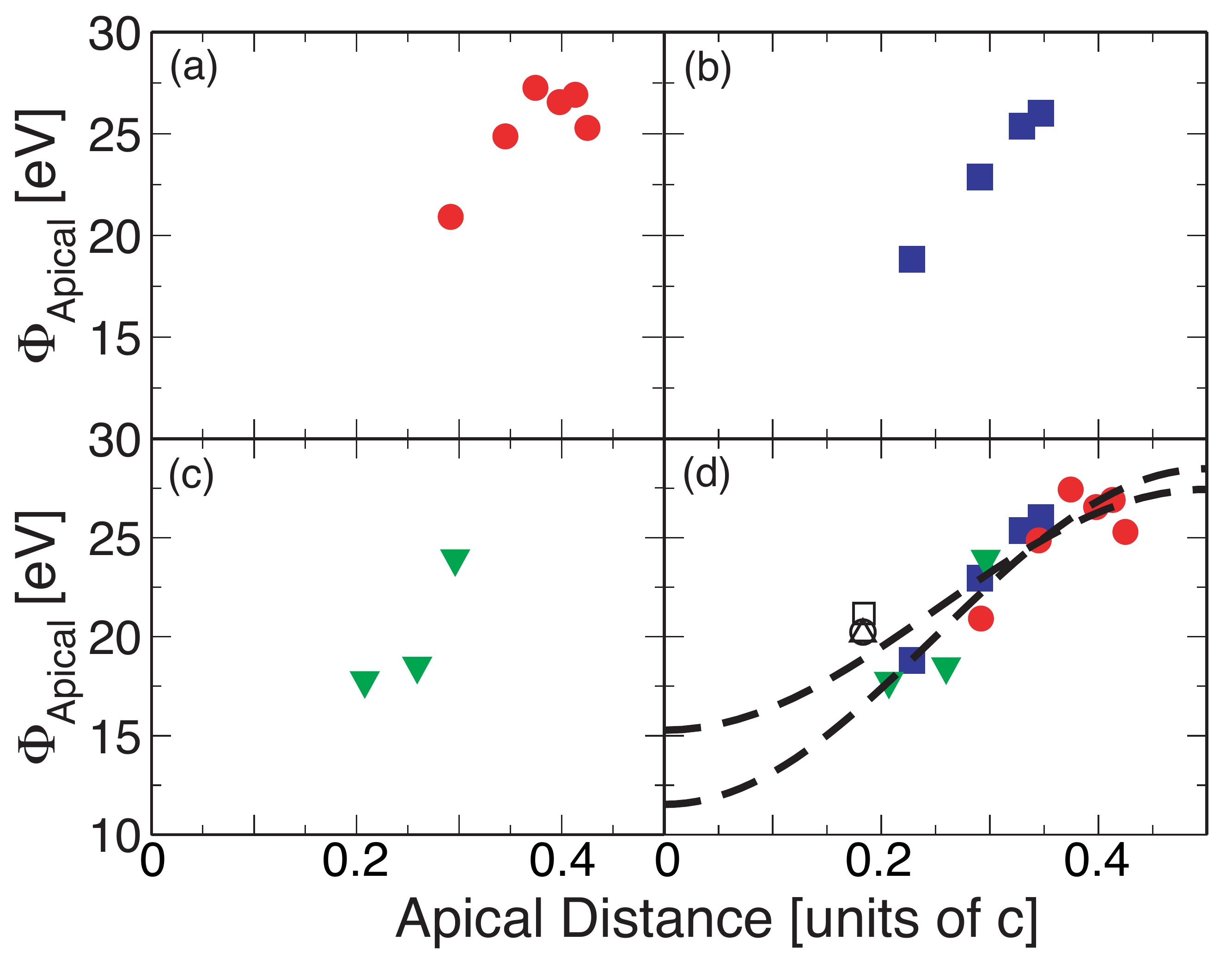}
 \caption{\label{Fig:MadelungApex}
 (Color online) The Madelung energy of the apical oxygen site as a function of 
distance from the plane of mirror symmetry: (a) 
HgBa$_{2}$Ca$_{n-1}$Cu$_{n}$O$_{2n+2}$ ($n$ = 1-6, red circles). 
(b) Tl$_{2}$Ba$_{2}$Ca$_{n-1}$Cu$_{n}$O$_{2n+2}$, ($n$ = 1-4, blue 
squares). (c) Bi$_{2}$Sr$_{2}$Ca$_{n-1}$Cu$_{n}$O$_{2n+2}$ ($n$ 
= 1-3, green triangles). (d) An overlay of the Hg, Tl and Bi plots in 
addition of points obtained for LSCO and LBCO (black 
open symbols). The dashed and solid black lines 
are fitted curves of the form $a\Phi = A + B\cos(2\pi z/c)$. The 
dashed curve is obtained when all of the data is included while 
the solid curve is obtained when the La data points are excluded.
 }
\end{figure}

The Madelung energies of the planar and apical oxygen sites are plotted in 
Figs. \ref{Fig:MadelungB1g} and \ref{Fig:MadelungApex}, respectively, and 
plotted as a function of the distance between outermost CuO$_2$ plane (or apical oxygen) 
and the mirror plane located at the center of the unit cell.   
The calculated apical Madelung energies are in agreement with those obtained 
in Ref. \onlinecite{OhtaPRB1991} for similar structural data, while the Madelung 
energies for the planar sites are, to the best of our knowledge, presented here 
for the first time.  One can see that the Madelung energies for both the planar and 
apical sites increase when the outermost CuO$_2$ plane lies further from the center of the unit cell.  
This occurs both when the number of planes per unit cell is increased as well as 
when the layers are spaced closer to the charge reservoir due to the larger spacer 
ions between the CuO$_2$ layers.  Scaling the Madelung energies by the in-plane 
lattice constant $a$ collapses the data for each cuprate family onto a single curve, 
as shown in subpanel (d) of Figs. \ref{Fig:MadelungB1g} and \ref{Fig:MadelungApex}.  
By symmetry, the electric field falls to zero at a mirror plane implying that 
$\Phi(z/c)$ must have zero slope at these points.  If $z$ is measured relative 
to the mirror plane, such as those indicated in Fig. \ref{Fig:Structure}, 
then the simplest function that satisfies these boundary conditions is 
$a\Phi(z/c) = A + B\cos(2\pi z/c)$, and serves as an overall guide for 
describing the behavior of the data.  For the planar oxygen sites we obtain 
$A = 66.3$, $B = -24.0$ while for the apical oxygen sites $A = 81.2$, $B = -23.1$ 
($A = 76.0$, $B = -32.2$ when the points for the La family are excluded), in units 
of $eV\cdot$\AA.   

\begin{figure}[t]
 \includegraphics[width=\columnwidth]{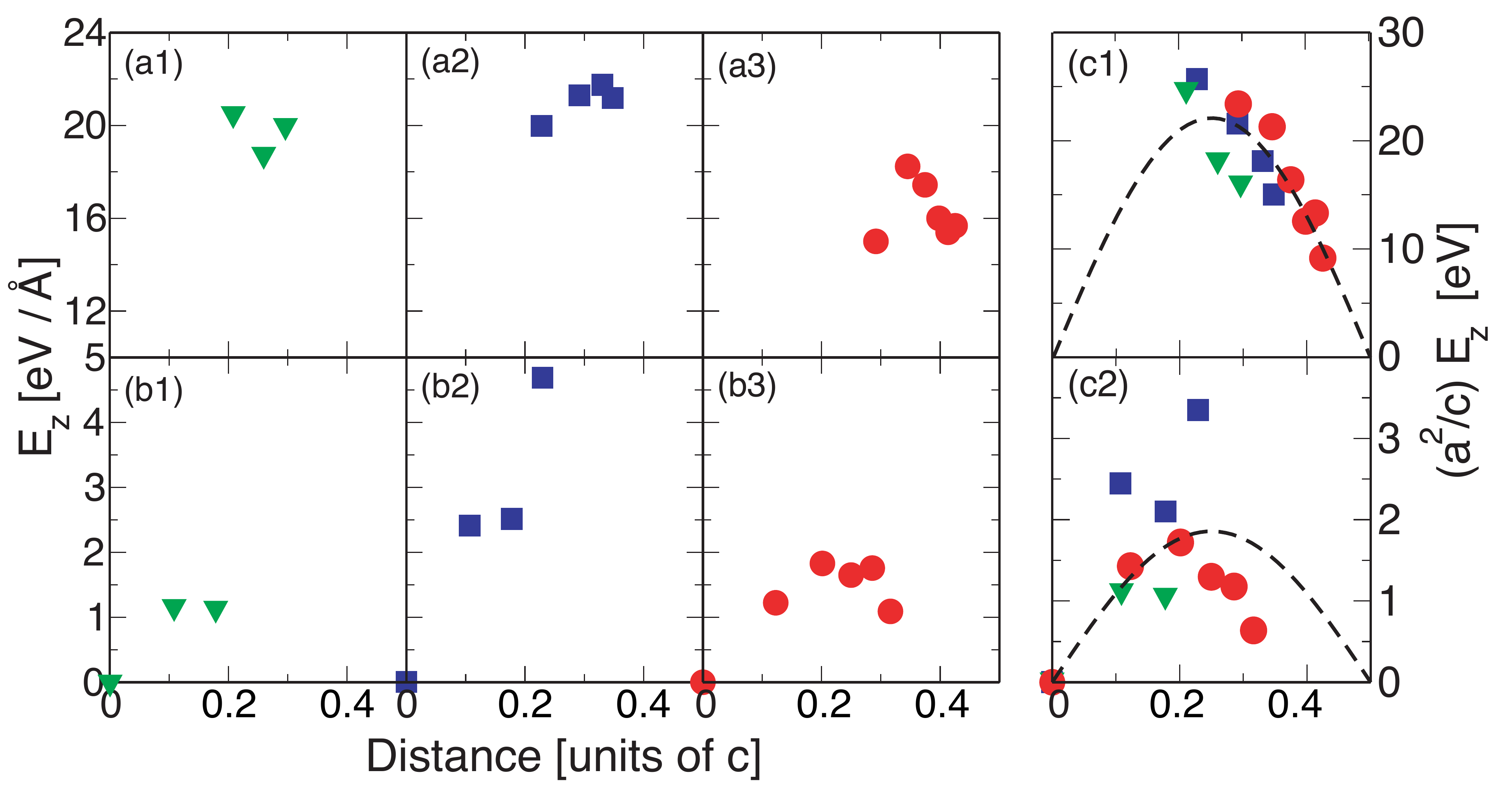}
 \caption{\label{Fig:Efield} (Color online) (a1)-(a3) The magnitude 
  of the electric field at the apical oxygen site of the 
  Hg, Tl and Bi families of cuprates, respectively. The data is 
  plotted as a function of the apex distance from the plane of 
  mirror symmetry located at the center of the CuO$_2$ planes. 
  (b1)-(b3) The magnitude of the electric field at 
  the planar oxygen site of the outermost CuO$_2$ plane for the set of same 
  compounds. (c1)-(c3) The strength of 
  the electric field scaled by the unit cell lattice parameters for 
  the planar and apical oxygen fields, respectively. The dashed 
  lines represents a fit $a^2 E(z/c)/c = A\sin(2\pi z/c)$, where A = 
  7.1 and 83.9 eV for the planar and apical sites, 
  respectively.  }
\end{figure}

Fig. \ref{Fig:Efield} shows the local $E$ fields at the apical and planar oxygen sites 
for the outermost CuO$_2$ plane in the unit cell for a number of cuprate families 
having different numbers of layers.  For the single layer cuprates the CuO$_2$ plane 
lies in a mirror plane located at the center of the unit cell resulting in a zero 
field at the planar oxygen sites.  Thus there is no el-ph first order coupling to 
the $A_{1g}$ and $B_{1g}$ branches in the single layer materials.  This is no longer 
the case for $n > 1$.  Examining the largest data set, that of the Hg family ($n=1-6$), 
the $E$-field at the planar oxygen site rises from $n=1$ and peaks for $n = 3$ where the 
location of the outer plane of the cell is $z/c \sim 0.2$.  As $n$ is increased beyond this 
point the magnitude of the crystal field begins to decrease.  
The data for the Bi- and Tl-families are consistent 
with this trend - generally the local fields rise and follow the associated Madelung 
energy as the outermost plane moves further from the center of the cell, and falls once the 
outermost plane moves closer to the edges of the unit cell where the Madelung energy 
profile flattens.    
Since the apical oxygen does not lie in a mirror plane for any cuprate, a finite field is allowed 
even for $n = 1$.  Behaviour similar to the planar oxygen site is also seen with a peak 
field strength generally occurring for $n = 3$ and decreasing beyond this point. 

The scatter in values is largely due to uncertainties in structural data as well as the 
oversimplified use of formal valences for all materials.  However, the results 
are similar among the different cuprates if the position of the outermost plane 
relative to the center of the cell is considered.  Figs. \ref{Fig:Efield}c1 and 
\ref{Fig:Efield}c2 show that the local field may be collapsed onto a sinusoidal curve 
as a function of $c$-axis distance when rescaled to account for the unit cell volume 
variations.  

The presented $E$-field values are sensitive to the structural data, 
particularly for the planar oxygen sites where the degree of static Cu-O bond buckling 
is relatively uncertain.  Since a static buckling further breaks mirror plane symmetry 
across the CuO$_2$ plane, small variations in the degree of buckling can have a large 
impact on the field values obtained while only having a minor effect on the Madelung 
energies.  For example, neglecting buckling on the order of $0.001c$ in 
the $n=2$ layer materials results in $E$-field variations on the order of 20-40 percent 
for the planar oxygen sites, while the variations in the corresponding Madelung energies 
vary only by a few percent.  We have found that both the Madelung energies and local 
fields for the apical oxygen are not as sensitive to the degree of static 
buckling in the CuO$_2$ plane.  

We now comment on the $E$-field values reported here in the context of previous works 
involving coupling to the $c$-axis modes.  
Due to the fact that Ba$^{2+}$ and Sr$^{2+}$ lie on either side of the 
CuO$_2$ plane in Bi-2212, previous expectations were that the planar field 
in this system would be smaller than that found in YBCO and work by one of the 
authors on the Raman derived $B_{1g}$ phonon lineshape confirmed 
these expectations.\cite{tpdSSC}  Since the T$_c$'s of these two materials are 
roughly equivalent, it was concluded that the el-ph coupling was irrelevant to the pairing 
mechanism.  However, a re-examination of the data set, in view of the greater sample 
inhomogeneity and impurity effects in Bi-2212 compared to YBCO, indicates empirically that 
the field may not be as small as previously considered.  Using the theory of Ref. 
\onlinecite{tpdPRB1995}, the Fano line-shape of the $B_{1g}$ phonon in optimally doped 
Bi-2212\cite{tpdSSC} can be fit equivalently with a value of the coupling, parameterized 
by $\lambda$, which is an order of magnitude different if one assumes a large, 
intrinsic broadening of the phonon line.  Since the values of $\lambda$ extracted from 
fitting Raman lineshapes are sensitive to disorder for values of $\lambda$ in the 
range of those found in YBa$_2$Ca$_3$O$_{7-\delta}$, a direct comparison with other 
compounds should again be considered qualitatively, with a view towards materials of 
comparable quality.  Our Ewald calculations confirm that the field at the planar 
oxygen site for Bi-2212 ($E_z = 1.16$ eV/\AA) is similar to 
YBa$_2$Cu$_3$O$_{7-\delta}$. We show in the next section that this value is 
likely underestimated once considerations for doping are taken into account.

\subsection{Doping Dependence of the Electric Field}

We now briefly comment on the doping dependence of the local 
crystal fields.  In discussing the trends in $E_z$ across the cuprates we have assumed  
formal valences for each lattice site, which is appropriate for undoped parent 
compounds. However, as was shown in Ref. \onlinecite{JohnstonEPL2009}, the doping process 
can significantly affect the strength of these fields resulting in a substantial  
local increase 
in $\lambda$'s.  From a symmetry point of view, the movement of charge from the 
CuO$_2$ plane to the charge reservoir layers with increased doping will 
enhance local symmetry breaking across plane.  To model this effect 
in Bi-2212 we repeated the calculation for $E_z$, but this 
time with the charge of the CuO$_2$ planes uniformly raised by $0.15e$, corresponding 
to optimal doping.  In order to maintain charge neutrality the charge of the 
BiO and SrO layers, where the O dopants are known to sit,\cite{HePRL2006} were 
uniformly reduced by the appropriate amount.  Using this crude model  
we find the $E_z$ field at the planar site is raised to $3.56$ eV/\AA\space while 
the field at the apical site is reduced to $16.33$ eV/\AA, in agreement with the 
general trends previously reported.\cite{JohnstonEPL2009} 
In section III we were interested in determining the values of $\lambda_{z,\phi}$ at 
optimal doping and therefore these are the values of 
the crystal field strength which were used in section \ref{Section:Lambda} to obtain 
the unscreened $\lambda_{z,\phi}(\bk)$ and total $\lambda_{z,\phi}$'s.  

Throughout this paper we have discussed a number of aspects of el-ph coupling 
in the cuprates which are 
affected by doping the system.  The local $E$-field strength, the ionicity 
of the compound, characterized by $\Omega_{pl}$, and the orbital character of the band 
at the Fermi level all modify the el-ph couplings and are all affected by 
doping.  Therefore, el-ph coupling in the cuprates is expected to 
exhibit a strong doping dependence.  We also note that many of these parameters are also 
expected to vary between materials comprising the various families and we 
turn our attention to these trends in the following section.  

\subsection{Material Dependence of $\lambda_{z,\phi}$ for the $B_{1g}$ branch}

We end this section by examining the systematic variation in the el-ph coupling 
strength expected from material to material.  Since our goal is to understand the 
materials variations in T$_c$ we focus our attention on coupling to the $B_{1g}$ branch, 
which provides the largest contribution to pairing and whose coupling is the strongest 
after the effects of screening are taken into account.  
As discussed previously, there is a empirical relationship between T$_c$ and the 
distance of the apical oxygen from the CuO$_2$ plane and this relationship this has 
generally been tied to the effective increase in the next nearest neighbour 
hopping $t^\prime$ through hybridization effects associated with the Cu 4$s$ orbital.  
These hybridization effects will also alter the character of the  
$pd$-$\sigma^*$ band crossing the Fermi level and can therefore affect the overall 
strength of the coupling to all of the phonon branches considered in this work.  
For the $B_{1g}$ branch, such changes are reflected in $A_O$ and, since this 
enters the total coupling as the fourth power, changes in hybridization can have 
a large affect on the total coupling strength. 

\begin{figure}[t]
 \includegraphics[width=0.7\columnwidth]{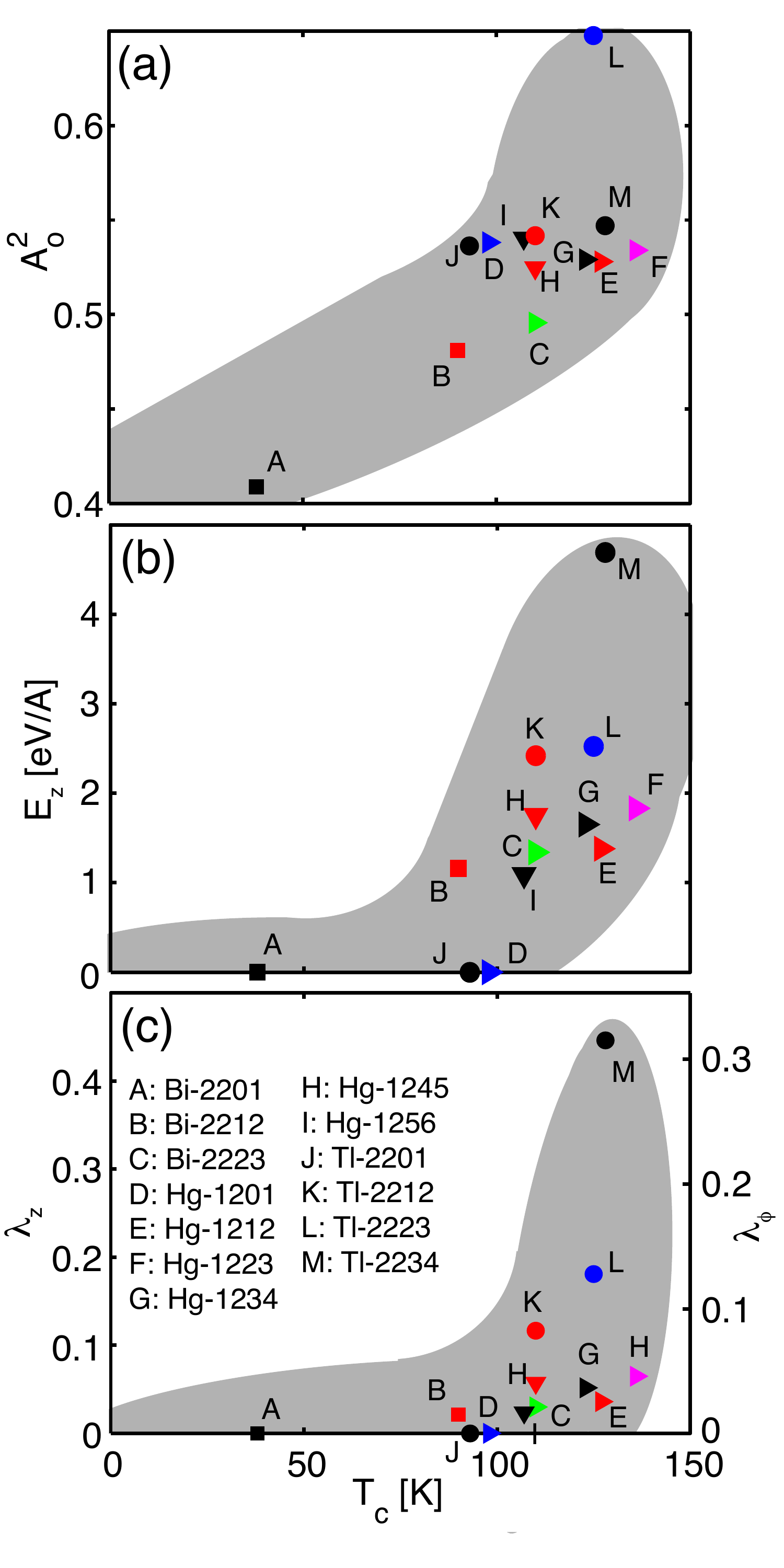}
 \caption{\label{Fig:OxygenCharacter} (Color online) (a) The planar oxygen character at the 
 Fermi surface $A_O^2$ as a function of the material's T$_c$.  (b) The corresponding 
 local $E$-field at the planar oxygen site of the outermost CuO$_2$ plane 
 in the parent compound. (c) The value of $\lambda_z$ for the $B_{1g}$ modes 
 of the outermost plane for each material. }
\end{figure}

In Fig. \ref{Fig:OxygenCharacter}a the systematic 
variation of $A_O$ is presented as a function of $T_c$ for the same materials considered 
in the Fig. \ref{Fig:Efield}.  To estimate  
$A_O$ we consider the same five-band model used in section \ref{Section:Lambda}, 
but assign the site energies based on the Madelung calculations presented  
here and the energy level scheme of Ref. \onlinecite{OhtaPRB1991}.  
Using hole language, we set $\epsilon_d = 0$, 
$\epsilon_p = \Delta$, $\epsilon_a = \Delta + \Delta\Phi_A/\epsilon(\infty)$ and 
$\epsilon_s = \epsilon_d - 7$.  Here, $\Delta$ is the charge transfer energy, 
which is related to the difference in Madelung energies between the Cu and planar 
O sites $\Delta\Phi_M = \Phi_{O} - \Phi_{Cu}$ and is given by
\begin{equation}
\Delta = \frac{\Delta\Phi_M}{\epsilon(\infty)} - I_{Cu}(2) + A_O(2) -\frac{e^2}{d_p}
\end{equation}
where $I_{Cu}(2)$ and $A_O(2)$ are the second ionization and second electron 
affinity energies for the Cu and O sites, respectively. The factor of 
$e^2/d_p$ represents the contribution of the Coulomb interaction between 
the introduced electron-hole pair. Following Ref. \onlinecite{OhtaPRB1991}, 
we take the dielectric constant $\epsilon(\infty) = 3.5$ and 
$I_{Cu}(2)-A_O(2) + e^2/d_p = 10.9$ eV.  In determining $\epsilon_a$, 
$\Delta\Phi_A = \Phi_A - \Phi_O$ denotes the Madelung energy difference 
between the apical and planar oxygen sites.   
In addition to varying the site energies between materials, the 
values of the hoppings $t^\prime_{pp}$ and $t^\prime_{ps}$ are also adjusted using 
the prescription of Ref. \onlinecite{Harrison}.  This allows us to  
account for changes in hybridization due to the different 
apical distance given in the experimental structural data.  
Furthermore, we neglect any variation of the $a$ and $b$ lattice constants and hold 
the in-plane hybridizations fixed throughout.   
Finally, correlations are handled at the mean field level.  The    
overall effect of the mean field corrections is to shift the 
site energies 
\begin{eqnarray}\label{Eq:Meanfield} \nonumber
 \tilde{\epsilon}_{d}&=& U_{dd}\langle n^d_{i}\rangle/2
                      + 4U_{pd}\langle n^p_i \rangle
                      + U_{pd}\langle n^a_i \rangle \\
 \tilde{\epsilon}_{p}&=&\epsilon_p + U_{pp}\langle n_i^p \rangle/2
                      + 2U_{pd}\langle n^p_{i} \rangle \\ \nonumber
 \tilde{\epsilon}_{a}&=& \epsilon_a + U_{pp}\langle n^a_i \rangle/2 
                      + U_{pd}\langle n^d_i \rangle  
\end{eqnarray}
and $\tilde{\epsilon}_s$ is fixed to maintain the difference 
$\tilde{\epsilon}_d - \tilde{\epsilon}_s = 7$ eV.  The Hubbard repulsions 
are set to canonical values with (in eV) $U_{dd} = 8$, $U_{pp} = 4$ and $U_{pd} = 1$.   
In Eq. (\ref{Eq:Meanfield}), $n_i^\alpha$ denotes the total number 
operator for orbital $\alpha$ and site $i$ and the paramagnetic 
solution for $\langle n^d_{i}\rangle$ has been assumed.  

The results for $A_O$, $E_z$ and $\lambda_{z,\phi}$ are shown 
in Fig. \ref{Fig:OxygenCharacter}.  The E-fields presented here  
are the same as those used in 
Fig. \ref{Fig:Efield}, and were obtained by assigning formal 
valences for each ion.  These field values  
reflect best the crystal field of the undoped parent compounds and 
therefore Fig. \ref{Fig:OxygenCharacter} does not include the effects 
of doping.  As we have already seen, even 
in this simple model, the redistribution of charge from the 
charge reservoir to the CuO$_2$ plane can alter the crystal 
field values resulting in large changes in the value of $\lambda_{z,\phi}$.  
In the case of Bi-2212 this produced an order of magnitude change 
in the value of $\lambda_{z,\phi}$ for the factor 4 change in $E_z$ in 
the doped lattice.   Furthermore, the ionic point charge model 
neglects any covalent nature of the bonds which will smooth 
the charge distributions.  Therefore, the 
field strengths in real materials may be quite different 
from the values reported here, especially in the doped systems.  
However, by using the same method for each material we can 
obtain valuable comparative information about the crystal field 
variations between materials.     

As shown in Fig. \ref{Fig:OxygenCharacter}a,  
there is a correlation between the material's maximum T$_c$ at optimal doping 
(apical distance) and the planar oxygen 
character $A_O$.  Again, this can be understood in terms of the increase in the 
effective planar O-O hopping {\it via} the Cu 4$s$ as the apical oxygen is further removed 
from the CuO$_2$ plane.  
A similar correlation, shown in Fig. \ref{Fig:OxygenCharacter}b, occurs 
between $E_z$ at the planar oxygen site and T$_c$ for the reasons 
previously discussed.   Since the total couplings  
$\lambda_{z,\phi}$ for the $B_{1g}$ branch are proportional to the 
product $A^4_oE_z^2$ a clear correlation between the material's T$_c$ and  
the $B_{1g}$ coupling naturally arises.  This is shown in Fig. \ref{Fig:OxygenCharacter}c.  

\begin{figure}[tr]
 \vskip 0.5cm
 \includegraphics[width=0.8\columnwidth]{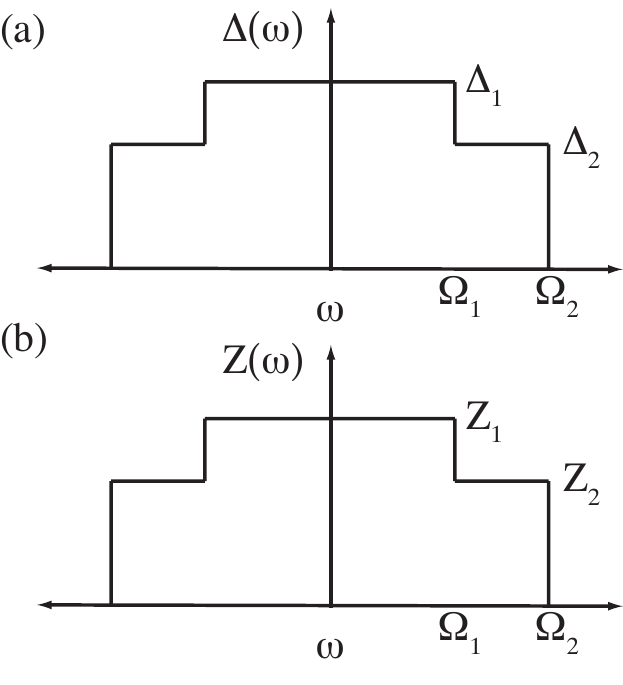}
 \caption{\label{Fig:DoubleWell} The double square well model for $\Delta(\omega)$ and 
 $Z(\omega)$.}
\end{figure}

The correlation between T$_c$ and the strength of the coupling to the $B_{1g}$ branch, 
which provides the largest contribution to $d$-wave pairing, provides 
a natural framework for thinking about materials variations in T$_c$. 
Although it is clear that the overall strength of the  
el-ph interaction is too small to support HTSC on its own, 
\cite{HeidPRL2008, BohnenEPL2003,GiustinoNature2008} phonons are not precluded 
from contributing to pairing, enhancing the pairing correlations from 
another dominant interaction.   
Such a possibility has been considered in previous works that have 
examined the contribution of el-ph coupling in 
conjunction with antiferromagnetic spin fluctuations.\cite{NunnerTc, BangPRB2008} 
In such a scenario, a baseline T$_c$ is set by the dominant pairing mechanism, 
typically identified with antiferromagnetic spin fluctuations.  
Such an interaction is likely to be governed by the properties of the CuO$_2$ plane, 
such as the charge transfer energy $\Delta$ or antiferromagnetic exchange energy $J$,  
and therefore the strength of this interaction is likely independent of the material.   
T$_c$ is then further enhanced by the weaker contribution from the el-ph 
interaction which, as we have shown, is strongly materials dependent. 
This picture naturally explains the variations in T$_c$ that are observed across 
materials and provides a microscopic connection between the structure and composition 
of the unit cell and the T$_c$ of the material.    

In the next section we continue examine such a multi-channel model and present simplified 
considerations for T$_c$ in order to assess the degree to which 
phonons can enhance a dominant interaction at high energy.  
As we will show, moderate values of $\lambda_{z,\phi}$, such as those given in 
Fig. \ref{Fig:OxygenCharacter}c, can lead to sizeable enhancement to T$_c$ which is 
greater than the T$_c$ that would result from phonons alone.  
Phonons are therefore capable of providing a sufficiently large enhancement 
to T$_c$ despite the relatively low value of their total coupling $\lambda_{z,\phi}$.  
  
\section{Considerations of Multiple Pairing Channels: T$_c$ and the Isotope Effect}

In this section we turn our attention to considerations for T$_c$ by considering 
contributions to pairing from multiple channels.  T$_c$, one of the most 
difficult quantity to calculate correctly, varies across the cuprate families 
by two orders of magnitude even though the same CuO$_2$ building block is present.  
However, as we have demonstrated, the total el-ph coupling for the $B_{1g}$ branch shows 
a strong materials dependence which is empirically correlated with the observed 
value of T$_c$ at optimal doping.  Therefore it is of interest to 
explore how el-ph interactions may provide a material dependence to T$_c$ arising from 
the material conditions along the $c$-axis and poor screening. 

To calculate T$_c$ we consider a multi-channel boson exchange model for pairing, 
which is a straightforward generalization of the Eliashberg equations.  
A similar approach has recently been used\cite{JohnstonPRB2010} 
to account for the qualitative structures in the phonon-modulated density of 
states of the cuprates as observed by STM.\cite{LeeNature2006} 
Implicit in this calculation is the assumption that the dominant pairing channel 
can be described in terms of a boson exchange mechanism and such a scenario is the 
simplest one which can be adopted in examining the idea of phonon-enhanced HTSC. 
However, such a model can be perhaps be justified based on several recent works. For example, 
studies of the penetration depth in electron-doped 
cuprates using a model with coexisting AFM and SC order has produced 
good agreement with experiment indicating the importance of AFM in 
these systems.\cite{DasPRL2007}   
Furthermore, a DMFT study of optical conductivity 
in a moderately correlated single-band Hubbard model,\cite{ComanacNatPhys2008} 
has concluded that AFM with moderate correlations is the correct picture 
for describing the underdoped cuprates rather than strongly correlated Mott physics. 
Finally, a study of the $t$-$J$ and Hubbard models  
has shown that the $d$-wave pairing in these models have a significant 
contribution from a retard component 
with an energy set by a few times $J$ or a fraction of $U$ in the $t$-$J$ and Hubbard models, 
respectively.\cite{MaierPRL2008}  
This result indicates that the pairing interaction in these models  
is largely retarded in nature with an energy scale set by the dynamics of the magnetic 
excitation spectrum.     
These theoretical studies lend some support to the spin-fluctuation mediated picture 
adopted here although the details may differ.  Nevertheless, this model 
provides a simple means to evaluate the role of phonons in modifying T$_c$ in 
a multi-channel model.

For this calculation we assume that each channel can be described in terms of a 
boson exchange mechanism. 
Individual channels are indexed by $\nu$ and the el-boson contributions to 
the mass renormalization $Z(\omega)$ and anomalous self-energy $\phi(\theta,\omega)$ 
are parameterized by $\lambda_{\nu,z}$ and $\lambda_{\nu,\phi}$, respectively. 
Making the standard approximations of a structureless band and cylindrical Fermi surface, 
the multi-channel Eliashberg equations are
\begin{widetext}
\begin{eqnarray}\label{Eq:Eliashberg} \nonumber
\omega[1-Z(\omega)]&=& \sum_{\nu} \frac{\lambda_{\nu,z}\Omega_\nu}{2} \int_0^\infty d\omega^\prime
 \left\langle \frac{\omega^\prime }{\sqrt{\omega^{\prime2} - \Delta^2(\omega)\cos^2(2\theta)}}\right\rangle K_{+}^\nu(\omega^\prime,\omega)\\
\Delta(\omega)Z(\omega)&=& \sum_{\nu} \frac{\lambda_{\nu,\phi}\Omega_\nu}{2} \int_0^\infty d\omega^\prime
 \left\langle \frac{\Delta(\omega^\prime)\cos^2(2\theta)}{\sqrt{\omega^{\prime2} - \Delta^2(\omega^\prime)\cos^2(2\theta)}}\right\rangle K_{-}^\nu(\omega^\prime,\omega).
\end{eqnarray}

\noindent Here we have assumed a separable form for the anomalous self-energy 
$\phi(\theta,\omega) = \phi(\omega)\cos(2\theta)$ and  
$\Delta(\omega) = \phi(\omega)/Z(\omega)$ is the complex gap function. 
The kernels in Eq. (\ref{Eq:Eliashberg}) are given by 

\begin{equation} \nonumber 
K^\nu_{\pm}(\omega^\prime,\omega) = [f(-\omega^\prime) + n(\Omega_\nu)] 
\bigg[
\frac{1}{\omega+\Omega_\nu+\omega^\prime}\pm 
\frac{1}{\omega - \Omega_\mu - \omega^\prime} 
\bigg] 
+[f(\omega^\prime)+n(\Omega_\nu)]
\bigg[
\frac{1}{\omega-\Omega_\nu+\omega^\prime}\pm 
\frac{1}{\omega+\Omega_\nu-\omega^\prime}
\bigg]
\end{equation}

\noindent
where $f$ and $n$ are the Fermi and Bose occupation factors, respectively.  Following 
Ref. \onlinecite{NunnerTc} we adopt a double square-well model for the frequency dependence of 
$\Delta(\omega)$ and $Z(\omega)$, as shown in Fig. \ref{Fig:DoubleWell}.  Here, we identify 
$\nu = 1$ with the $\Omega_1 \sim 35$ meV $B_{1g}$ phonon branch, and $\nu = 2$ is identified with 
antiferromagnetic spin fluctuations with $\Omega_2 = 2J \sim 260$ meV, reflecting the 
top of the magnon band in the case of a long-range antiferromagent.   Next, we 
follow a standard set of approximations\cite{CarbotteRMP} and obtain the coupled equations 
\begin{equation}\label{Eq:gap2}
Z_1\Delta_1=(\lambda_{1,\phi}+\lambda_{2,\phi})\int_0^{\Omega_1} d\omega^\prime 
\frac{\Delta_1}{\omega^\prime} \tanh\left(\frac{\omega^\prime}{2k_bT_c}\right) 
+ (\lambda_{1,\phi} + \lambda_{2,\phi})\int_{\Omega_1}^{\Omega_2} d\omega^\prime 
\frac{\Delta_2}{\omega^\prime} \tanh\left(\frac{\omega^\prime}{2k_bT_c}\right) 
\end{equation}
for $\Omega_1 > \omega \ge 0$ and  
\begin{equation}\label{Eq:gap1}
Z_2\Delta_2=\lambda_{2,\phi}\int_0^{\Omega_1}d\omega^\prime \frac{\Delta_1}{\omega^\prime} \tanh\left(\frac{\omega^\prime}{2k_bT_c}\right) + 
\lambda_{2,\phi}\int_{\Omega_1}^{\Omega_2}d\omega^\prime \frac{\Delta_2}{\omega^\prime} \tanh\left(\frac{\omega^\prime}{2k_bT_c} \right)
\end{equation}
for $\Omega_2 \ge \omega \ge \Omega_1$.
Here $Z_1 = 1 + \lambda_{1,z} + \lambda_{2,z}$ and $Z_2 = 1 + \lambda_{2,z}$.   
\end{widetext}
Eqs. (\ref{Eq:gap1}) and (\ref{Eq:gap2}) are then solved for the non-trivial solution 
for $\Delta_i$ and the resulting expression for T$_c$ is  
\begin{equation}
k_BT_c=1.134\hbar\Omega_1^{1-\gamma}\Omega_2^\gamma 
\exp\left(-\frac{1+\lambda_{1,z}+\lambda_{2,z}}{\lambda_{1,\phi}+\lambda_{2,\phi}} \right)
\end{equation}
where 
\begin{equation}\label{Eq:nu}
\gamma = \frac{\lambda_{2,\phi}}{1+\lambda_{2,z}}\frac{1+\lambda_{1,z}+\lambda_{2,z}}{\lambda_{1,\phi}+\lambda_{2,\phi}}
\end{equation}
%\begin{eqnarray*}
%k_BT_c&=&1.134\hbar\Omega_1\exp{\left( -\frac{Z_1}
%{\lambda_{1,\phi} + \lambda_{2,\phi}}\frac{Z_2 - 
%\lambda_{2,\phi}\ln\left(\frac{\Omega_{2}}{\Omega_{1}}\right)}{Z_2} \right)} \\
%&=&1.134\hbar\Omega_1^{1-\nu}\Omega_2^\nu \exp\left(-\frac{Z_1}{\lambda_{1,\phi}+\lambda_{2,\phi}} \right)
%\end{eqnarray*}
This expression recovers the McMillan result (we have neglected $\mu^*$) for the single 
well case in either limit when $\lambda_{1,z}$,$\lambda_{1,\phi}$ or 
$\lambda_{2,z}$,$\lambda_{2,\phi}$ are set to 
zero.  
%We also note that this expression differs slightly from the one obtained in 
%Ref. \onlinecite{NunnerTc} due to the frequency dependence of $Z$ assumed therein. 
Other expressions have been obtained for T$_c$ using different forms of the 
two-well model and neglecting the difference in contributions from each of the 
momentum channels (i.e. $\lambda_{z} = \lambda_{\phi}$) while taking the 
spin-fluctuation contribution to be repulsive.\cite{Tc,Tc2,DolgovPRL2005}  
We refer the reader to Ref. \onlinecite{DolgovPRL2005} for 
a more thorough discussion of the various proposed expressions. 

\begin{figure}[t]
 \includegraphics[width=0.8\columnwidth]{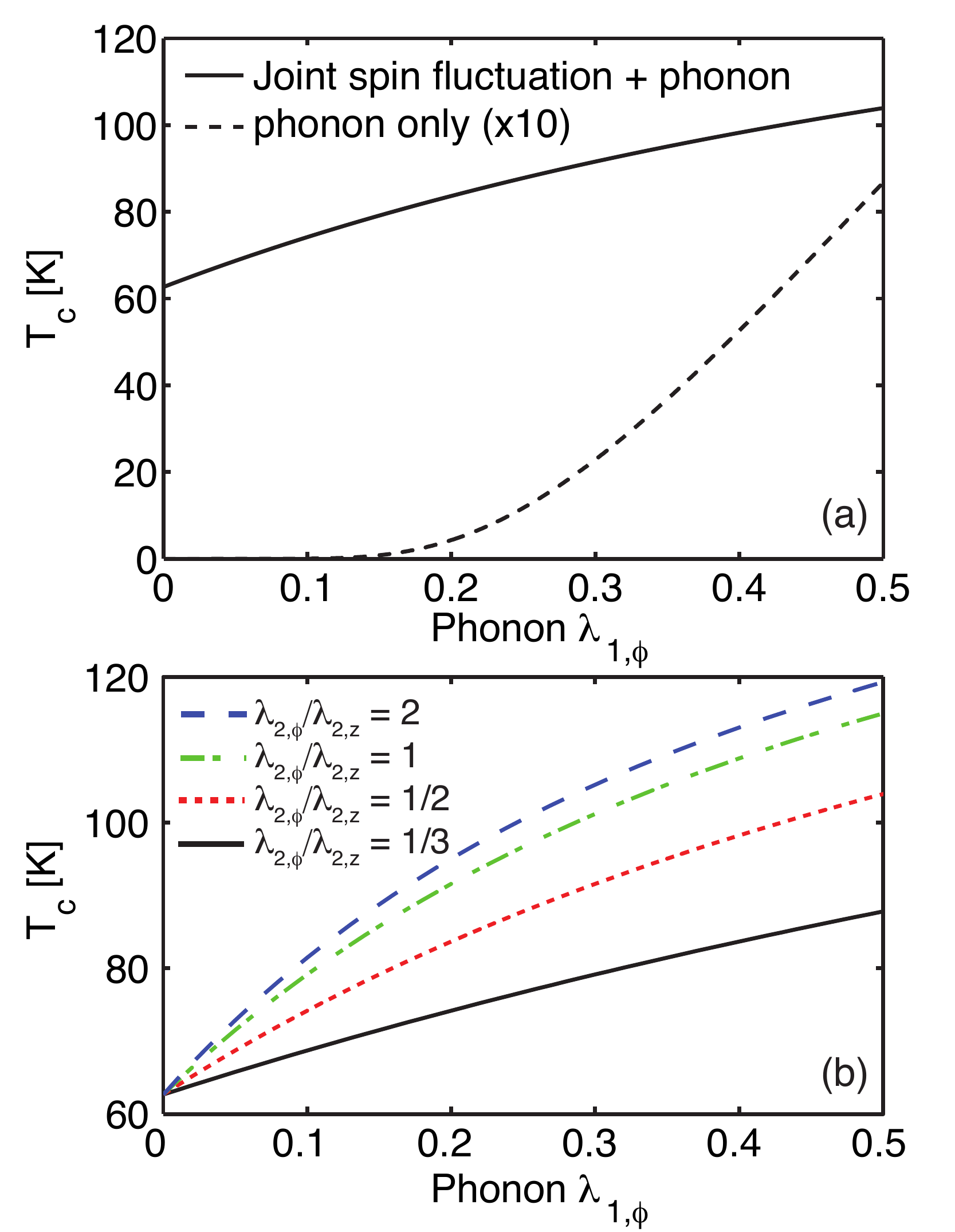}
 \caption{\label{Fig:Tc} The transition temperature T$_c$ as a function of el-ph coupling 
 in the $d$-wave channel $\lambda_{1,\phi}$, calculated in the two-square well model, 
 defined in the text.  The dashed line is T$_c$ expected for the phonon alone.}
\end{figure}

In order to assess the phonon contribution to T$_c$, we now assume that spin-fluctuations provide 
the dominant source for $d$-wave pairing and parameterize this mode with 
$\lambda_{2,z} = 2\lambda_{2,\phi} = 1$.  This choice results in a baseline 
T$_c = 62$ K in the absence of phonons.   
For the phonon mode we assume $\Omega_{1} = 36$ meV and hold $\lambda_{1,z} = 2\lambda_{1,\phi}$.  
In Fig. \ref{Fig:Tc}a the resulting T$_c$ as a function of el-ph coupling is presented.  
For reference, the value of T$_c$ obtained from phonons alone is also 
shown, multiplied by a factor of 10.  It is important to note that even small el-ph 
coupling can enhance T$_c$ considerably over the single-channel spin fluctuation model.  
Furthermore, the degree of enhancement depends on the ratio of 
$\lambda_{2,\phi}/\lambda_{2,z}$ assumed for the spin fluctuations. In Fig. \ref{Fig:Tc}b we  
plot T$_c$ for the two-mode model while varying ratios of $\lambda_{2,\phi}/\lambda_{2,z}$.  
In each case we have adjusted $\lambda_{2,\phi}$ such that the baseline of $T_c = 62$ K 
is maintained for $\lambda_{1,\phi} = 0$. In general, as $\lambda_{2,\phi}/\lambda_{2,z}$ increases 
in the dominant channel, the overall gain in T$_c$ mediated by phonons also increases but with 
diminishing returns for larger $\lambda_{2,\phi}/\lambda_{2,z}$. 
An isotropic repulsion in the spin fluctuation channel is also expected to produce a similar effect.  
We also note that our results are limited by the questionable applicability of Migdal-Eliashberg theory 
in the cuprates and 
therefore T$_c$ itself is a poor metric of a theory to be compared to experiment.  Nevertheless, 
as this enhancement is generic to multi-channel couplings, it may serve as a guide for 
understanding the material dependence of T$_c$ itself.  

\begin{figure}[t]
 \includegraphics[width=0.8\columnwidth]{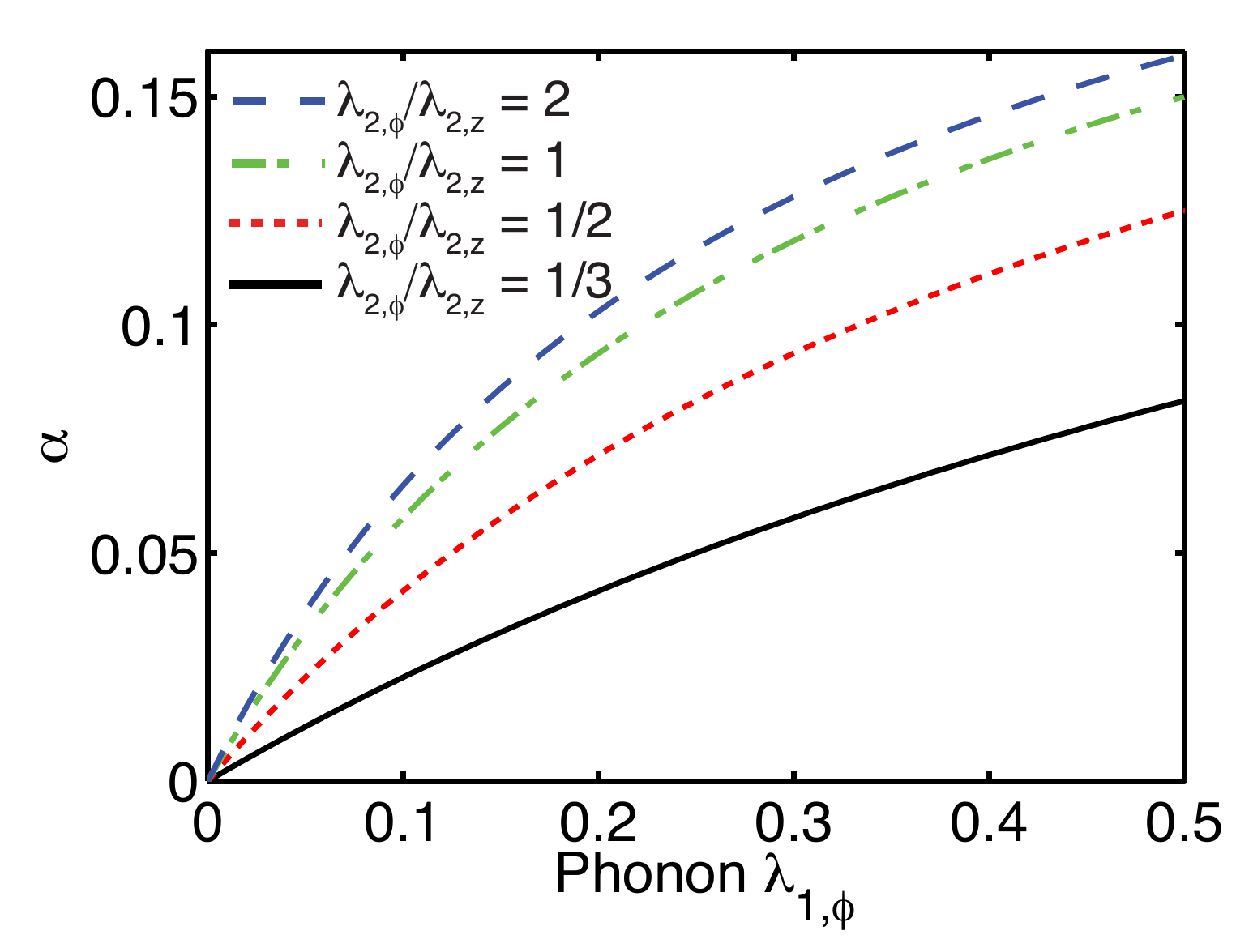}
 \caption{\label{Fig:alpha} The isotope exponent $\alpha$ as a function of 
 el-ph coupling $\lambda_{1,\phi}$ for the same parameter set used in Fig. \ref{Fig:Tc}.}
\end{figure}

In Fig. \ref{Fig:alpha} the isotope exponent $\alpha = -\partial \ln T_c/\partial \ln M$, 
where $M$ is the mass of the oxygen ion,  
is plotted using the same parameters as in Fig. \ref{Fig:Tc} with $\alpha = (1 - \gamma)/2$, 
and $\gamma$ defined by Eq. (\ref{Eq:nu}), 
reducing the overall magnitude of $\alpha$ from the value for phonons alone.  A similar 
reduction in $\alpha$ and enhancement of T$_c$ has been reported in Ref. 
\onlinecite{BangPRB2008} when the phonon contribution to pairing was taken to be 
attractive in the $d_{x^2-y^2}$ channel.    
Fig. \ref{Fig:alpha} shows that the overall isotope exponent is small even though 
the phonon mediated enhancement to T$_c$ is sizeable and demonstrates that 
small isotope exponents are possible despite large 
enhancements in T$_c$ due to the el-ph interaction.   

Some degree of caution is warranted in making a direct comparison to experiment 
where the isotope exponent is small at optimal doping and increases with 
underdoping,\cite{isotope} while our results show that the overall contribution of el-ph coupling 
to $d$-wave pairing is expected to increase with doping due to screening effects.   
Therefore one might expect that the isotope exponent should increase with doping, 
contrary to what has been observed experimentally.  However, doping dependent changes are also  
likely to occur for a spin-fluctuation mediated pairing mechanism.  Since it is the relative contributions 
from each channel that sets the value of $\alpha$ a direct calculation of $\alpha$ expected 
experimentally  is not possible until the details of the dominate pairing interaction are 
understood. Nevertheless, these calculations show that phonons with relatively low 
el-ph coupling strengths can, in principle, play a substantial role in determining T$_c$ while 
producing small signatures in traditional metrics such as the isotope exponent. 

We end with a few comments about these results in relation to 
prior works.  As we have already noted, the specific results presented are limited by 
the applicability of Eliashberg theory and the Fermi liquid description of the cuprates.  
Such a picture may not be completely applicable, especially in the underdoped region of the 
phase diagram where the role of correlations becomes more prominent.  
As the role of correlations and its interplay with the el-ph interaction is better understood 
the details of the picture presented here 
may change.  However, the phonon-mediated enhancement of T$_c$ is expected remain as 
a general phenomenon.  

We also note that the inclusion of correlations may 
further enhance the phonon contribution to T$_c$ through possible enhancements of the el-ph 
vertex or proximity to AFM.   For example, Recent DMFT work examining phonon-mediated pairing in the 
fullerides\cite{CaponeScience2002,CaponeRMP2009} has found that a phonon-mediated pairing 
mechanism, which produces an exponentially small T$_c$ when considered on its own, will be significantly 
enhanced in proximity to a metal-insulator transition.  The mechanism via which this occurs  
is a density of states enhancement due to a narrowing of the quasi-particle band and Mott physics.  
This is a different origin than the enhancement reported here where we are 
concerned with the overall effects of coherence and band width of the paired 
electrons.  Some aspects of the Mott physics is included in our considerations 
in the form of the breakdown of metallic screening along the $c$-axis (a Coulomb effect) however, 
we have not included the breakdown of carrier properties in the CuO$_2$ plane.  
This is an intriguing possibility but it is beyond the scope of our paper and we 
leave it for future work.  
% here which is due to a boot strapping 
%of the phonon contribution by the dominant mode.  This effect is independent of the 
%the origin of the dominant mode and does not depend on the proximity of AFM.  (This of course neglects any role that 
%the proximity to the AFM phase plays in establishing the existence of the dominant mode.) 
%It is therefore possible that these two effects could add in a non-trivial way. 

Finally, it has recently shown that high values of T$_c$ are also obtainable 
from el-ph coupling without invoking a second dominant pairing interaction 
by departing from the conventional Fermi liquid framework.   
In Ref. \onlinecite{ShePRB2009} it was shown that T$_c$'s on the order of 100 K 
are attainable for $\lambda_{z,\phi}$ values similar to those obtained here, 
provided the normal state behaves as a non-Fermi-liquid quantum critical metal.  
Although the simple approach outlined here does not hold in this scenario, 
much of the doping and materials-dependent variations in coupling still do. 

\section{Summary and Conclusions}
Although there appears to be a great deal of universality in the hole-doped cuprates,  
large changes in T$_c$ can be achieved through changes in the chemical composition and 
structure of the unit cell.  As we have shown here, these changes are reflected in a number 
of properties including the orbital character of the band crossing the Fermi level, 
the strength of local electric fields arising from structural-induced symmetry breaking, 
doping dependent changes in the underlying band structure, and ionicity of the 
crystal governing its ability to screen $c$-axis perturbations.  
Given the sensitivity of T$_c$ to the structural details of the 
crystal it is clear that the underlying mechanism(s) for HTSC must incorporate these 
elements in some way. In this work we have examined a number of aspects of coupling to 
oxygen modes and demonstrated that the overall coupling to these modes is 
quite sensitive to these factors.  Therefore, the inclusion of el-ph coupling to 
these modes provides a natural means of linking the electronic properties of the 
CuO$_2$ plane to the structural elements of the material lying off plane and 
the carrier concentration. This picture calls for oxygen phonons to play a role in HTSC. 

In extending the previous works by some of the authors on el-ph coupling in the cuprates, we 
have formulated a theory for poor screening in these materials.  Due to the quasi-2D nature 
of the cuprates with poor conductivity along the $c$-axis, we have shown that a window 
of small $\bq$ opens in which screening is inoperable.  This results in an overall enhancement 
of the total $d$-wave projected coupling which enhances the phonon's ability to mediate 
$d$-wave pairing.  With progressive doping, the total phonon contribution to $\lambda_\phi$ is 
enhanced while the total contribution from $\lambda_z$, which suppresses $d$-wave pairing, is 
screened away.  As a result, doping the system away from half-filling results in 
an increased contribution to $d$-wave superconductivity mediated by el-ph coupling.  

A systematic examination of the variation in el-ph coupling across the cuprate 
families was also performed. 
Using experimental structural data we determined the materials variation in the total 
electric field strength and orbital character of the $pd$-$\sigma^*$ band at the Fermi 
level and the expected variations in el-ph coupling strength to the $B_{1g}$ branch.
A direct correlation between the strength of the coupling and the optimal 
T$_c$ for each material was observed.  As a result, the materials variation in 
T$_c$ can be naturally accounted for through the inclusion of el-ph coupling to 
the presently accepted intrinsic planar models.        

To be clear, we do not suggest that el-ph coupling alone accounts for HTSC 
as both the bare and screened values for the el-ph coupling strength are insufficient to 
produce large T$_c$'s on their own.   Instead, we propose that the phonons work in conjunction 
with the presently unidentified dominant mechanism for HTSC.  
The simplified multi-channel model for T$_c$ presented here has demonstrated that moderate el-ph 
coupling, when combined with spin fluctuations, can produce enhancements to the 
total T$_c$ in excess of the contribution produced by phonons alone.  Furthermore, this 
enhancement is expected to be a generic phenomena and independent of the presently 
second mechanism.  Therefore, the 
inclusion of the el-ph interaction can account for the large variations in T$_c$ observed 
across the cuprate as a function of material and doping.  

There are some open issues regarding el-ph coupling in the cuprates.  
First of all, our examination of the field strength as a function of material neglected any 
doping-induced changes to the field strengths.  Our simplified treatment of the 
bi-layer system Bi-2212 showed that these effects can be substantial.  Therefore, this 
problem requires a  
systematic examination of the doping process across the families of cuprates, perhaps 
with more sophisticated approaches.  
This issue is complicated in the multi-layer systems with $n > 2$ where  
there is the possibility for inequivalent dopings for the CuO$_2$ planes.\cite{YChenPRL2009} 
This can further modify charge distributions altering the strength of 
the crystal fields in each plane.\cite{JohnstonCondmat}  Any systematic treatment of the doping 
induced fields should take this into account.  
There is also the additional question of the role of covalency in 
determining the overall field strength.   

Finally, we have investigated the role of el-ph coupling in the cuprates, taking 
into account the long range Coulomb interaction and its role in screening.  
However, the role of the short-range Coulomb interaction and its affect on the el-ph 
interaction remains an open and intriguing question.  We believe that the findings 
presented in this paper serve to further highlight the need for progress in this area.

\acknowledgments
The authors would like to acknowledge useful discussions with A. Balatsky, O. Gunnarsson,  
G. Sangiovanni, I. Mazin, R. Hackl and D. J. Scalapino.  
T. P. D. would like to acknowledge support from NSERC, ONR grant N00015-05-1-0127 and the 
A. von Humboldt Foundation.  The Stanford work is also supported by DOE contract 
DE-AC03-765F00515, NSF grant DMR-0304981 and ONR grant N00014-01-1-0048.  
S. J. would like to acknowledge support from NSERC and SHARCNET. N. N. is supported 
by the Grant-in-Aids for Scientific Research (No. 17105002, 19019004, 
19048008, 19048015, and 21244053) from the Ministry of Education, 
Culture, Sports, Science and Technology of Japan, and also by 
Funding Program for World-Leading Innovative R\&D on Science and Technology (FIRST Program)．

\appendix
\section{Modifications of the Zhang-Rice Singlet}\label{Sec:Cluster}
Throughout this work the role of electronic correlations was neglected in deriving 
the anisotropic el-ph couplings arising from different charge transfer processes.  In this 
appendix an analysis via a different route is undertaken by examining directly the role played by 
correlations.  Previous studies on Hubbard clusters have illustrated the usefulness of exact diagonalization 
for exploring low energy excitations, such as the Zhang-Rice singlet (ZRS),\cite{ZR} as a function of hole number.  
In particular, these studies have demonstrated a strong dependence of the ZRS energy and mobility, as well 
as the magnetic exchange $J$, on the planar charge transfer energy $\Delta = \epsilon_p - \epsilon_d$.
\cite{EskesPRB1993}  Ref. \onlinecite{OhtaPRB1991} pointed out that the stability of the ZRS 
increases if the apical atom lies further away from the CuO$_2$ plane, which reduces the mixing 
of the ZRS with other low energy states of the cluster.  Since the phonons under consideration 
directly modulate $\Delta$ and the apical Madelung energy, strong modifications should be seen in 
the ZRS and magnetic exchange.  The cell perturbation approach in Ref. \onlinecite{PiekarzPRB1999}, 
which included in-plane oxygen breathing and buckling modes, showed that an effective simple 
Hubbard-like Hamiltonian can be deduced with renormalized parameters.  

\begin{figure}[t]
 \includegraphics[width=0.75\columnwidth]{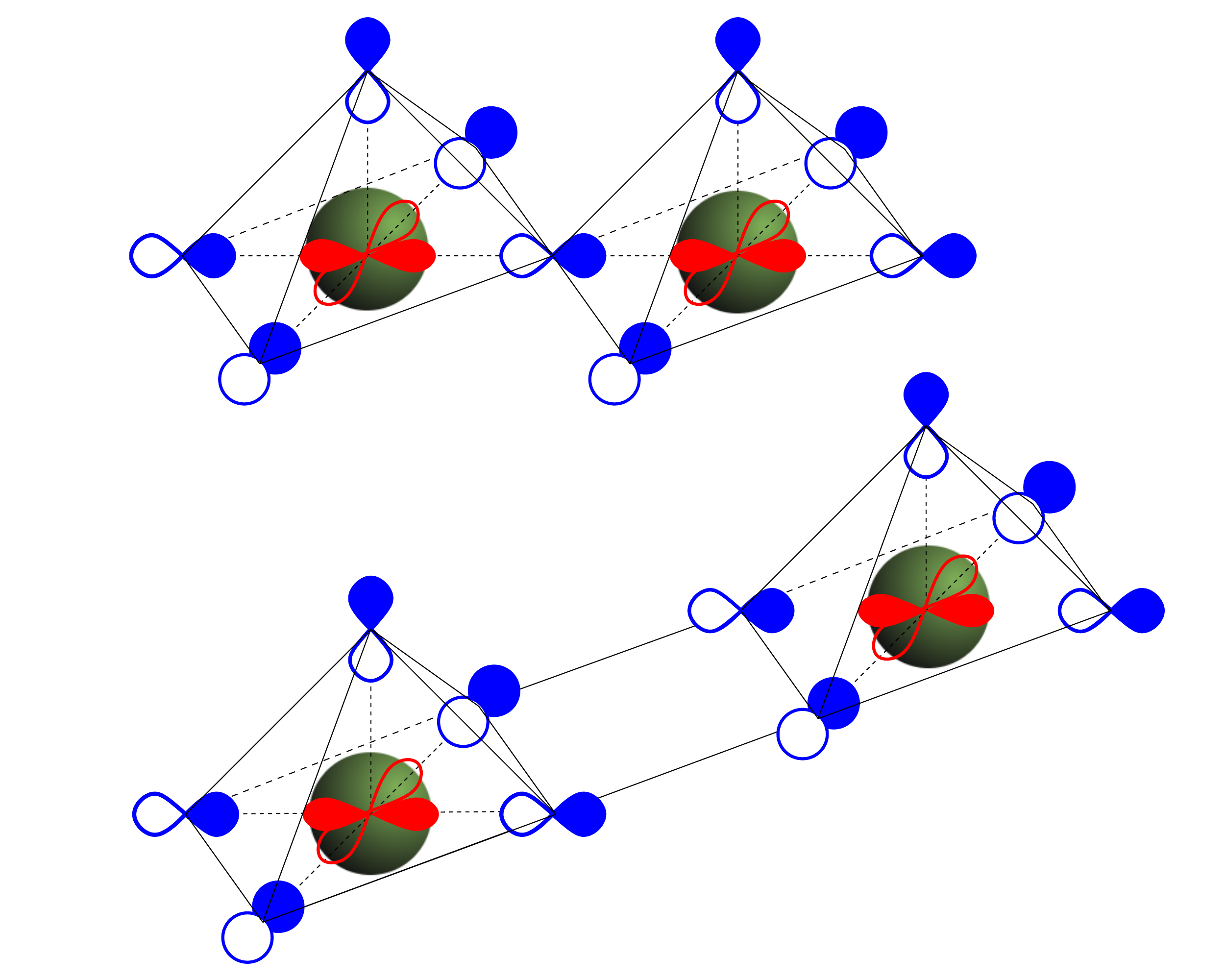}
 \caption{\label{Fig:Cluster} 
 (Color online) Cu$_2$O$_9$ and Cu$_2$O$_{10}$ clusters used to compute the 
 modifications of the effective Zhang-Rice singlet hopping $t$ and $t^\prime$ and magnetic coupling 
 $J$ and $J^\prime$.}
\end{figure}

Here, exact diagonalization studies of multi-band Hubbard clusters are used to investigate how the 
low energy sector is modified by local, static atomic displacements in a pattern given by various 
phonon eigenvectors.  The modifications of the ZRS and $J$ due to el-ph coupling receive close 
attention in order to derive renormalized parameters for single-band Hubbard to $t$-$J$ model 
Hamiltonians.

The clusters show in Fig. \ref{Fig:Cluster} contain two sets of orbitals at the Cu site: 
3$d_{x^2-y^2}$ and 4$s$.  By symmetry, the hopping amplitude from the apical oxygen site to the 
Cu 3$d_{x^2-y^2}$, $t_{ad}$ vanishes, while $t_{as}$ is non-zero.  Thus, a general Hubbard model 
can be written as $H = H_{kin} + H_{pot}$ where the kinetic piece is
\begin{eqnarray}\nonumber
H_{kin}&=& t_{pd}\sum_{i,\sigma} Q_\delta[d^\dagger_{i,\sigma}p_{i,\delta,\sigma} + h.c.] \\\nonumber
&+& t_{pp}\sum_{i,\delta,\delta^\prime,\sigma} Q^\prime_{\delta,\delta^\prime} 
p^\dagger_{i,\delta,\sigma}p_{i,\delta^\prime,\sigma} \\\nonumber
&+& t_{ap}\sum_{i,\delta,\sigma} Q^{\prime\prime}_{\delta}[a^\dagger_{i,\sigma}p_{i,\delta,\sigma} + h.c.] \\\nonumber
&-& t_{as}\sum_{i,\sigma}[a^\dagger_{i,\sigma}s_{i,\sigma} + h.c.] \\ 
&-& t_{ps}\sum_{i,\sigma}Q^{\prime\prime}_{\delta}[s^\dagger_{i,\sigma}p_{i,\delta,\sigma} + h.c.]
\end{eqnarray}
with phase factors $Q_{\pm x} = -Q_{\pm y} = \mp 1$, $Q^\prime_{\pm x,\pm y} = 1 = Q^\prime_{\pm x,\mp y}$, 
$Q^{\prime\prime}_{\pm x} = Q^{\prime\prime}_{\pm y} = \pm 1$, 
and the potential piece is given by
\begin{eqnarray} \nonumber
H_{pot}&=&U_{dd}\sum_i \hat{n}^d_{i,\uparrow}\hat{n}^d_{i,\downarrow} + 
U_{pp} \sum_{i,\delta} \hat{n}_{i,\delta,\uparrow}^p \hat{n}^p_{i,\delta,\downarrow} \\\nonumber
&+& \epsilon_d\sum_{i,\sigma}\hat{n}_{i,\sigma}^d + \epsilon_s\sum_{i,\sigma}\hat{n}^s_{i,\sigma} \\\nonumber 
&+&\sum_{i,\sigma}(\epsilon_a + \vec{U}_a\cdot\vec{E_a})\hat{n}^a_{i,\sigma} \\
&+&\sum_{i,\delta,\sigma}(\epsilon_p + \vec{U}_{i,\delta}\cdot\vec{E_p})\hat{n}^a_{i,\sigma}.
\end{eqnarray}
Here, $a$, $p_\delta$, $d$, $s$ ($a^\dagger$,$p_\delta^\dagger$, $d^\dagger$, $s^\dagger$) annihilate (create) a 
hole on apical oxygen, in-plane oxygen ($\delta = x,y$), copper $d$- and copper $s$-orbital, respectively, 
$\vec{U}$ is the displacement of a given ion and $\vec{E}$ is the associated local electronic field.  
This type of coupling is different than that considered in previous studies, where the buckling modes were 
represented as an electrostatic modulation of the Cu site energies.\cite{PiekarzPRB1999}  Since the 
4$s$ orbitals are rather extended we neglect the on-site Coulomb repulsion as well as inter-orbital 
interactions with the 3$d_{x^2-y^2}$ orbitals.  

There are a total of 13 and 14 orbitals for the Cu$_2$O$_9$ and Cu$_2$O$_{10}$ clusters, respectively.  
In both cases, the clusters contain seven holes in order to investigate the formation and delocalization of 
a ZRS to a nearest neighbour copper-oxide plaquette (4 holes being almost exclusively located in the Cu 4$s$ 
orbitals).  Exact diagonalizations in the $S_{tot}^z = 1/2$ sector, which contains the ground state, 
are performed since the Hamiltonian conserves $S_{tot}^z$.  Parameters similar to those of Ref. 
\onlinecite{OhtaPRB1991} for the Hubbard model are used, with values representative of the electric 
field for the planar and apical oxygen sites in bilayer cuprates (in eV and eV/\AA for the fields):
\begin{eqnarray*}
t_{pd} = 1.13 \quad\quad t_{ps} = 2\\
t_{pp} = 0.49 \quad\quad t_{ap} = 0.29\\
t_{as} = 2    \quad\quad \epsilon_s = -7 \\
\epsilon_{p} = \epsilon_a = 2.9 \quad\quad U_{pp} = 4.1  \\
U_{dd} = 8.5 \quad\quad E_p = 1.6 \\
E_{a} = 16
\end{eqnarray*}

The diagonalization of the Cu$_2$O$_9$ cluster yields the effective hopping parameter $t$ of the 
ZRS as well as the magnetic exchange $J$: the splitting between the groundstate and first excited 
state gives $2t$, whereas the singlet-triplet splitting of the 6-particle problem gives $J$.  
The parameters $t^\prime$ and $J^\prime$ are analogously defined from the Cu$_2$O$_{10}$ cluster.  
In the absence of el-ph coupling we obtain the following values:  $|2t| = 0.63$ eV, 
$|2t^\prime| = 0.28$ eV, $J = 0.17$ eV, and $J^\prime$ = 15 meV.

\begin{figure}[tl]
 \includegraphics[width=\columnwidth]{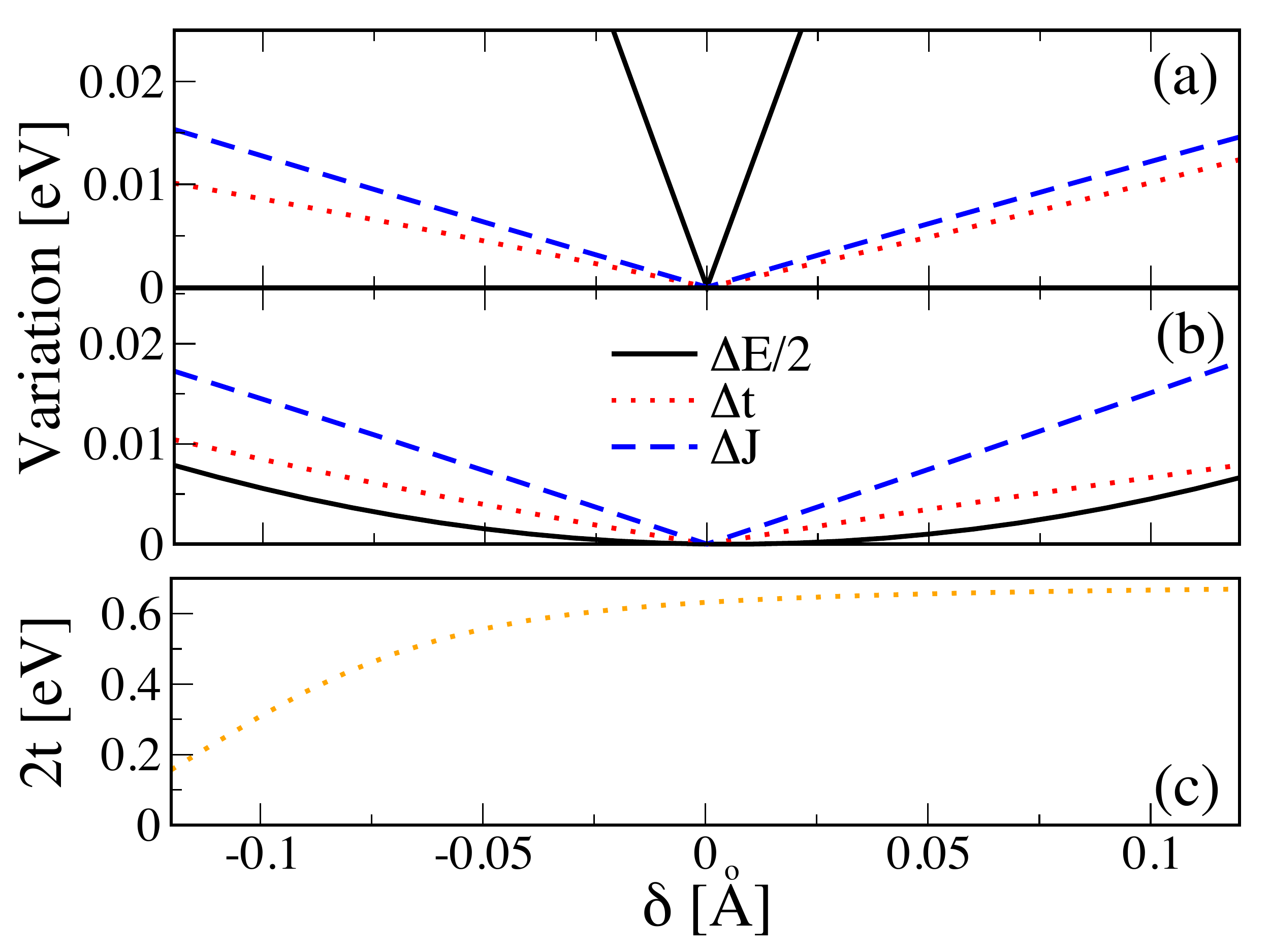}
 \caption{\label{Fig:EDresults} (Color online) The effect of different modes on various Zhang-Rice parameters for 
 the Cu$_2$O$_9$ cluster: plots (a) and (b) show the effect of a given mode on the ground-state 
 energy and on the effective ZRS hopping $t$ with $\Delta t = |t(\delta)-t(\delta=0)|$ and 
 $\Delta E = E(\delta) - E(\delta=0)$.  (a) The effect of the $A_{1g}$ modes.  (b) The 
 effect of the $B_{1g}$ mode.  (c) The effect of the apical mode on $2t$. }
\end{figure}

Fig. \ref{Fig:EDresults} displays the effect of different phonon modes on the ground state 
energy, $t$ and $J$ as a function of static displacement $\delta$.  Although only static 
atomic displacements have been investigated in configurations corresponding to the phonon 
eigenvectors, the differences between the relevant time scales associated with the electron 
hopping and atomic motion justify such an analysis for the purpose of viewing a snap-shot of 
the modifications of the ZRS parameters.  (We also note that Ref. \onlinecite{JohnstonEPL2009} 
considered a similar calculation with fully quantized atomic motion for the $B_{1g}$ and 
$A_{1g}$ phonons at $\bq = 0$ and reported similar results for $J$.)  One clearly sees, for 
small displacements, the main effect on $t$ is due to the $A_{1g}$ and $B_{1g}$ modes.  
For the $B_{1g}$ mode, the effective hopping of the ZRS is modulated to first order in the 
displacement, while the energy is only modified at second order, in agreement with 
perturbative analysis. For the $A_{1g}$ mode the first-order correction to the energy does not 
vanish and both $\Delta E$ and $\Delta t \propto \delta$.  The magnitude of the modulation in 
$t$ is comparable to the $B_{1g}$ mode with the same local field.  The effect of the 
apical mode on $t$ is very weak for $\delta > 0$.  This is not surprising since the ZRS 
has little apical character if the orbital lies away from the CuO$_2$ plane.  For $\delta < 0$ 
the hybridization of the apical-$p$ and Cu-$d_{3z^2-r^2}$ orbitals destabilizes the ZRS and 
thus the effective hopping $t$ drops as in an earlier work.\cite{OhtaPRB1991}  Also, both 
the ground state energy $E$ and the effective exchange $J$ are not affected much by this mode.  

\begin{figure}[tr]
 \includegraphics[width=\columnwidth]{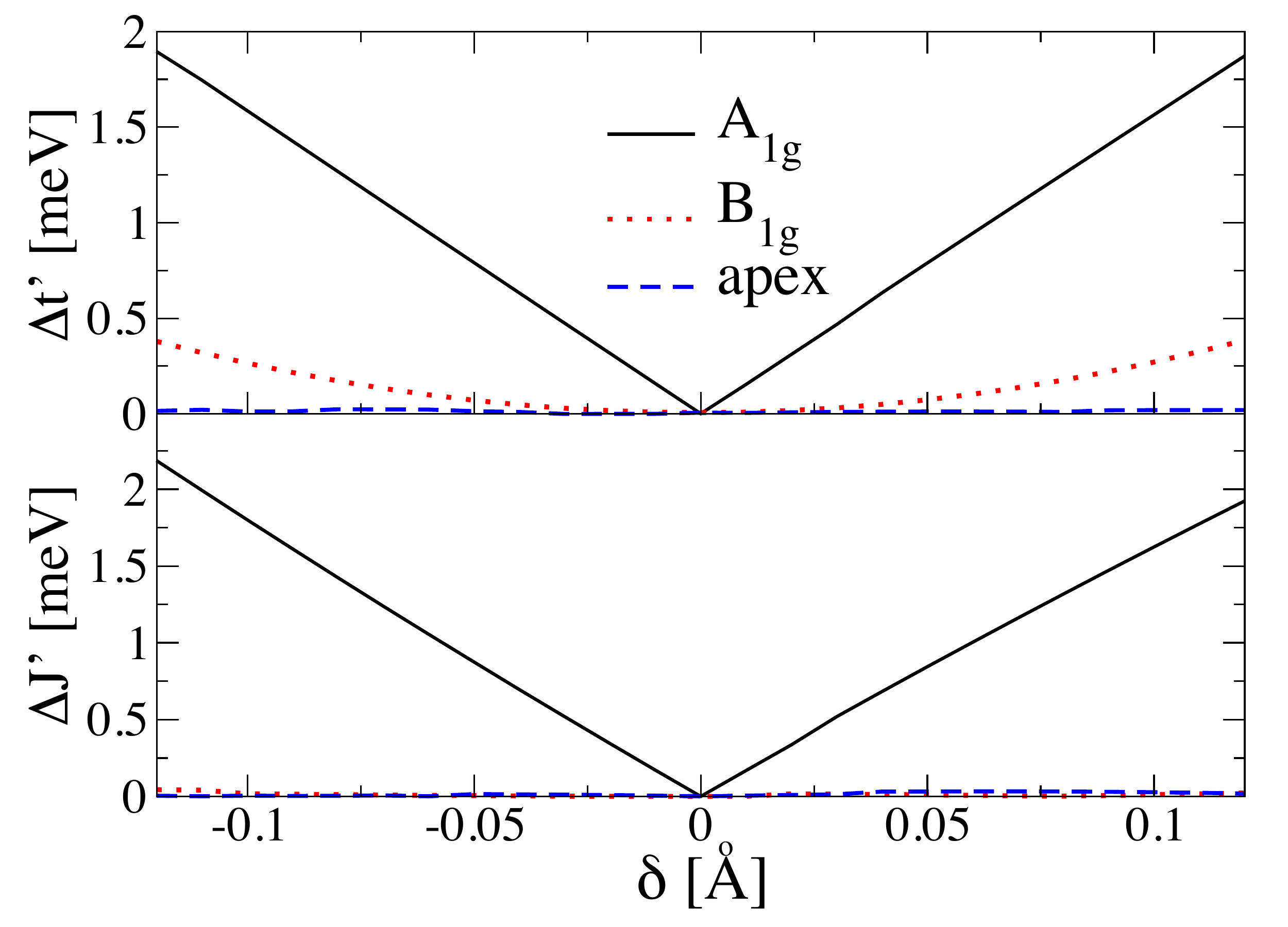}
 \caption{\label{Fig:EDresults2} 
 (Color online) 
 Variation of the effective Zhang-Rice hopping $t^\prime$ and 
 magnetic exchange $J^\prime$ to next-nearest neighbor.  The calculation performed for the 
 Cu$_2$O$_{10}$ cluster for $A_{1g}$-, $B_{1g}$- and apex-modes.}
\end{figure}

It is clear, based on the calculations performed, that the three modes modify the 
effective hopping of the ZRS.  However, the mechanism by which the apical mode affects the 
hopping is very different since the formation of the ZRS no longer takes place in 
the CuO$_2$ plane if the displacement of the apical oxygen is towards the plane.  
The effective next-nearest neighbour quantities $t^\prime$ and $J^\prime$ (see Fig. 
\ref{Fig:EDresults2}) are affected less by the two planar modes.  

Summarized in Figs. \ref{Fig:EDresults} and \ref{Fig:EDresults2}, our results indicate 
that for both the $A_{1g}$ and $B_{1g}$  phonons, $\Delta t$ as well as $\Delta J$ 
corrections have a linear dependence on displacement of about the same magnitude.  For the 
apical phonon the magnitude is different due to the destabilization of the ZRS.  The on-site 
energy is modified mainly by the $A_{1g}$ ep-ph coupling; which can be quantified by looking 
at the corresponding modification of the ground-state energy.  Modifications to parameters 
$\Delta J^\prime$ and $\Delta t^\prime$ indicate again that the dominant contribution comes 
from the $A_{1g}$ mode.  However, this conclusion is based on a naive cluster approach and 
should be corroborated by other means of investigation.  

In terms of treatments of effective single-band models, such as Hubbard or $t$-$J$, 
incorporating phonons, we note that our results indicate that the effective couplings for 
these modes are neither purely Holstein-like nor can they be treated as bond phonons 
simply modifying the effective ZRS hopping $t$:  the effective parameters, such as $t$ and $J$, 
and as a consequence, the Mott-Hubbard effective interaction $U$ in a single-band approach, 
are all modified.  The actual modification of the effective $U$ for static displacements 
corresponding to these phonon modes requires an investigation of different three-band 
clusters containing Cu $d^8$, $d^9$ and $d^{10}$ and oxygen ligand configurations.  This points 
out the complications involved in ascribing real phonon modes to any corresponding model 
phonons in a single-band approach, and illustrates the important interplay between el-ph 
coupling and electronic correlations.


\begin{thebibliography}{99}
\bibitem{SiteEnergy}
To estimate the magnitude of the apical coupling we have determined, for Bi-2212, 
the Madelung energy difference $\Delta\Phi = 10.09$ eV, the dielectric constant 
$\epsilon(\infty) = 3.5$, giving $\epsilon_{apex} = \epsilon_p + \Delta\Phi/\epsilon(\infty) 
= \Delta + 2.88$ eV.  Mean field corrections to the site energies then shift these values. 
This is derived in section VI.
\end{thebibliography}
\end{document}